\documentclass[manuscript,10pt]{acmart}

\setcopyright{none}
\renewcommand\footnotetextcopyrightpermission[1]{}
\usepackage{booktabs}   
\usepackage{caption}
\usepackage{subcaption} 
\usepackage{xspace}
\usepackage{wrapfig}
\usepackage{listings}

\long\def\ignore#1{}

\usepackage{amssymb,bbold}
\usepackage{mathtools,thm-restate}
\usepackage{extpfeil}
\usepackage{shuffle}
\usepackage[scaled=.84]{helvet}

\usepackage{pdfpages}
\usepackage{hyperref}

\usepackage[utf8x]{inputenc}

\usepackage{mynatproof}
\makeatletter
\newbox\sf@box

\lstdefinelanguage{SOP}
{morekeywords={and, case, do, done,
    end, else, if, in, last, let, new, object, pre, returns,
    then, true, when, where, with},
  morecomment=[l]{\%},
  numbers=none,
  frame=none
}

\newcommand{\tcode}[1]{{\small \texttt{#1}}}

\newcommand{\mi}[1]{\ensuremath{\mathit{#1}}\xspace}

\lstdefinelanguage{Esterel}
{morekeywords={abort, and, await, call,
case, combine, constant, do,
    each, emit, else, elseif, end, every, exec, exit, function, halt,
    handle, if, immediate, in, init, input, inputoutput, loop, module, not,
    nothing, or, output, pause, positive, pre, present, procedure,
    relation, repeat, return, run, sensor, signal, suspend, sustain,
    task, then, times, trap, type, unsigned, var, weak, when, with},
  morecomment=[l]{\%},
  emphstyle=\bfseries,
}

\lstdefinelanguage{SCL}
{language=C,
alsolanguage=Esterel,
morekeywords={bool, true, false, fork, join, input, output, par, module, pause,
if, then, else},
morecomment=[l]{//},
deletestring=[b]',
}

\lstnewenvironment{code}
    {\lstset{}%
      \csname lst@SetFirstLabel\endcsname}
    {\csname lst@SaveFirstLabel\endcsname}
    {}
\lstnewenvironment{spec}
    {\lstset{}%
      \csname lst@SetFirstLabel\endcsname}
    {\csname lst@SaveFirstLabel\endcsname}
    {}

\lstdefinestyle{hs}{
      basicstyle=\small\ttfamily,
      flexiblecolumns=false,
      morecomment=[l]\--,
      basewidth={0.5em,0.45em}, escapechar=@,
      literate={+}{{$+$}}1 {/}{{$/$}}1 {*}{{$*$}}1 {=}{{$=$}}1
               {>}{{$>$}}1 {<}{{$<$}}1 {\\}{{$\lambda$}}1
               {\\\\}{{\char`\\\char`\\}}1
               {->}{{$\rightarrow$}}2 
               {<=}{{$\leq$}}2 {=>}{{$\Rightarrow$}}2 
               {\ .}{{$\circ$}}2 {\ .\ }{{$\circ$}}2
               {|}{{$\mid$}}1 
               {-<}{{$\leftY$}}1
               {_}{{$\_$}}1 
               {\$}{{\$}}1
               {!!!}{{\textcolor{red}{!!!}}}3
    }

\definecolor{orange}{RGB}{255,127,0}
\definecolor{quest}{RGB}{0,155,0}
\definecolor{mark}{RGB}{138,43,226}

\newcommand{\green}[1]{}

\newcommand{\tsf}[1]{\textsff{#1}}

\newcommand{\domain}{\ensuremath{\mathit{dom}}\xspace}

\newcommand{\textsff}[1]{\textsf{#1}}

\newcommand{\ttpause}{\ensuremath{\mathop{\tcode{pause}}}\xspace}
\newcommand{\ttkill}{\ensuremath{\mathop{\tcode{kill}}}\xspace}
\newcommand{\ttTrue}{\ensuremath{\mathop{\tcode{True}}}\xspace}
\newcommand{\ttFalse}{\ensuremath{\mathop{\tcode{False}}}\xspace}

\newcommand{\ttreturn}{\ensuremath{\mathop{\tcode{return}}}\xspace}
\newcommand{\ttdone}{\ensuremath{\mathop{\tcode{done}}}\xspace}
\newcommand{\ttatomic}{\ensuremath{\mathop{\tcode{atomic}}}\xspace}
\newcommand{\ttmethod}{\ensuremath{\mathop{\tcode{method}}}\xspace}
\newcommand{\ttretry}{\ensuremath{\mathop{\tcode{retry}}}\xspace}

\newcommand{\ttpar}[2]%
    {\ensuremath{\mathrel{{}_{#1}\mathtt{||}{}_{#2}}}\xspace}
\newcommand{\ttfork}%
    {\ensuremath{\mathrel{\tcode{|\!|\!|}}}\xspace}

\newcommand{\ttseq}%
    {\ensuremath{\mathrel{\lhs{>>>}}}\xspace}
\newcommand{\ttbind}%
    {\ensuremath{\mathrel{\lhs{>>>=}}}\xspace}

\newcommand{\ret}{\ensuremath{\tcode{ret}}\xspace}

\newcommand{\abswr}%
  {\ensuremath{\mathrel{\raisebox{0.5pt}{\sfi}\!\!\texttt{=}}}\xspace}

\mathchardef\rcol="303A         
\newcommand{\rrcol}{\ensuremath{\mathrel{::}}\xspace}

\newcommand{\preste}[3]%
   {\ensuremath{\tcode{pres}\, #1\, \tcode{then}\, #2\, \tcode{else}\, #3}\xspace}

\newcommand{\ttif}{\ensuremath{\tcode{if}}\xspace}

\newcommand{\ttthen}{\ensuremath{\tcode{then}}\xspace}
\newcommand{\ttelse}{\ensuremath{\tcode{else}}\xspace}

\newcommand{\cseq}{\ensuremath{\texttt{;}}\xspace}
\newcommand{\cpar}{\ensuremath{\mathrel{\mid}}}

\newcommand{\sfi}{\ensuremath{\textsf{i}}\xspace}
\newcommand{\derives}[2]{\ensuremath{\mathrel{\rightarrow^{#1}_{#2}}}}

\newcommand{\lDerives}[2]{\mathrel{\xRightarrow{#1}_{#2}}}
\newcommand{\lderives}[2]{\mathrel{\xrightarrow{#1}_{#2}}}
\newcommand{\Derives}[2]{\ensuremath{\mathrel{\Rightarrow_{#2}^{#1}}}}

\newcommand{\spstep}[1]%
  {\ensuremath{\mathrel{{\relbar}{#1}{\shortrightarrow}}}}
\newcommand{\wpstep}[1]%
  {\ensuremath{\mathrel{\xRightarrow{#1}}}}

\newcommand{\indep}[1]{\ensuremath{\mathrel{\diamond_{#1}}}}

\newcommand{\init}{\ensuremath{\mathit{init}}\xspace}
\newcommand{\term}{\ensuremath{\mathit{term}}\xspace}
\newcommand{\tick}{\ensuremath{\mathit{tick}}\xspace}

\newcommand{\sem}[1]{\ensuremath{[\![#1]\!]}}

\newcommand{\bbP}{\ensuremath{\mathbb{P}}\xspace}

\newcommand{\bbE}{\ensuremath{\mathbb{E}}\xspace}

\newcommand{\mmarg}[1]{}
\newcommand{\mmargr}[1]{}
\newcommand{\lvar}{\ensuremath{\lhs{LVar}}\xspace}
\newcommand{\ivar}{\ensuremath{\lhs{IVar}}\xspace}
\newcommand{\mvar}{\ensuremath{\lhs{MVar}}\xspace}

\newcommand{\pvar}{\ensuremath{\lhs{PSMVar}}\xspace}
\newcommand{\pvars}{\ensuremath{\lhs{PSMVars}}\xspace}
\newcommand{\tvar}{\ensuremath{\lhs{TVar}}\xspace}
\newcommand{\ptvar}{\ensuremath{\lhs{PTVar}}\xspace}

\newcommand{\lvars}{\ensuremath{\lhs{LVars}}\xspace}
\newcommand{\ivars}{\ensuremath{\lhs{IVars}}\xspace}
\newcommand{\mvars}{\ensuremath{\lhs{MVars}}\xspace}
\newcommand{\tvars}{\ensuremath{\lhs{TVars}}\xspace}

\newcommand{\stm}{\ensuremath{\lhs{STM}}\xspace}

\newcommand{\psm}{\ensuremath{\lhs{PSM}}\xspace}
\newcommand{\io}{\ensuremath{\lhs{IO}}\xspace}

\DeclareFontFamily{U}{bigshuffle}{}
\DeclareFontShape{U}{bigshuffle}{m}{n}{
  <5-8> s*[1.3] shuffle7
  <8->  s*[1.3] shuffle10
}{} \DeclareSymbolFont{BigShuffle}{U}{bigshuffle}{m}{n}

\usepackage[smaller,nolist,nohyperlinks]{acronym}
\begin{acronym}
    \acro{kiel}[KIEL]{Kiel Integrated Environment for Layout}
\end{acronym}
\newboolean{lv} 

\newcommand{\lvsv}[2]{\ifthenelse{\boolean{lv}}{#1}{#2}}
\newcommand{\onlylv}[1]{\ifthenelse{\boolean{lv}}{#1}{}}
\newcommand{\onlysv}[1]{\ifthenelse{\boolean{lv}}{}{#1}}

\newcommand{\tF}{\lhs{tF}\xspace}
\newcommand{\tR}{\lhs{tR}\xspace}
\newcommand{\tC}{\lhs{tC}\xspace}
\newcommand{\tA}{\lhs{tA}\xspace}
\newcommand{\tB}{\lhs{tB}\xspace}

\newcommand{\farmer}{\lhs{farmer}\xspace}
\newcommand{\aWorker}{\lhs{aWorker}\xspace}
\newcommand{\bWorker}{\lhs{bWorker}\xspace}
\newcommand{\tSync}{\lhs{tSync}\xspace}
\newcommand{\zA}{\lhs{zero}\xspace}

\newcommand{\dA}{\lhs{dec}\xspace}
\newcommand{\iA}{\lhs{inc}\xspace}

\newcommand{\inc}{\lhs{inc}\xspace}
\newcommand{\dec}{\lhs{dec}\xspace}
\newcommand{\zero}{\lhs{zero}\xspace}
\newcommand{\new}{\lhs{new}\xspace}

\newcommand{\hlst}{\lstinline[style=hs,mathescape]}

\newcommand{\lhs}[1]{%
\mbox{\protect\lstinline[%
  language=Haskell,
  escapechar=@,
  mathescape=true,
  literate=
     {_}{$\,$}1
     {::}{{$\,::\,$}}1
     {\\}{{$\lambda$}}1
     {init}{{init}}4
     {do}{{do}}2
     {in}{{in}}2
     {fst}{{fst}}3
     {snd}{{snd}}3
     {True}{{True}}4
     {False}{{False}}4
     {IO}{{IO}}3
     {Int}{{Int}}3
     {Bool}{{Bool}}3
     {String}{{String}}3
     {pred}{{pred}}4
     {data}{{data}}4
     {new}{{new}}3
     {let}{{let}}3
     {map}{{map}}3
     {newtype}{{newtype}}7
     {putStr}{{putStr}}6
     {union}{{union}}5
     {Nothing}{{Nothing}}7
     {Just}{{Just}}4
     {Maybe}{{Maybe}}4
     {Monad}{{Monad}}4
     {take}{{take}}4
     {zero}{{zero}}4
     {read}{{read}}4
     {return}{{return}}6
     {repeat}{{repeat }}6
     {cycle}{{cycle}}5
     {catch}{{catch}}5
     {type}{{type}}4
     {getLine}{{getLine}}4
]{#1}}}

\newcommand{\EVAL}{\ensuremath{\mathit{EVAL}}\xspace}
\newcommand{\BIND}{\ensuremath{\mathit{BIND}}\xspace}
\newcommand{\RETRY}{\ensuremath{\mathit{RETRY}}\xspace}
\newcommand{\KILL}{\ensuremath{\mathit{KILL}}\xspace}
\newcommand{\DONE}{\ensuremath{\mathit{DONE}}\xspace}
\newcommand{\PAUSE}{\ensuremath{\mathit{PAUSE}}\xspace}

\newcommand{\TICK}{\ensuremath{\mathit{TICK}}\xspace}
\newcommand{\COND}{\ensuremath{\mathit{COND}}\xspace}

\newcommand{\READ}{\ensuremath{\mathit{READ}}\xspace}
\newcommand{\WRITE}{\ensuremath{\mathit{WRITE}}\xspace}
\newcommand{\NEW}{\ensuremath{\mathit{NEW}}\xspace}

\newcommand{\ANEW}{\ensuremath{\mathit{ANEW}}\xspace}
\newcommand{\AADMIN}{\ensuremath{\mathit{AADMIN}}\xspace}
\newcommand{\ADMIN}{\ensuremath{\mathit{ADMIN}}\xspace}

\newcommand{\ARET}{\ensuremath{\mathit{ARET}}\xspace}
\newcommand{\METHOD}{\ensuremath{\mathit{METHOD}}\xspace}

\newcommand{\eqdef}{\ensuremath{\mathrel{{=}_{\mathit{\scriptsize df}}}}}

\usepackage{fancyvrb}
\usepackage{stmaryrd}

\DefineShortVerb{\|}

\makeatletter\if@ACM@journal\makeatother
\acmJournal{PACMPL}
\acmYear{2023}
\startPage{1}
\else\makeatother
\acmConference[???]{???}{???}{???} 
\acmISBN{978-x-xxxx-xxxx-x/YY/MM} \acmDOI{10.1145/nnnnnnn.nnnnnnn}
\startPage{1} \fi

\setcopyright{none}             

\begin{document}

\title[PSM]{PSM: Policy Synchronised Deterministic Memory}





\author{Michael Mendler}
\affiliation{
  \institution{University of Bamberg}            
  \city{Bamberg}
  \country{Germany}
}
\email{michael.mendler@uni-bamberg.de}          



\author{Marc Pouzet}

\affiliation{
 \institution{ENS}
 \city{Paris}
 \country{France}
}
\email{marc.pouzet@ens.fr}  


\begin{abstract} 

Concurrency and determinacy do not go well with each other when resources must be shared.
\lvsv{The conflict is reflected in Haskell's terminology that distinguishes between \emph{parallel} programming abstractions (\ivar, \lvar in the \lhs{Par} monad) and \emph{concurrent} abstractions (\mvar in \io, \tvar in \stm).
The former is determinate but has no destructive updates and the latter has destructive updates but does not guarantee determinacy.
}
{
Haskell provides \emph{parallel} programming abstractions such as \ivar and \lvar in the \lhs{Par} monad and \emph{concurrent} abstractions such as \mvar and \tvar in the in \io and \stm monads, respectively.
The former are determinate but have no destructive updates and the latter have destructive updates but do not guarantee determinacy.
}
Programming patterns that are both concurrent and determinate, such as those provided by Kahn or Berry require memory abstractions at a higher level than is currently available.
\lvsv{
In this paper we describe a new type context \psm for \emph{policy synchronised memory} in Haskell that serves this goal and combines the benefits of \lhs{Par}, \stm and \io. 
Like \stm and \io, but unlike \lhs{Par}, the computations in \psm can access persistent state and, as a side-effect, update the memory in imperative style.
Like the \lhs{Par} and \io monads, but unlike \stm, the \psm context supports concurrent threads and communication through shared state. 
However, in contrast to \io, our \psm contexts are race-free since 
concurrent accesses 
are \emph{policy coordinated} which guarantees determinacy. 
Determinacy of evaluation is a feature of
\lhs{Par} and \stm, yet these do not offer imperative memory updates.
Well-typed transactions in the \psm context can accommodate abstract data structures (ADS) that are imperative, concurrently shareable between concurrent threads and behave deterministically, by construction.}
{
In this paper we describe a new type context \psm for \emph{policy synchronised memory} in Haskell. 
Like \stm and \io, the computations in \psm can access persistent state and, as a side-effect, update the memory in imperative style.
Like the \lhs{Par} and \io monads, \psm supports concurrent threads and shared state. 
However, in contrast to \io, our \psm contexts are race-free since 
concurrent accesses are \emph{policy coordinated} which guarantees determinacy. 
Well-typed transactions in the \psm context can accommodate abstract data structures that are imperative, concurrently shareable and still behave deterministically, by construction.
}

\end{abstract}


%
%


\maketitle


\ignore{ 

\begin{code}

{-# LANGUAGE GADTs #-}
{-# LANGUAGE AllowAmbiguousTypes #-}
{-# LANGUAGE MultiParamTypeClasses #-}
{-# LANGUAGE TypeFamilies #-}
{-# LANGUAGE TypeFamilyDependencies #-}
{-# LANGUAGE ExistentialQuantification #-}

module PSM13

where

import Prelude hiding (init)
import Control.Concurrent
import Control.Concurrent.STM
import Control.Concurrent.STM.TMVar
import Control.Monad
import Control.Exception.Base
import System.IO
import System.IO.Unsafe
import Data.Map.Strict as Map
import Data.Maybe
-- only for comparison
import Data.IORef
import qualified Data.Set as Data (singleton,empty,union,elems,fromList)

import TTree4

\end{code}

-- original experiments under ghc-8.0.2 / lts-8.5
-- Feb 2023: ghc-8.8.4 / lts-16.27

} 

\textbf{Editorial Note:}
  \textit{This report summarises work on coding the theory of policy-synchronised memory~\cite{AguadoMPRH18} in Haskell. This was developed for a graduate level course on Functional Reactive Programming taught at Bamberg University by the first author during 2020-2023. An early version of the PSM library had been presented at the SYNCHRON Workshop (Aussois, France), November 2019.
}

\section{Introduction}
\label{sec:motivation}

Concurrent programming is a difficult task.
Data races on shared objects jeopardise memory consistency and the determinacy of program behaviour~\cite{Lee06,AsadollahSEHA16}.
Some distributed applications are intrinsically asynchronous so that the result of a computation must depend on the scheduling order, e.g., mutual exclusion or consensus algorithms. 
However, for many critical reactive applications
determinacy is a crucial requirement as it ensures predictability and controllability for testing and debugging.
The two most well-known universal programming models for concurrent systems that achieve determinacy by construction are Kahn's data-flow model of parallel stream processing~\cite{Kahn74,LeeEtAl:14:Dataflow} and Berry's constructive control-flow model for the synchronous, hierarchical composition of Mealy machines~\cite{Berry00,LeeEtAl:14:Synchronous}.
Many specialised programming languages and tools have been built for these domains~\cite{CaspiP96,EkerJL+03,Esterelv705,BerryNS:Plastic11,Ptolemaeus:14:SystemDesign,Colaco17,SmythMRvH+17}\onlylv{ to name but a few}. 
These higher-level concurrency abstractions have not found resonance in main-stream languages, yet. General-purpose concurrent languages are traditionally biased to support free-style asynchronous programming. 
Typically, some rudimentary synchronisation primitives are provided, like semaphores, spin locks and mutexes, to be used at the discretion of the programmer.
More recent languages like Rust~\cite{JungJKD:POPL17} take more responsibility for safety, such as protecting memory from aliasing problems (ownership). Yet, Rust does not systematically help with determinacy. 
To protect shared memory from non-determinism requires higher-level synchrony and global scheduling, in addition to ownership control. 
Such mechanisms must still be handcrafted from the available synchronisation primitives.

\ignore{
Turn the theory into a methodology to implement race-free data structures in functional programming. 
We use Haskell as as demonstrator not primarily because of it lazy engine. Simply because we are familiar with it and because our concept most naturally extend the software transactional memory \lhs{STM} that is available out of the box in Haskell's concurrency model. Also, its rich type and class system make PSM smoothly integrate on top of the existing libraries.  
}

Haskell provides several useful abstractions for concurrency~\cite{Marlow:CEFP12} that allows distributed access to shared memory. 
\ivar memory cells~\cite{MarlowNPJ:Haskell11} implement strictly deterministic behaviour because
they can be written at most once and all reads (\lhs{get}) must wait for
the writing (\lhs{put}) to be completed. They are
implemented in the 
\lhs{Par} monad~\cite{MarlowNPJ:Haskell11} to benefit from the performance speed-up through distributed computations,
without losing determinacy and referential transparency of the $\lambda$-calculus.
\ivars have been generalised to \lvars~\cite{KuperN:2013} which
accumulate monotonic memory accesses
via lattice structures. A systematic limitation of these parallel abstractions is that they do not permit iterated and alternating reading and writing with destructive memory updates as in standard imperative programming. 
For non-monotonic writes and destructive updates, Haskell provides other concurrent abstractions which live inside the \io monad. 
For instance, \mvars~\cite{Peyton-JonesGF:POPL14} give referential memory with destructive accesses (\lhs{takeMVar}, \lhs{putMVar}) that are synchronised in a hand-shake fashion to work like $1$-place buffers. 
Free unsynchronised memory access is available using \tvars.
The synchronisation of reading (\lhs{takeTVar}) and writing (\lhs{putTVar}) on \tvars is handled in the associated
\stm monad which wraps software transactional memory~\cite{HarrisMPJH:PPoPP05}.
The operator \lhs{atomically} encapsulates an
arbitrary sequence of \tvar accesses to be executed as an atomic memory transaction. 
An STM transaction can block at any time with the \lhs{retry} operator.
\tvars in STM allow arbitrary distributed data structures to be implemented in Haskell in a compositional fashion.
However, free destructive updates under \io concurrency makes the \stm transactions behave inherently non-deterministically.
It is left to the programmer's skills to synchronise the code for determinacy via judicious use of the \lhs{retry} and \lhs{atomically} primitives. 
\onlylv{%
Suppose we wanted to ensure that computations referencing a variable \lhs{a :: TVar Int} remain determinate.
The trivial, though useless, restriction would be to decree that \lhs{a} can be used at most once.
Non-trivial programs will perform at least one read and one write access.
Here, we must distinguish between sequential variable accesses inside \stm and concurrent accesses outside in \io: 
\begin{align}
  & \text{\lstinline[style=hs]
      {do a <- newTVarIO 0; atomically (do writeTVar a 42; readTVar a)}} 
  \label{eqn:seq} \\
  & \text{\lstinline[style=hs]{do a <- newTVarIO 0; forkIO(atomically $ writeTVar a 42); atomically $ readTVar a}}
  \label{eqn:par} 
\end{align}
The former~\eqref{eqn:seq} is determinate, the latter~\eqref{eqn:par} is not.
}
\onlylv{%
It would be a serious effort to deeply code 
the vernacular of an imperative synchronous language such as Berry's Esterel~\cite{Esterelv705} using \stm.
Likewise, \lhs{MVars} can be used in \io to build queues and data-flow channels from which Kahn-style data-flow programs can be assembled. Again, determinacy is not guaranteed.
It remains a property of a concrete application program using the primitives, such as obeying single-writer, single-reader restrictions.
}

Determinacy is not a compositional property of Haskell's language primitives but arises from the way they are used.  
Since the primitives must be hand-assembled in the \io monad, the type checker does not give any determinacy guarantees. 
This is unfortunate, considering that Kahn-style data flow and Berry-style control flow are essentially declarative functional models. 
They enjoy semantic coherence and functional determinacy \onlylv{from properties of computational confluence,} similar to the Church-Rosser property in the pure $\lambda$-calculus. 
Of course, they can be embedded into the ``pure sequential'' part of Haskell \emph{outside} of {\io}.
However, this is not compositional and defeats the point, which is to exploit the principles of Kahn and Berry for concurrent application programming.
In particular, this requires interaction with the outside world and 
type-safe access to deterministic concurrency \emph{inside} \io, which is currently missing.

\ignore{

\begin{code}
p1 = do 
  a <- newTVarIO 0
  atomically (do writeTVar a 42; readTVar a)

-- id d is small we still see 0; when d is bigger 42 is seen
p2 d = do 
  a <- newTVarIO 0 
  forkIO $ do threadDelay 0; atomically $ writeTVar a 42
  threadDelay d
  atomically $ readTVar a

\end{code}

}

\paragraph*{Contributions}

We show how concurrency abstractions such as by Kahn~\cite{Kahn74} and Berry~\cite{Berry00} can (i) be generalised to provide race-free abstract data structures of arbitrary composite types and (ii) be implemented generically in Haskell.  
The resulting library, called \psm for \emph{Policy Synchronised Memory}, is
the first of its kind to abstract deterministic concurrency with destructive (i.e., non-monotonic) updates within Haskell's type system.
This is our attempt to
lift the separation between parallelism and concurrency in functional programming, hinted at by Simon Marlow:
\begin{quote} 
\emph{``So concurrency is a structuring technique for effectful code; in Haskell, that means code in the IO monad.'' [...]
While it is possible to do parallel programming using concurrency, that is
often a poor choice, because concurrency sacrifices determinism.''}~\cite{Marlow:CEFP12}
\end{quote}
PSM provides a generic class of memory abstractions, called \emph{policy-synchronised variables} (\lhs{PSMVar}), that are built on top of \stm. These combine the parallel determinacy of \ivars and \lvars and the concurrent effectful updates of \mvars and \tvars.
\lhs{PSMVar}s naturally extend to general composite data structures. These may be built from Haskell's concurrency primitives and pre-compiled foreign functions.

PSM is inspired by the idea of clock-synchronised shared memory (CSM)~\cite{AguadoMPRH18} which itself borrows from synchronous programming languages~\cite{BenvenisteCEH+03} the concept of a \emph{clock}. 
A clock acts as a synchronisation barrier that orchestrates concurrent memory accesses under a (distributed) cyclic scheduling pattern. This principle is well-established, e.g., in VHDL hardware modelling~\cite{KloosB95} or parallel programming~\cite{Valiant:CACM90}.
Each \emph{cycle} 
comprises a 
burst of concurrent computations that are synchronised with each other, conformant with priority-based scheduling interfaces of the shared data structures, represented as \emph{policy automata}.

$\bullet$ 
While CSM has been developed only as a mathematical theory, \psm now provides the first concrete encoding in a (functional) programming language. 

$\bullet$ 
  We present an operational semantics (Sec.~\ref{sec:psm-semantics}) and show how the theory of CSM can be mapped to the synchronisation model of STM. This offers further evidence of the versality of STM to build non-trivial forms of higher-level concurrency abstractions. 

$\bullet$ 
  We realise the scheduling policies 
  using barrier synchronisers (predictions stored as counters)
  to follow what we call the 
  \emph{Cox-Rower Protocol} (Sec.~\ref{sec:psm-idea}).  
  These barrier counters permit a simplified coding
  of the more complex enabling conditions defined in the semantics of CSM.  

  $\bullet$
  The theory of CSM assumes an abstract kernel language in which CSM variables are global, fixed and statically allocated. 
  This is a major limitation from a practical perspective.  
  \psm supports generic libraries of \pvar{s} using Haskell's class system, from which a program can dynamically instantiate (lexically-scoped, thread-local) abstract data structures.

\section{The Idea behind PSM}
\label{sec:psm-idea}

When multiple threads access a shared data structure, destructively 
updating parts of it, we must synchronise these accesses to avoid data races. The simplest possible data structure is a TVar on the heap, say \hlst{x $\rrcol$ TVar a}. Concurrent threads can atomically read and write its content executing
\begin{tabbing} 
  $\quad$ 
   \= \hlst{wr x $v$} 
   \= = \hlst{atomically (writeTVar x $v$) $\rrcol$ IO ()} \\
   \> \hlst{rd x} 
   \> = \hlst{atomically (readTVar x) $\rrcol$ IO a}
\end{tabbing}
where \hlst{$v \rrcol$ a} is a new value. 
Obviously, the value returned by \hlst{rd x} depends on whether it is executed before or after the write \hlst{wr x $v$}. We can avoid the read-write race by enforcing a precedence order between the accesses. If \lhs{x} is a memory register, it would be natural to make the read take precedece over the write. If \lhs{x} acts as a dataflow variable the opposite makes sense. Then we want the write to take effect before the read. 

The question now is how to implement such precedence orderings systematically and robustly, generalising \lhs{x} for arbitrary data structures, and generalising \hlst{wr x} and \hlst{rd x} for arbitrary families of method calls. In addition, we want multiple threads perform sequences of data accesses, not just a single \hlst{wr x} and a single \hlst{rd x}.
Our answer draws inspiration from a coordination pattern that is well-established in synchronous programming (e.g.,~\cite{Esterelv705,vonHanxledenDM+14}). We explain it here by analogy with coxswain rowing: A central synchroniser (the cox) issues a clock signal that makes the threads (rowers) proceed in cyclic lock-step fashion. During each clock cycle, also called a \emph{macro-step}, each thread performs one round of contribution which consists of a sequence of \emph{micro-steps} (drive phase). 
The micro-steps may access shared resources and avoid interference by adhering to a pre-defined precedence policy that orders the method calls for a deterministic overall effect. Finally, all threads return to a safe neutral position (recovery phase) where they wait for a clock signal to start the next cycle. 
Such synchronous clocking with local access policies will achieve determinacy for multi-threading in analogy to the coxswain's role for maintaining balance, power and controlled direction in rowing, compensating variations in the performance of rowers, uncontrollable wind conditions or competing boats.

Technically, we implement a \emph{Cox-Rowers Protocol}\footnote{This is a special instance of the director-actors model of Ptolemy~\cite{Ptolemaeus:14:SystemDesign}.} to synchronise a single \lhs{cox} thread with multiple \lhs{rower} threads we use barrier counters of \lhs{type TSync = TVar Int}. 
The methods that we need on these barrier counters are 
$$\lhs{incTSync, decTSync, zeroTSync :: TSync -> IO()}$$
for incrementing, decrementing and testing for zero.
The \lhs{retry} mechanism of the \lhs{STM} monad is used to implement \lhs{zeroTSync} test, so the thread blocks until the counter reaches value $0$. 

In order to organise the succession of macro-steps, each of which is split into a drive phase and a recovery phase, we need two counters, \lhs{rC,cC::TSync}.  
The name of the counter indicates which thread's progress is measured by it: \hlst{rC} is decremented by the \lhs{rower}s to unblock \hlst{cox}, while \hlst{cC} blocks the \hlst{rower}s and is decremented by \hlst{cox}. To implement the micro-steps in a race-free manner, we use prediction counters \hlst{pred $r$ $m$ :: TSync}, one for each shared data structure $r$ and each method $m$. We use them to make individual method calls wait for each other. 

Let us now explain how to coordinate the protocol 
with the help of the \hlst{rC}, \hlst{cC} and \hlst{pred $r$ $m$} counters. We first introduce some useful notation. 
Let us write \hlst{$n$<-cnt} to state that the current value of \hlst{cnt::TSync} is \hlst{$n$::Int}.
We further assume a generic function \hlst{method $m$ $r$} to access shared memory, for a method name $m$ and data structure $r$. 

The protocol starts in the \emph{recovery phase} of the macro-step, with counters \hlst{$n$ <- rC}, where $n$ is the number of \hlst{rower}s, \hlst{$0$ <- cC} and \hlst{$0$ <- pred $r$ $m$} for each data structure $r$ and method $m$. 
The \hlst{cox} waits for \hlst{rC} be $0$ and so is blocked immediately. The \hlst{rower}s are free and increment the counters \hlst{pred $r$ $m$} as a prediction of how many times each of them is going to execute $m$ on $r$ during the ensuing micro-steps of the imminent macro-step. They may increment concurrently, in any order. Once a \hlst{rower} has finished booking its predictions, but not before, it increments \hlst{cC}. Then, it decrements \hlst{rC} and waits for \hlst{cC} to become $0$ again. When all of the $n$ \hlst{rower}s have reached this point, but not before, the counter \hlst{rC} has value $0$.
Now, the waiting \hlst{cox} thread is released. It knows that all \hlst{rower}s have recovered and the family of prediction counters \hlst{pred} have been recharged. The \hlst{cox} now first resets \hlst{rC} to $n$ (doing $n$ increments) and then resets \hlst{cC} to $0$ (doing $n$ decrements). Then it blocks and waits for \hlst{rC} to reach $0$ again. This is because we now enter into the second part of the macro-step, the \emph{drive phase}. It starts as soon as \hlst{cox} decrements \hlst{cC} to reach $0$, which wakes up the waiting \hlst{rower}s.  
During the drive phase, the \hlst{rower}s execute their working code, concurrently, but using the \hlst{pred} counters to synchronise each other as follows: 

$\bullet$  
Before \hlst{rower} can execute any \hlst{method $m$ $r$}, it checks for all methods $m'$ taking precedence over $m$ on $r$, that \hlst{$0$<-pred $r$ $m'$}. If this is the case, method $m'$ is not going to be executed on $r$ again during the current macro-step, and thus $m$ is free to go. Otherwise, the method call blocks.
After executing \hlst{method $m$ $r$}, the \hlst{rower}  decrements \hlst{pred $r$ $m$}. 

When a \hlst{rower} has finished its working code for the current drive phase, it first increments \hlst{cC}, then decrements \hlst{rC}, and then waits for the next recovery phase by blocking on \hlst{cC} to become $0$. 
When \hlst{rC} reaches $0$, all \hlst{rower}s are done. Then \hlst{cox} unblocks, sets \hlst{rC} back to $n$ and clears \hlst{cC}. Now, all start over again with the next recovery phase.     

The above protocol encapsulates the core of synchronous programming translated into shared memory multi-threading and mapped to Haskell's \lhs{STM} monad. 
In theory, the Cox-Rowers Protocol, as described, can of course be implemented from scratch for every application on a case-by-case basis. However, it is highly non-trivial and determinacy is easily lost if the precedences of the method calls are not carefully tuned with their behaviour on the shared data structures. We propose \lhs{PSM} as a new concurrency abstraction that implements Cox-Rowers compositionally, extending \lhs{STM}, so it permits destructive memory and IO, but guarantees determinacy by construction, while supporting arbitrary abstract data structures with a conventional class-based interface between program and memory.

\ignore{%
%
%
\begin{code}
newtype MSync = MSync (MVar [MVar()])

newMSync :: IO(MSync)
newMSync = fmap MSync $ newMVar [] 

mFork :: IO() -> MSync -> IO ThreadId
mFork io (MSync mSync) = do
  m <- newEmptyMVar
  children <- takeMVar mSync
  putMVar mSync $ m:children
  forkFinally io $ \_ -> putMVar m ()

mJoin :: MSync -> IO()
mJoin (MSync mSync) = do
  cs <- takeMVar mSync
  case cs of
    []   -> return ()
    m:ms -> do
      putMVar mSync ms
      takeMVar m
      mJoin $ MSync mSync
\end{code}

\begin{code}

type AState = ()
type BState = () 
aNextState = id 
bNextState = id 
aInit = ()
bInit = ()

aWorker :: IORef AState -> IO()
aWorker aState = do 
  state <- readIORef aState
  writeIORef aState $ aNextState state

bWorker :: IORef BState -> IO()
bWorker bState = do 
  state <- readIORef bState
  writeIORef bState $ bNextState state

exchange :: IORef AState -> IORef BState -> IO()
exchange aState bState = return ()

farmer :: IO()
farmer = do
  aState <- newIORef aInit
  bState <- newIORef bInit
  mSync <- newMSync
  let cycle = do
        mFork (aWorker aState) mSync
        mFork (bWorker bState) mSync
        mJoin mSync
        exchange aState bState
        cycle
  cycle
\end{code}

}

\onlylv{

This fork-join synchronisation pattern is implemented in Fig.~\ref{fig:fork-join-mvar} using Haskell's \lhs{MVars}.
The code in Lst.~\ref{lst:fork-join-mvar-1} defines a type of synchonizers \lhs{MSync} and functions \lhs{mFork}, \lhs{mJoin} for spawning threads.
These are used in Lst.~\ref{lst:fork-join-mvar-2}
to run two worker threads \lhs{aWorker} and \lhs{bWorker} concurrently and in
lock-step from a main \lhs{farmer} thread. The workers operate on distinct local
state references \lhs{aState} and \lhs{bState} which get updated by
the farmer in the \lhs{exchange} action. This action can only
happen \emph{after} both worker threads have finished their computations
and \emph{before} they are started again. This strict sequentialisation together
with the isolation ensures overall determinacy, i.e., that the sequence of 
states computed does not depend on the order in which the workers are scheduled. 

\begin{figure*}[ht]
\small
\begin{minipage}{.95\linewidth}
 \begin{subfigure}[b]{.5\linewidth}
  \centering
  \lstinputlisting[style=hs, label=lst:fork-join-mvar-1, 
      caption={\lhs{MSync} Fork-Join Synchoniser}]
       {fork-join-mvar-1.lst}  
 \end{subfigure}
 \begin{subfigure}[b]{0.5\linewidth}
  \centering
  \lstinputlisting[style=hs, label=lst:fork-join-mvar-2, 
      caption={Farmer-Worker Communication}]
       {fork-join-mvar-2.lst}  
 \end{subfigure}
\end{minipage}
\caption{Safe cyclic fork-join thread synchronisation using \lhs{MVar}s modelled after Hackage Control.Concurrent online documentation on GHC's implementation of Concurrrency in section ``Terminating the Program''~\cite{ConcHackageWeb}.}
\label{fig:fork-join-mvar}
\end{figure*}

The puzzle with the isolation argument is that farmer and workers do in fact need to operate on a shared memory structure $\lhs{mSync} :: 
\lhs{MVar}[\lhs{MVar}()]$ in order to achieve sequentialisation of control flow.
Through \lhs{mSync} the workers communicate termination to the
farmer.
For each worker an empty \lhs{MVar()} list entry \lhs{m} is created by the farmer with \lhs{putMVar mSync (m:children)} in \lhs{mFork}, before the child is started.
Upon termination, each worker fills an $\lhs{MVar()}$ cell in the
\lhs{mSync} list by executing \lhs{putMVar mvar ()} with its \lhs{forkFinally} call-back.
The farmer, after spawning
all workers, waits in \lhs{mJoin mSync} for all \mvars in the list to be filled.
So, the threads are not really isolated at all. They all concurrently and destructively update the \lhs{mSync} structure. 

So, how can this be race-free? 
The first part of the argument lies in the
fact that \mvars are controlled by an access protocol which strongly
synchronises the reading and writing. An \lhs{MVar} has two implicit
synchronisation states, \emph{empty} and \emph{filled}. The read action is only admissible
when the \lhs{MVar} is filled and the write action only when it is empty. The writing, called \lhs{putMVar}, fills the
\lhs{MVar} with a value and the reading, called \lhs{takeMVar}, empties it again. This is a universal
communication model which makes \lhs{MVars} behave like 1-place data-flow buffers.
If these buffers are used with at most one reader and one writer thread, then determinacy follows from Kahn process theory~\cite{KahnM77}. 
The core point is that any data race between a writer filling a value and a reader removing the value is resolved by the synchronous blocking of the \lhs{MVar} buffer. Depending on the state of the \lhs{MVar} either the reader or the writer will be blocked.  
Indeed, a closer inspection of the code of Fig.~\ref{fig:fork-join-mvar} reveals that the single-reader and single-writer pattern can be applied to the shared object \lhs{mSync} as a list of lists. 
Each individual \lhs{m::MVar} cell inside the \lhs{mSync} list is created by the farmer who passes it (in \lhs{forkFinally}) to be filled by a unique worker.
This cell \lhs{m}, however, is emptied (in \lhs{mJoin}) only by the farmer. 
The outer \lhs{MVar} reference of \lhs{mSync}, in contrast, is created, filled and emptied only by the farmer.
So, we find that each of the \emph{inner} cells in the \lhs{mSync} list satisfies the single-reader, single-writer condition.
The remaining part of the determinacy argument lies in the fact that the destructive updates of the \emph{outer} structure of \lhs{mSync} are single-threaded in the farmer and thus trivially race-free.

\begin{figure*}[ht]
\begin{minipage}{.9\linewidth}
 \begin{subfigure}[c]{.5\linewidth}
  \centering
  \lstinputlisting[style=hs, label=lst:fork-join-tvar-1, 
      caption={\lhs{TSync} Synchroniser Interface}]
       {fork-join-tvar-1.lst}  
 \end{subfigure}
 \hfill
 \begin{subfigure}[c]{0.47\linewidth}
  \centering
  \lstinputlisting[style=hs, label=lst:fork-join-tvar-2, 
      caption={Farmer-Worker using \lhs{TSync}}]
       {fork-join-tvar-2.lst}  
 \end{subfigure}
\end{minipage}
\caption{Safe (determinate) cyclic fork-join thread synchronisation using a single \lhs{TVar} in \lhs{STM}.}
\label{fig:fork-join-tvar}
\end{figure*}
} 

\ignore{%

%
%

\begin{code}

newtype TSync = TSync(TVar Int)

newTSync :: STM TSync
newTSync = fmap TSync $ newTVar 0

incTSync, decTSync, zeroTSync :: TSync -> STM()
incTSync (TSync tSync) = do
  cnt <- readTVar tSync
  writeTVar tSync (cnt + 1)

-- decTSync :: TSync -> STM()
decTSync (TSync tSync) = do
  cnt <- readTVar tSync
  writeTVar tSync (cnt - 1)
  
-- zeroTSync :: TSync -> STM()
zeroTSync (TSync tSync) = do
  cnt <- readTVar tSync
  if cnt == 0 then return () else retry

\end{code}

\begin{code}

tFork :: IO() -> TSync -> IO ThreadId
tFork io tSync = do
  atomically $ incTSync tSync 
  forkFinally io $ 
    \_ -> atomically $ decTSync tSync

tJoin :: TSync -> IO()
tJoin = atomically . zeroTSync
  
tFarmer :: IO()
tFarmer = do
  aState <- newIORef aInit
  bState <- newIORef bInit
  tSync  <- atomically newTSync
  let cycle = do
        tFork (aWorker aState) tSync
        tFork (bWorker bState) tSync
        tJoin tSync
        exchange aState bState
        cycle
  cycle 

\end{code}

}

\onlylv{ 
The argument is still a bit rough as it ignores the dynamic aspect of memory allocation in the \emph{nested} list structure. 
We can eliminate this complication if we unravel \lhs{MSync} of Fig.~\ref{fig:fork-join-mvar} in the \stm monad of Haskell, which builds on \tvars.
In fact, we can implement the fork-join synchronisation pattern via a single 
\lhs{TVar Int} that counts the number of active children. The \tvar is set to
$0$ initially and incremented every time before the farmer spawns
a new worker. Each worker thread decrements the \tvar counter when it terminates.
The farmer thread blocks until the counter reaches $0$. If we implement the increment
and decrement as atomic operations, then the counter behaviour is determinate and the
fork-join executed race-free, independent of the scheduling. Packaging this in a new type, called \lhs{TSync},
we have the interface:
\begin{spec}
  newtype TSync = TSync(TVar Int)
  newTSync :: STM TSync
  incTSync, decTSync, zeroTSync :: TSync ->  STM()
\end{spec}
where \lhs{newTSync} creates a new sharable synchroniser and \lhs{incTSync}, \lhs{decTSync}, \lhs{zeroTSync} are the methods to operate on it as suggested above.
We can then refactor the implementation of the farmer-worker scenario with \lhs{TSync} as shown in
Fig.~\ref{fig:fork-join-tvar}.
The determinacy argument for Fig.~\ref{fig:fork-join-tvar} is based on the following three principles, applied to the only shared structure \lhs{tSync::TSync}\footnote{We assume that the workers, which are running in \lhs{IO}, only access their local memory \lhs{aState, bState}.}:
\begin{itemize}
  \item[A] (\emph{Atomicity}) All accesses are executed atomically;
  \item[C] (\emph{Confluence}) Whenever two concurrent accesses are enabled, then both commute with each other, i.e., the order of execution is unobservable in the resulting
  memory\footnote{The execution order may at most influence unobservable parts of the system, such as external log files.}.
  \item[S] (\emph{Stability}) Once an access becomes enabled, it remains enabled, irrespective of
  the activities of concurrent threads.
\end{itemize}
Atomicity is guaranteed by the \stm context and its
\tsf{atomically} operator.
Next, observe that  
\lhs{incTSync} and \lhs{decTSync} accesses in \lhs{STM} are non-blocking and thus are always enabled. They also commute with
each other, irrespective of the counter value.\footnote{With 2-complement's cyclic wrap-around \lhs{(maxBound + (1::Int)) - 1 == (maxBound - (1::Int)) + 1}.} 
Hence, it remains to verify Stability of \lhs{zeroTSync} and Confluence for each of \lhs{incTSync} and \lhs{decTSync} with respect to \lhs{zeroTSync}.
Here we use the following two observations about the ``sequential happens-before'' ordering of the program's method execution:
\begin{itemize}
  \item[(i)] All executions of \lhs{incTSync} actions happen strictly before the execution of \lhs{zeroTSync} in the farmer thread.
  \item[(ii)] At any point in an execution history, at least as many \lhs{incTSync} actions as \lhs{decTSync} actions must have been executed. 
  This holds because the farmer first performs \lhs{incTSync} on \lhs{tSync} \emph{before} it spawns a worker and this child gets a chance to execute \lhs{decTSync}. 
\end{itemize}
Condition (i) implies that we have Confluence between \lhs{incTSync} and \lhs{zeroTSync} simply because they are only executed by the farmer thread and thus never concurrent. 
Condition (ii) implies that if \lhs{decTSync} is to be executed by a worker then the value of \lhs{tSync} must be greater than $0$.
Hence, at that point \lhs{zeroTSync} is blocked in the \lhs{retry} of the \lhs{STM}. This means that there is no reachable configuration in which both
\lhs{decTSync} and \lhs{zeroTSync} are concurrently enabled. 
Thus we have Confluence between \lhs{decTSync} and \lhs{zeroTSync}. 
Conditions (i) and (ii) together imply that once the value of \lhs{tSync} has become $0$, it remains $0$.
Hence, whenever \lhs{zeroTSync} is enabled it remains enabled, so we have Stability for \lhs{zeroTSync}. 

Atomicity, Confluence and Stability (ACS) of method accesses are the general building blocks to eliminate data-races and thus ensure Determinacy of program behaviour. Note that the ACS argument naturally also works in the scenario of Fig.~\ref{fig:fork-join-mvar}. Atomicity for the outer structure of \lhs{mSync} corresponds to the assumption that the execution of actions by the farmer are sequentially consistent. Further, if accesses to an object are made only by a single thread, then these are trivially confluent and stable (among each other). For the inner structure of \lhs{mSync} we appeal the single-writer, single-reader condition. First, we assume that accesses \lhs{readMVar} and \lhs{putMVar} are atomic in \lhs{IO}. Second, they are never enabled at the same time (either the \lhs{MVar} is empty or it is full) and thus trivially confluent. Stability of \lhs{putMVar} follows from the single-writer and Stability of \lhs{readMVar} from the single reader assumption.

} 

\ignore{

\begin{code}

aWorker' :: IORef AState -> TSync -> IO()
aWorker' aState tSync = cycle where
  cycle = do
    atomically $ incTSync tSync 
    state <- readIORef aState
    writeIORef aState $ aNextState state
    atomically $ decTSync tSync
    cycle

bWorker' :: IORef BState -> TSync -> IO()
bWorker' bState tSync = cycle where 
  cycle = do
    atomically $ incTSync tSync 
    state <- readIORef bState
    writeIORef bState $ aNextState state
    atomically $ decTSync tSync
    cycle

farmer' :: IORef AState -> IORef BState -> TSync -> IO()
farmer' aState bState tSync = cycle where
  cycle = do
    atomically $ zeroTSync tSync
    exchange aState bState
    cycle

main' :: IO()
main' = do 
  aState <- newIORef aInit
  bState <- newIORef bInit
  tSync  <- atomically newTSync

  forkIO $ aWorker' aState tSync
  forkIO $ bWorker' bState tSync 
  forkIO $ farmer' aState bState tSync
  return ()
\end{code}

}

\onlylv{ 

A precise formalisation of the ACS argument such as for the centralised Farmer-Worker pattern of Figs.~\ref{fig:fork-join-mvar} and~\ref{fig:fork-join-tvar} is highly non-trivial. It becomes particularly tangled if  thread creation and ownership have to be taken into account. Novices of concurrent functional programming will find it hard to cross the t's and dot the i's of ACS for free-style code handcrafted in the \lhs{IO} monad.
So, can we find a practically useful concurrency abstraction and library that eliminates races by construction? 
Since testing for the presence of races is undecidable, this will force us to restrict the available concurrency primitives in some systematic way. 
} 

\onlylv{ 
\begin{figure*}[ht]
\begin{minipage}{\linewidth}
\centering
\begin{minipage}[t]{.5\linewidth}
 \begin{subfigure}[c]{\linewidth}
  \centering
  \lstinputlisting[style=hs, numbers=left, numberstyle=\tiny,
      caption={Farmer exchanging workers' memory \lhs{aState} and \lhs{bState}, synchronising on \lhs{tSync}.}, label={fig:farmer-race}]
       {fork-join-tvar-f.lst}  
 \end{subfigure}
 \begin{subfigure}[c]{\linewidth}
  \centering
  \lstinputlisting[style=hs, numbers=left, numberstyle=\tiny,
      caption={Workers with local memory \lhs{aState}, \lhs{bState} and synchroniser \lhs{tSync}.}, label={fig:workers-race}]
       {fork-join-tvar-w.lst}  
 \end{subfigure}
\end{minipage}
 \hfill
\begin{minipage}[t]{.45\linewidth}
 \begin{subfigure}[c]{\linewidth}
 \centering
  \lstinputlisting[style=hs, numbers=left, numberstyle=\tiny,
      caption={Farmer-Worker main program.}, label={fig:main-race}]
       {fork-join-tvar-m.lst}  
 \end{subfigure}
 \begin{subfigure}[c]{\linewidth}
  \centering
  \lstinputlisting[style=hs, numbers=left, numberstyle=\tiny,
      caption={Synchronising Barrier Counter.}, label={fig:counter-race}]
       {fork-join-tvar-bs.lst}  
 \end{subfigure} 
\end{minipage}
\end{minipage}
\caption{The distributed (not race-free) farmer-worker scenario in \io with a \lhs{TVar} barrier synchroniser in \stm}
\label{fig:fork-join-dist}
\end{figure*}
} 

\onlylv{%

\lvsv{The general problem simplifies in a natural way when threads and shared memory are statically allocated. 
Consider the static reformulation of the farmer-worker example in Fig.~\ref{fig:fork-join-dist}.}{This scenario is seen Fig.~\ref{fig:fork-join-dist}.}
The \lhs{main} program (Lst.~\ref{fig:main-race}) is a composition of a \lhs{farmer} and two workers \lhs{aWorker}, \lhs{bWorker}, forked concurrently in \io.
Each \lhs{xWorker} (Lst.~\ref{fig:workers-race}) continuously iterates in a cycle to update its local \lhs{xState} (lines~5--6). The \lhs{farmer} (Lst.~\ref{fig:farmer-race}) runs in a cycle where it \lhs{exchange}s the local states between the workers (line~5). The \lhs{exchange} function may be any (single-threaded) manipulation connecting the local \lhs{xState}s, which are implemented as \lhs{IORef} variables. 
If we would let workers and the farmer update and exchange the local states in free asynchronous interleaving, we would lose the cyclic alignment of computations. 
In order to make the computation cycles proceed in lock step we use a shared barrier counter \lhs{tSync}. 
The counter \lhs{tSync} is implemented using a \lhs{TVar} in Haskell's software transactional memory \stm (Lst.~\ref{fig:counter-race}). 
The abstract \lhs{newtype TSync} provides \lhs{newTSync} to create a local synchroniser, the methods \lhs{incTSync} and \lhs{decTSync} for increment and decrement, as well as \lhs{zeroTSync} to block until $0$. 

At the start of each cycle \lhs{xWorker} first increments the counter with \lhs{incTSync} (Lst.~\ref{fig:workers-race}, line~4) and when it has finished decrements it again (line~7). The \lhs{farmer} at each cycle first waits for the barrier to reach zero with \lhs{zeroTSync} (Lst.~\ref{fig:farmer-race}, line~4) and then exchanges the states (line~5).
The idea is that \lhs{tSync}'s counter value, at each moment, indicates to \lhs{farmer} how many of its children are still working within the current transaction cycle. 

\lvsv{Fig.~\ref{fig:fork-join-dist} is a more compositional solution to the farmer-worker problem as compared to the main-thread version of Figs.~\ref{fig:fork-join-mvar} and~\ref{fig:fork-join-tvar}. It is also simpler to analyse for ACS:
The static structure \lhs{tSync} is shared by three threads \lhs{farmer}, \lhs{aWorker} and \lhs{bWorker}. The \lhs{farmer} thread calls \lhs{zeroTSync} and the workers call \lhs{incTSync} on \lhs{tSync}.}%
{Our distributed synchronisation scheme is generic for an arbitrary number of workers. Yet, as it stands, it is still a little too naive. 
} 
There is a race condition in the counter which is shared. When \lhs{tSync} has value $0$ then \lhs{zeroTSync} in \lhs{farmer} and \lhs{incTSync} in \lhs{xWorker} are both enabled. But if \lhs{incTSync} happens first, it will block \lhs{zeroTSync} in the \lhs{retry}. This can generate non-determinism.
If \lhs{aWorker}, say, is very fast and races through its cycle before \lhs{bWorker} has even incremented \lhs{tSync}, then the counter value is back at $0$ so that the \lhs{farmer} is unblocked at \lhs{zeroTSync} to exchange prematurely, before \lhs{bWorker} has finished operating on its local state.  As a result, \lhs{bWorker} may see different states depending on its relative speed with respect to \lhs{aWorker}. 

The naive composition of Fig.~\ref{fig:fork-join-dist} does not work for two interlinked reasons.
First, every time a cooperation cycle is started, we need to make \lhs{zeroTSync} in \lhs{farmer} wait for (at least one of) the \lhs{incTSync} operations to have happened.   
This is a \emph{precedence} constraint  
establishing an ordering relation between \emph{concurrent} actions.
This cannot be implemented inside \stm which does not have concurrency. One might think that \lhs{farmer} could wait for \lhs{tSync} to have reached a value $\geq 1$ before it calls \lhs{zeroTSync} to wait for the end of the cycle. This does not work, because then we have another race, viz. between the \lhs{xWorkers} incrementing and decrementing \lhs{tSync} very quickly and \lhs{farmer} testing $\geq 1$ too late.  
The point is that implementation of the precedence constraint requires a prediction of the future actions  
in the concurrent workers. This can only be done in \lhs{IO}. 
Secondly, the precedence blocking must only look at the potential concurrent \lhs{incTSync} actions in the current cycle. 
This is solved by implementing a separate clock barrier and having \lhs{farmer} and \lhs{xWorker}s proceed in lock-step on this barrier. 
} 

\ignore{
Note that Condition (ii) depends on sequential consistency of the execution
architecture. If the program-order between \lhs{incTSync} and
\lhs{zeroTSync} was not preserved then we would lose condition (ii). If the last
\lhs{incTSync} was delayed by effects of a weak memory model, or bad implementation of the 
\lhs{tSync} structure, then it would be possible that the program might
reach the \lhs{zeroTSync} with the \lhs{tSync} being $0$, while there is still
an \lhs{incTSync} from the last worker pending in the memory architecture. 
This \lhs{incTSync}-\lhs{zeroTSync} race could make the farmer thread continue prematurely. 
}

\section{The PSM Evaluation Context}
\label{sec:psm-language}

The \lhs{PSM} type operator wraps a safe evaluation context for concurrent functional transactions on race-free, deterministic shared data structures. Every such data structure is accessed though a policy synchronised variable of type \lhs{PSMVar_a}. It is polymorphic in values  of type \lhs{a} and encapsulates information for the scheduling of method accesses, such as a unique identifier and prediction information. 
The type \lhs{a} must instantiate the \lhs{PSMType} class\lvsv{}{ which defines the policy interface} (see Sec.~\ref{sec:pvar}). \onlylv{The scheduling of function application is barrier-synchronised and controlled by policies as described above.}
An expression $e::\lhs{PSM_b}$ is an action that computes a value of type \lhs{b} while accessing arbitrary \lhs{PSMVars}\onlylv{ and the external world}. 
The value returned by $e$ on termination is determinate, i.e., independent of thread scheduling. 

\subsection{Core Operators}

The core operators of \lhs{PSM} are given in Fig.~\ref{fig:core-operators}. 

\begin{figure}[ht]
\begin{spec}
newPSMVar :: PSMType a => PSMStatic a -> 
               (PSMVar a -> PSM b) -> PSM b
pause :: PSM () 
kill  :: PSM a
(|||)   :: PSM a -> PSM b -> PSM (a,b) 
(>>>)   :: PSM a -> PSM b -> PSM b
(>>>=)  :: PSM a -> (PSMVal a -> PSM b) -> PSM b
ifte    :: PSMVal Bool -> PSM a -> PSM a -> PSM a 
switch  :: PSMVal (Either a b) -> (PSMVal a->PSM c) 
             -> (PSMVal b->PSM c) -> PSM c 
done    :: a -> PSM a 
return  :: a -> PSMVal a 
(>>=)   :: PSMVal a -> (a -> PSMVal b) -> PSMVal b
val     :: PSMVal a -> PSM a
evalSTM :: PSM (STM a) -> PSM a
runPSM  :: PSM a -> IO ThreadId
\end{spec}
\caption{The control operators of PSM.}
\label{fig:core-operators}
\end{figure}

The expression  
\hlst{newPSMVar init (\x->e)} 
creates and binds a fresh thread-local 
variable \lhs{x::PSMVar_a} from a parameter \lhs{init::PSMStatic}, for access inside the action \lhs{e::PSM_b}. 
Type families ensure that each instance \onlylv{\lhs{PSMType a} }of the \lhs{PSMType} class has  an associated type \lhs{PSMStatic a} for initialisation.

A \psm thread may \emph{complete} in three ways, via \lhs{done}, \lhs{pause} or \lhs{kill}. When a thread wants to \textit{terminate} it executes \lhs{done}, whereupon control passes to the next expression in sequential program order. If it wants to synchronise with the clock, called \emph{pausing}, it executes \lhs{pause}, whereby it completes the drive phase and publishes its prediction for recovery.
The thread blocks to be reactivated for the next macro-step cycle. Finally, a thread can \emph{exit} by calling \lhs{kill} to raise a global exception which forces termination of all active threads up to the root thread. This is a weak form of preemption in the sense that it becomes effective only at the end of the cycle in which \lhs{kill} is raised.

\onlylv{%
PSM is provides fork-join multi-threading $M \ttfork N$ rather than general asynchronous \lhs{forkIO}. 
This keeps the scheduling theory simple and structural. It also suffices to capture the bulk of synchonous programming paradigms.
}
Binary fork-join is \lhs{e1 ||| e2} which runs  
both actions \lhs{e1} and \lhs{e2} in concurrent threads. The calling thread sleeps and resumes when both actions have terminated.
All the \lhs{PSMVar} accesses of the children are synchronised 
conformant to the policies associated with shared objects.
A fork-join action terminates when both its children have terminated. It pauses when one of them pauses.
Sequential composition of actions without value binding is \lhs{(>>>)}, and \lhs{(>>>=)} with value binding. Specifically, \onlylv{the sequential composition} \lhs{e1_>>>_e2} iterates action \lhs{e1} through an arbitrary number of macro-steps until it terminates and then passes control to the continuation process \lhs{e2}. The return value from \lhs{e1} is ignored. When the continuation depends on it, then the binding form \lstinline[style=hs]{e1 >>>= $ \x->e2} can be used which passes the return value as \lhs{x} to \lhs{e2}. 
Note that if \lhs{e1::PSM a} provides a value of type \lhs{a}, then all references to this value in \lhs{e2} must be typed as \lhs{x::PSMVal a}. 
The type \lhs{PSMVal} is a monad~\cite{Wadler:POPL92} that wraps thread-local, unsynchronised values in \psm.  Its generic monad methods are \lhs{return} and \lhs{(>>=)}.
In this way all thread-local data in \psm are separated from mixing with the ambient Haskell.  \onlylv{Function evaluation is either fully inside the \psm context or outside.}

Expressions \lhs{ifte c e1 e2} implement branching of control flow.
Depending on the value received with \lhs{c::PSMVal Bool} either the action \lhs{e1} or \lhs{e2} is executed.
The \lhs{switch} operator generalises this to a value-binding form. 

\onlylv{
The functions \lhs{var} and \lhs{val} as well as \lhs{evalSTM} are discussed below.
}
The function \lhs{val} wraps \lhs{PSMVal} as constant \psm actions, analogous to how \lhs{return} wraps Haskell values as \lhs{PSMVal}. Both \lhs{val} and \lhs{>>>=} equip \psm with a monad-like structure. 
It 
subsumes the \stm monad via the operator 
\lvsv{
\begin{spec}
  evalSTM :: PSM (STM a) -> PSM a
\end{spec}
}{\lhs{evalSTM},}
lifting atomic evaluation \lhs{atomically::STM a -> IO}~\lhs{ a} into \lhs{PSM}.
This is possible, because the atomic execution in \lhs{STM} is single-threaded and thus determinate.

Finally, analogous to \lhs{runSTM:: STM a -> IO ThreadId}, applying \lhs{runPSM e} closes action \lhs{e::PSM a}, installs the run-time context and runs \lhs{e} through a, possibly unbounded, number of macro-steps until it terminates with a return value of type \lhs{PSMVal a}.

\ignore{ 

\begin{spec}

instance Functor PSM where 
  fmap f (PSM(pred, act)) = PSM(pred, act') where
    act' ctxRef cntPred = do 
      x <- act ctxRef cntPred
      return $ f x -- $

instance Applicative PSM where 
  PSM(pred1, act1) <*> PSM(pred2, act2) = 
    PSM(pred, act) where
      pred = unionWith (unionWith addPred) pred1 pred2 
      act ctxRef cntPred = do 
        f <- ioRef ctxRef cntPred
        x <- ioRef ctxRef cntPred
        return (f x)

  pure x = PSM(fromList [(CmplId 0, empty)], 0, \_ _ -> return x)

\end{spec}
} 

\ignore{

\begin{code}
instance Functor PSM where 
  fmap f (PSM(pred, trp, act)) = PSM(pred, trp, act') where
    act' ioRef ctxRef cntPred = do 
      x <- act ioRef ctxRef cntPred
      return $ f x -- $

instance Applicative PSM where 
  PSM(pred1, trp1, act1) <*> PSM(pred2, trp2, act2) = 
    PSM(pred, trp, act) where
      pred = unionWith (unionWith addPred) pred1 pred2 
      trp = max trp1 trp2
      act ioRef ctxRef cntPred = do 
        f <- act1 ioRef ctxRef cntPred
        x <- act2 ioRef ctxRef cntPred
        return (f x)

  pure x = PSM(pred, 0, \_ _ _ -> return x) where 
    pred = fromList [(CmplId 0, empty)]
    -- we register through path with empty prediction

\end{code}
}

\onlylv{ 

\begin{example}
The transcription of our farmer-worker system from Fig.~\ref{fig:fork-join-dist} 
into \lvsv{the \psm context}{\psm} using \lhs{pSync::PTSync} instead of \lhs{tSync::TSync} is given in Fig.~\ref{fig:pvar-farmer-dist}.
Let us explain the main differences of syntax. 
First, \lhs{PSM} is not (quite) a monad, in contrast to \lhs{STM} or \lhs{IO}. Therefore, the \onlylv{assembly of \lhs{PSM} actions in Fig.~\ref{fig:pvar-farmer-dist} does not use the \lhs{do} notation of monads. 
Instead, the} sequential composition of \lhs{PSM} actions is written \lhs{(>>>)}and \lhs{(>>>=)}, which cannot be overloaded with the monadic binders \lhs{(>>)} and \lhs{(>>=)}
that give rise to the syntactic sugar of the \lhs{do} notation use in Fig.~\ref{fig:fork-join-dist}. 
Second, we use \lhs{pause} as a global synchronisation barrier to make the three threads wait for each other before they repeat their cyclic behaviour (lines~6,25).
Third, the creation of the child threads by \lhs{forkIO} in Fig.~\ref{fig:fork-join-dist} is replaced by the (binary, associative) operator \lhs{(|||)} (lines~34,35). 
\onlylv{This implements the cyclic fork-join of our farmer-worker scenario with policy-synchronised scheduling. It enforces a stack-like behaviour of thread-creation and termination which is also better behaved for run-time optimisations and static program analysis than async-wait \lhs{forkIO}.} 
\onlylv{Note that concurrent composition \lhs{(|||)} is not available in \stm.}
Fourth, the declaration of variables is formally a second-order binder. So we write \lstinline[style=hs]{newPTSync $ \pSync -> e} in \psm (line~29), rather than \lstinline[style=hs]{do tSync <- atomically newTSync; e} as in Lst.~\ref{fig:main-race} (line~5).
Fifth, since we want determinacy by construction, we also replace the volatile \lhs{IORefs} for the workers' local \lhs{xState} by policy controlled \lhs{PSMVars}. 
These are extensions of \mvar{s} with atomic update that are race-free by enforcing a precedence of the update method \lhs{updPMVar} over the read method \lhs{takePMVar}  (see Sec.~\ref{sec:psmvars-assortment}). The \lhs{xStates} are filled by \lhs{main} (lines~32--33) before the \lhs{xWorkers} are spawned. The memory exchange by the \lhs{farmer} (line~24) uses combination functions \lhs{combA} and \lhs{combB} (lines~9--12) on states.
\end{example}
} 

\onlylv{
A clever compiler could find out that, in this special case, the concurrent accesses to \lhs{xState} are already synchronised by the user program's use of the barrier synchroniser \lhs{pSync}, and replace both \lhs{PTVars} by \lhs{IORefs}.}

\subsection{The PSM Type Class}
\label{sec:pvar}

For each \lhs{r::PSMVar_a} the type argument \lhs{a} permits a polymorphic parameterisation of the abstract data structure (ADS) that is wrapped by variable \lhs{r}. To define the actual ADS, the type \lhs{a} must be linked with associated types and access methods as specified by the \lhs{PSMType} class definition in Fig.~\ref{fig:psm-type-class}. 
We will discuss the different components of the class and, as we go along, instantiate them in Examples~\ref{ex:PTSync-2}--\ref{ex:PTSync-5} to implement the barrier counters discussed above in Sec.~\ref{sec:psm-idea}. Since are going to wrap them as first-class \lhs{PSMVar}s we shall call their type \lhs{PTSync} rather than \lhs{TSync}.

\begin{figure}[ht]
\begin{spec}
class PSMType a where 
  type PSMStatic a
  data PSMState a
  init   :: PSMStatic a -> IO (PSMState a)
  tick   :: PSMState a -> IO ()
  term   :: PSMState a -> IO ()
  data family Method a b c
  method   :: Method a b c -> 
                PSMState a -> b -> STM c
  methodId :: Method a b c -> MethodId
  prec     :: PSMState a -> MethodId -> 
                STM (Maybe [MethodId])
\end{spec}
\caption{The PSM type class. 
The implementation of the methods via \lhs{method} must be \emph{coherent} (Def.~\ref{def:coherence}) with respect to the policy \lhs{prec} under the method indexing \lhs{methodId}.}
\label{fig:psm-type-class}
\end{figure}

\noindent 
To begin with, the type \lhs{PSMStatic}, associated with class \lhs{PSMType}, defines the type of static values from which a \lhs{PSMVar} is initialised. The associated type \lhs{PSMState} describes the structure of the actual ADS, i.e., the (stateful) values that each \lhs{PSMVar} structure has at run-time. 

All instances of a \lhs{PSMType} must further support the special methods \lhs{init}, \lhs{tick} and \lhs{term} for initialisation, clock synchronisation and termination, respectively. 
The function \lhs{init} creates the initial memory structure
for the \lhs{PSMVar}. This may also involve accessing the operating system or external resources, such as for opening communication channels, files and retrieving  UTC time.
The \lhs{tick} function implements the \lhs{pause} method.
It performs a clock tick on the \lhs{PSMVar}, every time a new cycle is started.  
All active running threads are synchronised on \lhs{tick} so that its effect is consistently observed by all of them. 
A typical use of the \lhs{tick} method is to reset the state of a \lhs{PSMVar} at the beginning of each cycle. It may also perform external communication, e.g., retrieving UTC time, synchronise input/output with a user interface or access the file system. 
The \lhs{term} function is called when the thread leaves the scope of the \lhs{newPSMVar} binding. It can be used for deallocating external resources.

\onlylv{ 
This function does not need to be controlled by a policy since no other thread has access to the \lhs{PSMType} until this function has returned. 
}

\onlylv{%
The three functions \lhs{init}, \lhs{tick} and \lhs{term} must be synchronous in the sense that when they return, then the memory they manipulate is stable, i.e., there are no demonic threads left working on the memory.
The typical way to do this is to generate them as atomic (single-threaded) executions from evaluations in the \lhs{STM} monad.
}
The methods \lhs{init}, \lhs{tick} and \lhs{term}
are used only by the synchroniser (cox) not by the \lhs{PSM} threads (rowers). 
Since they are strictly sequentialised with all concurrent accesses to the ADS, 
no scheduling policy is needed to protect for races between them.

\onlylv{%
Moreover, \lhs{tick} is executed by the synchronising farmer thread strictly sequentially between two cycles, i.e., after all active workers have completed and before they resume for the next cycle. This implies that the effect of \lhs{tick} on the memory may non-deterministic, such as from a call to a user interface. It is still guaranteed that all worker threads affected by \lhs{tick} will see the \emph{same unique} behaviour. The situation is analogous to sequential hardware circuits that use synchronisers to read in external asynchronous signals arriving any time between two clock ticks. 
}

\begin{example}
  \label{ex:PTSync-2}
  We index our barrier counters as a (non-parametric) ADS with a singleton type \begin{center} 
    \lhs{data PTSyncId = PTSyncId} 
  \end{center}
  and instantiate the associated type families in the \lhs{PSMType} class as
  \begin{center} 
    \lhs{PSMStatic PTSyncId=()} and \lhs{PSMState PTSyncId=TVar Int}.
  \end{center}
  We abbreviate \lhs{PTState = PSMState PTSyncId = TVar Int}. 
  The method \lstinline[style=hs]{init::()->IO (TVar Int)} creates a counter with initial state \lhs{0::Int}.
  The counters can be used asynchronously, they do not react to the clock or allocate resources. So, \lhs{tick} and \lhs{term} are trivial, i.e., \lstinline[style=hs]{tick _ = term _ = return ()}.
  \qed
\end{example} 

Continuing our explanation of Fig.~\ref{fig:psm-type-class} we come to look at the methods. 
The names of the ADS-specific methods of a \lhs{PSMType} are collected in the family \lhs{Method a b c}. They can be executed via the \lhs{method} function. 
Each \hlst{$m$::Method a b c} of the data family defines an access method $m$ for the ADS, with \lhs{b} as the argument type and \lhs{c} as the return type.
An expression \hlst{method $m$ $r$ $v$::STM c} corresponds to calling method $m$ on a state \hlst{$r$::PSMState a} with method argument \hlst{$v$::b}. To achieve atomicity this is done in the STM monad.

\begin{figure}[ht]
 \begin{center}
 \includegraphics[scale=0.5]{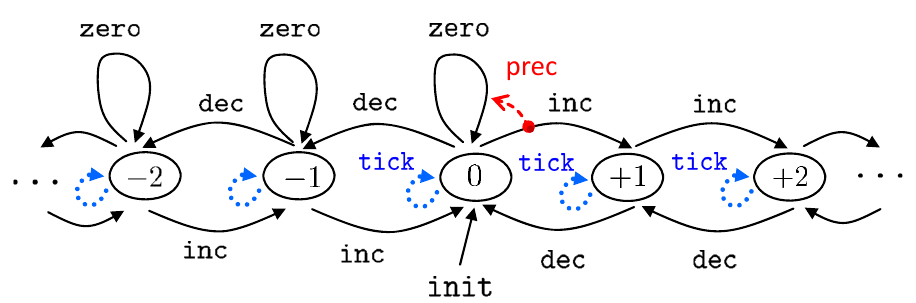}
 \end{center}
 \caption{Access policy for \lhs{PTSync}.} 
 \label{fig:psync-policy}
\end{figure}

\begin{example}
\label{ex:PTSync-3}
  The method names for \lhs{PTSync} are 
  \hlst{Inc}, \hlst{Dec} and \hlst{Zero} of type \hlst{Method PTSyncId () ()}. 
  The type parameters \lhs{()} reflect the fact our methods need no arguments nor do they return values. The automata graph in Fig.~\ref{fig:psync-policy} shows the effect of the methods for \hlst{method Inc $s$}, \hlst{method Dec $s$}, \hlst{method Zero $s$}  
  on the counter state \hlst{$s$::PTState},
  using the 
  indices \lhs{inc,dec,zero::MethodId}, associated to the methods by function 
  \lhs{methodId}, respectively.
  If \hlst{$s$::PTState} (see Ex.~\ref{ex:PTSync-2}) and \hlst{$n$<-atomically (readTVar $s$)} let us say that $s$ ``has value'' $n$. 
  Note that \hlst{method Zero $s$} is not \emph{admissible} when $s$ has value $n \geq 1$, i.e., there is no transition labelled \zA. This is where the method blocks in the \lhs{retry} of \lhs{STM}. The reason for this and the dashed arrow ``prec'' are explained below.
  Also, notice that Fig.~\ref{fig:psync-policy} displays the effect of methods \lhs{init} and \lhs{tick}. 
\qed
\end{example}

Finally, the \lhs{PSMType} interface (Fig.~\ref{fig:psm-type-class}) defines a state dependent policy \lhs{prec} that encapsulates \emph{admissibility} and \emph{precedence} for the ADS methods in \lhs{Method_a_b_c}. It determines the synchronisation of concurrent accesses to the \lhs{PSMState}, assuming that each method $m$ has a unique method index \hlst{methodId $m$}.
For each state \hlst{$s$::PSMState a} and method $m$, 
if the application \hlst{prec $s$ (methodId $m$)} evaluates to \lhs{Nothing} then $m$ is not admissible in state $s$. If it evaluates to \lhs{Just ps}, then $m$ is \emph{admissible}, written $s \Vdash m$, and 
all indices in the list \hlst{$ps$::[MethodId]} take concurrent \emph{precedence} over $m$, written $s \Vdash m' \prec m$ for all $m'$ with \hlst{methodId $m' \in ps$}. 
We write $s \Vdash a_1 \indep{} a_2$ in case both $a_i$ are admissible, i.e., $s \Vdash a_1$ and $s \Vdash a_1$, but $s \Vdash a_1 \not\prec a_2$ and $s \Vdash a_2 \not\prec a_1$.
Then we call $a_1$ and $a_2$ \emph{concurrently independent}.

\medskip 

The Determinacy Theorem (Thm.~\ref{thm:determinacy} below) depends on the assumption that the implementation of all \pvar methods are \emph{coherent} (Def.~\ref{def:coherence} below). 
For a type \lhs{a} in the \lhs{PSMType} class this means that the policy coded in \lhs{prec} must be a sound over-approximation of the concurrent independence of all methods in the family \lhs{Method a b c}.
Specifically, whenever two method ids $a_1$ and $a_2$ are admissible in a state \lhs{s::PSMState a} such that $s \Vdash a_1 \indep{} a_2$, then the execution order of the associated methods must be immaterial. 
Executing $a_1$ first and then $a_2$, or $a_2$
first and then $a_1$ must result in the same return values and observable memory state.

\begin{example}
\label{ex:PTSync-4}
  The precedences imposed by coherence are illustrated in Fig.~\ref{fig:psync-policy} using dashed arrows between the transitions admissible at a given state. 
  As seen, there is only one precedence, namely $s \Vdash \iA  \prec \zA$, when $s$ has value $0$. This is necessary since $\iA$ and $\zA$ are both admissible but their execution does not commute. We may call $\zA$ and then $\iA$, reaching value $0$. But if $\iA$ is called first and the counter reaches value $1$, method $\zA$ is not admissible any more.
  For the implementation this means \hlst{prec $s$ zero} must return \lhs{Just [inc]}.  
  When $s$ has value $n \geq 1$, then \hlst{prec $s$ zero} must return \lhs{Nothing}, 
  i.e., \zA is not admissible. 
  If $s$ has state $n < 0$ then \hlst{prec $s$ zero}  returns \lhs{Just []}.  
  We also have \hlst{Just [] <- prec $s$ mId} for $\lhs{mId} \in \{ \iA, \dA \}$, implying concurrent independence $s \Vdash \iA \indep{} \dA$.
  This is indeed coherent, since in each state \hlst{$s$::PTState} the (atomic) execution of $\iA$ followed by $\dA$ yields the same return values and state as executing first $\dA$ and then $\iA$. 
  Similarly, $s \Vdash a \indep{} a$ for all $a \in \{ \iA, \dA, \zA \}$, implies that if several threads execute the same method concurrently, the final outcome only depends on the number not the order of the calls. 
  On the other hand, we have $s \nVdash \zA \indep{} \iA$ for all $s \geq 0$. Hence, while the value is non-negative, the policy does not require commutation of $\zA$ with $\iA$. 
  Instead, concurrent calls of $\zA$ and $\iA$
  must be prohibited by policy-conformant scheduling. This happens in two different ways: In states $s > 0$ this is achieved because $\zA$ is not admissible, and in state $s = 0$ the policy gives method $\iA$ precedence over $\zA$.
  \qed
\end{example}

\subsection{PSMVar and PSMVal: Variables and Values}
\label{sec:methods-enabling}

Every instantiation of the \lhs{PSMType} class implements an ADS for use in the \lhs{PSM} evaluation context. 
Each \lhs{PSMType} method \lhs{method m::PSMState a -> b -> STM c} operates on an isolated \lhs{PSMState}, but exposes via \lhs{methodId} and \lhs{prec} a policy for which it is coherent. It thereby guarantees race-free behaviour, provided the concurrent scheduling of method calls conforms to the policy.  
To this end we must lift \lhs{method m} to become a scheduled \lhs{PSM} action 
\onlylv{of type \lhs{PSMVar a -> PSMVal b -> PSM c}} to operate on a variable \lhs{PSMVar_a} that is marshalled in \lhs{PSM} so that concurrent accesses to the \lhs{PSMState} satisfy the policy. 
For this we provide 
\begin{spec}
 runSTMMtd :: PSMType a => Method a b c -> 
   (PSMState a -> b -> STM c) -> 
      PSMVar a -> PSMVal b -> PSM c.
\end{spec}
Given \lhs{method m} as an STM transaction on state, we form 
\begin{spec}
 runSTMMtd m (method m) :: PSMVar a->PSMVal b->PSM c
\end{spec}
to build the associated \lhs{PSM} action. Note that the method name \lhs{m} now occurs twice, once in \lhs{method m} for selecting the method body from the ADS and also as a parameter to \lhs{runSTMMtd} to identify the method in the policy for scheduling.

\onlylv{
A similar lifting of functions \lhs{f::a -> b -> IO (d)} 
exists using the operator \lhs{runIOMtd} with the same type as \lhs{runSTMMtd} but with \stm replaced by \io. 
This uses an embedding \lhs{unsafeEvalIO :: PSM (IO d) -> PSM d} 
so that \lhs{myMtd = runIOMtd mId Mtd f} is only safe if the function \lhs{f} does not produce side-effects inside of \psm. Since this the case for standard input and output, we can implement \lhs{PSMVars} that connect to the external user interface. 
For safety, we do not export \lhs{runIOMtd} to the user of the \lhs{PSM} library.
}

\begin{figure}[ht]
  \begin{spec}
    newPTSync  :: (PTSync->PSM b)->PSM b -- init
    incPTSync  :: PTSync->PSM ()  -- inc
    decPTSync  :: PTSync->PSM ()  -- dec
    zeroPTSync :: PTSync->PSM ()  -- zero
  \end{spec}
 \caption{The policy-enriched method interface of \lhs{PTSync}.} 
 \label{fig:psync-methods}
\end{figure}

\begin{example} 
\label{ex:PTSync-5}
  The type \lhs{PTSync} for our barrier counters can be defined as \hlst{type PTSync = PSMVar PTSyncId} from the implementation \hlst{instance PSMType PTSyncId}.
  The policy enriched interface for \lhs{PTSync} is then obtained canonically from the methods of \lhs{PTSyncId} as seen in Fig.~\ref{fig:psync-methods}. For instance, the \lhs{PSM} action \lhs{zeroPTSync} is obtained by 
  \begin{spec}
    zeroPTSync r = 
      runSTMMtd Zero (method Zero) r (return ())
  \end{spec}
  where \lhs{return_()::PSMVal_()} is added to saturate the trivial method argument. It is then possible to call the method as \lhs{zeroPTSync r} rather than \lhs{zeroPTSync_r_(return_())}. 
  By wrapping \lhs{method Zero} as above, we put it under policy control. All accesses by \lhs{zeroPTSync r} on \lhs{r} execute the transaction \lhs{method Zero} according to the policy \lhs{prec} under method name \lhs{Zero} and method index \lhs{methodId Zero = zero}.
  The other methods \lhs{incPTSync} and \lhs{decPTSync} for \lhs{PTSync} are registered in exactly the same way.
  \qed
\end{example}

\onlylv{%

\subsection{PSM Values and PSM Variables}
\label{sec:PSMValVar}

Before moving on to sketch the implementation of \psm and the mechanisms to define arbitrary \lhs{PSMVars}, like \lhs{PTSync} or \lhs{PMVar}, it is useful to clarify the difference between values and variables in \psm.
First note that the type operator \lhs{PSM} is an applicative functor\onlylv{ with class methods
\begin{spec}
  (<*>) :: PSM (a -> b) -> PSM a -> PSM b
  pure :: a -> PSM a
\end{spec}
}
by which we can conduct ambient Haskell evaluations within \lhs{PSM}. 
\onlylv{%
In addition, \lhs{PSM} gives access to a restricted part of Haskell's \lhs{IO} for destructive memory update and policy-scheduled concurrent threads.
Due to its internal synchronisation, however, \lhs{PSM} makes determinacy guarantees not available in \lhs{IO}.
Naturally, \psm subsumes the \stm monad via the operator
\begin{spec}
  evalSTM :: PSM (STM a) -> PSM a
\end{spec}
which lifts \lhs{STM}'s atomic evaluation \lhs{atomically :: STM a -> IO a} into \lhs{PSM}. 
This is possible, because
the atomic execution in \lhs{STM} is single-threaded and thus determinate. 
Concurrent composition \lhs{(|||)}
in \lhs{PSM} makes sure determinacy is preserved, in contrast to 
thread forking in \lhs{IO} which does not prevent data races.
Since evaluations in \lhs{IO} are not determinate in general, there cannot be a natural forgetful embedding \lhs{PSM(IO a) -> PSM a}. 
}
\lvsv{It is interesting to observe that the}{The} applicative structure \lhs{pure} and \lhs{(<*>)} in \psm is derived from more primitive combinators involving the type operators \lhs{PSMVal} and \lhs{PSMVar} for \emph{un-synchronised} and \emph{synchronised} function application, respectively. The former emulates the normal Haskell evaluation of thread-local data that cannot be shared by multiple threads while the latter is our new policy-controlled value access to data that can be shared. 
\onlylv{These two mechanisms are discussed in the following two subsections.} 

\ignore{
\begin{code}
val :: PSMVal a -> PSM a
val = callVal $ pure id
\end{code}
}

Since \lhs{PSM} is not a monad, values of type \lhs{a} computed by an expression of type \lhs{PSM a} cannot escape into the Haskell context and be used as input to a function of type \lhs{a -> PSM b}. 
Instead, we must abstract thread-local values stemming from inside \lhs{PSM}, called \emph{weak values}, by the type operator \lhs{PSMVal} outside in the ambient Haskell context. 
The embedding function
\begin{spec}
val :: PSMVal a -> PSM a
\end{spec}
makes weak values available inside \psm. 
Values of type \lhs{PSM a} can be bound as input to a function of type \lhs{PSMVal a -> PSM b}, by the \lhs{(>>>=)} operator.
This allow us to use \lhs{PSM} almost like a monad but ``relative'' to \lhs{PSMVal}.

\onlylv{
The \lhs{val} embedding refines the applicative structure of \lhs{PSM} since it
makes the embedding \lhs{pure :: a -> PSM a} derivable from the
embedding \lhs{pure :: a -> PSMVal a} of the monad \lhs{PSMVal}. The application
of a function to a value may then given by the derived combinator
\begin{spec}
   callVal :: PSM (a -> b) -> PSMVal a -> PSM b
\end{spec}
which can be obtained as \lhs{\\ f x -> f <*> (val x)} from the embedding \lhs{val} and the functorial properties of \lhs{PSM}. An application \lhs{callVal f e :: PSM b}
encapsulates inside \lhs{PSM} the normal application of function \lhs{f} to an argument \lhs{e} without any synchronisation. 
The type wrapper \lhs{PSMVal}
internalises all definable data of Haskell as weak values inside the \lhs{PSM} context. 
}

\onlylv{

The type context \lhs{PSM} is not monadic since there is no universal bind operation \lhs{(>>=)} which would have the type
\lhs{PSM a -> (a -> PSM b) -> PSM b}. 
To understand the reason we observe that evaluation of an action \lhs{e::PSM a} consists of two phases, a \emph{prediction} and a \emph{value} phase. The former is used 
to block concurrent threads from accessing resources used inside \lhs{e}. The latter obtains the actual value of type \lhs{a} obtained by running \lhs{e} to termination. For the scheduling it is crucial that the prediction is available independently and before the value is computed. 
The problem now is that in a composition \lhs{e >>= g} the function \lhs{g::a -> PSM b} may depend in arbitrary ways on its input value, which is supposed to be computed by upstream \lhs{e}. In particular, the prediction in \lhs{g} may depend on the value from \lhs{e}. 
But then the prediction for the composite \lhs{e >>= g} could not be obtained solely from the predictions of \lhs{e} and \lhs{g}. It would necessarily involve the value phase of \lhs{e}.
Contrast this with our weaker ``bind'' operator \lhs{(>>>=)} where the reveiver has type \lhs{g::PSMVal a -> PSM b}.
Here, the input of \lhs{g} is only a weak value with the effect that \lhs{g} 
can use its input only in its value phase. Predictions in \psm cannot depend on weak values. 
}

\ignore{ 
The bind \lhs{(>>>=)} is the sequential composition operator of \lhs{PSM} from which the applicative \lhs{(<*>)} is derivable as 
\begin{spec}
  \ f x -> x >>>= callVal f :: PSM (a -> b) -> PSM a -> PSM b.
\end{spec}
and \lhs{pure:: a -> PSM a} is derivable as \lhs{\\ x -> val (pure x)} from the monad structure of \lhs{PSMVal}.
}

The second way to call an argument in \lhs{PSM} is by a shareable synchronised reference. Outside of \lhs{PSM}, such references carrying values of type \lhs{a} are wrapped by the type constructor \lhs{PSMVar}.
Each reference \lhs{r::PSMVar a} to values of type \lhs{a} has a unique identity and maintains predictive information about all concurrent threads sharing \lhs{r}.
In order to be synchronisable as \lhs{PSMVar a}, the type \lhs{a}  must be an instance of the type class
\begin{code}
class PSMType a where 
  type PSMStatic a
  data PSMState a 
  init :: PSMStatic a -> IO (PSMState a)
  tick :: PSMState a -> IO ()
  term :: PSMState a -> IO ()
  data family Method a b c
  prec :: PSMState a -> MethodId -> STM (Maybe [MethodId])
  methodId  :: Method a b c -> MethodId
  method :: Method a b c -> PSMState a -> b -> STM c
\end{code}
which defines a state-dependent \emph{precedence relation} \lhs{prec} that encapsulates the access policy for \lhs{a}.
It determines the synchronisation of concurrent calls to references of type \lhs{PSMVar a}, assuming that
each method application has a unique \emph{method index} specified by the type \lhs{MethodId}.
\ignore{
\begin{code}
data AnyMtd a = forall b c. PSMType a => AnyMtd (Method a b c)

data MethodId = MethodId Int String
instance Eq MethodId where 
  (==) (MethodId m1 _) (MethodId m2 _) = m1 == m2
instance Ord MethodId where 
  compare (MethodId m1 _) (MethodId m2 _) = compare m1 m2
instance Show MethodId where 
  show (MethodId _ s) = s

data ThrdId = ThrdId ThreadId String
instance Eq ThrdId where 
  (==) (ThrdId t1 _) (ThrdId t2 _) = t1 == t2
instance Ord ThrdId where 
  compare (ThrdId t1 _) (ThrdId t2 _) = compare t1 t2
instance Show ThrdId where 
  show (ThrdId _ s) = s

\end{code}
}
For each value \lhs{x::a} and method index \lhs{mId::MethodId}, \lhs{prec} codes admissibility and blocking in the following way:
If \lhs{prec x mId} evaluates to \lhs{Nothing} then \lhs{mId} is not admissible in state 
\lhs{x}. If \lhs{prec x mId} evaluates to \lhs{Just ps}, then \lhs{mId} is admissible and 
all method indices in the list \lhs{ps::[MethodId]} take concurrent precedence over \lhs{mId}.

Consider our barrier synchroniser (Fig.~\ref{fig:psync-interface}) as an instance \lhs{PSMType PTSync}. 
Suppose the methods \lhs{incPTSync}, \lhs{decPTSync} and \lhs{zeroPTSync} (Fig.~\ref{fig:psync-methods}) have indices \lhs{mI,mD,mZ::MethodId}, respectively. Since  (Fig.~\ref{fig:psync-policy}) \lhs{incPTSync}, \lhs{decPTSync} take precedence over \lhs{zeroPTSync} when \lhs{s} is in state $0$, we would have \lstinline[style=hs]{Just [mD,mZ] <- prec s mI}. When \lhs{s} is in any other state, then \lhs{zeroPTSyn} is not admissible, i.e., \lstinline[style=hs]{Nothing <- prec s mI}.

Synchronised references \lhs{PSMVar a} are embedded in the \lhs{PSM} context via the function
\begin{spec}
  var :: PSMType a => MethodId -> PSMVar a -> PSM a.
\end{spec}
Each application \lhs{var mId r} represents an action accessing \lhs{r} where the method index \lhs{mId} specifies the precedence of this access in competition to all other simultaneous evaluations predicted from other \emph{concurrent} threads.
The access \lhs{var mId r :: PSM a} will block if there are any other 
concurrent calls registered in \lhs{r} with method indices that take precedence over \lhs{mId}. 
The blocking is necessary if the value stored in \lhs{r} could be destructively modified by concurrent threads, creating data races with the thread evaluating \lhs{var mId r}.
\onlylv{%
The derived binding combinator 
\begin{spec}
  bindVar :: PSMType a => MethodId -> PSMVar a -> (PSMVal a -> PSM b) -> PSM b
  bindVar mId r f = var mId r >>>= f
\end{spec}
lifts functions \lhs{f::PSMVal a -> PSM b} on weak values, to functions 
\lhs{PSMVar a -> PSM b}
on synchronised references \lhs{r}. 
A \lhs{PSM} \emph{action} in general  
is a function of type
\lstset{mathescape=true}
\begin{spec}
  proc :: (PSMType $a_1$, $\ldots$, PSMType $a_n$) => 
            PSMVar $a_1$ ->  $\ldots$ -> PSMVar $a_n$ ->
                PSMVal $b_1$ -> $\ldots$ -> PSMVal $b_m$ -> PSM $c$
\end{spec}
\lstset{mathescape=false}
where $a_1$, ..., $a_n$ are the synchronised shared structures which are passed as synchronised references, and $b_1$, ..., $b_m$ are pure (weak) value parameters. 
The construction of actions is restricted to a set of primitive methods on predefined \lhs{PSMVars} exported by the \psm library. Examples are \lhs{PSync} (Fig.~\ref{fig:psync-interface}) or the \lhs{PSMVars} mentioned in Sec.~\ref{sec:psmvars-assortment}.  
}

} 

\subsection{PSM Scheduling}
\label{sec:psm-scheduling}

\onlylv{
Our \lhs{PSM} context provides an implementation of the CSM theory~\cite{AguadoMPRH18} on top of Haskell's \stm monad to generate race-free shared abstract data structures in a compositional fashion.
The main idea 
is that the ACS argument can be factorised through (concurrent access) \emph{policies}, acting as synchronisation contracts between data structures and run-time environment. Our observation here is that these CSM contracts when translated into the \lhs{STM} model correspond to a separation between \emph{admissibility blocking}, \emph{precedence blocking} and \emph{clock barriers}.
In this section we sketch the basic principles, by way of our farmer-worker example, before we describe the \lhs{PSM} language interface in Sec.~\ref{sec:psm-language}. 
}

After we have registered our methods on an abstract data type, let us look at how these are scheduled. 

\begin{example} 
\label{ex:PTSync-6}
  Consider the simplified scenario in Fig.~\ref{fig:cox-rower}. The \lhs{main} program is a concurrent composition of a \lhs{cox} and two \lhs{rower}s. The \lhs{cox} continuously iterates in a cycle to write a shared command \hlst{cmnd::State}, while each \lhs{rower}
  reads it. 
  Each cycle over again, there is a read-write race on \lhs{cmnd}, which we assume unsynchronised. To make sure all \lhs{rower}s read the same command in each cycle, we use a barrier counter \hlst{p::PTSync}.
  Let the \lhs{cox} and \lhs{rower} threads be called $\tC$, $\tR_1$, $\tR_2$, respectively.
  The $\tR_i$ call \iA before they \lhs{read cmnd} and \dA afterwards. The $\tC$ calls \zA to test \hlst{p}, before it writes \lhs{cmnd}. Since \zA blocks by precedence, \tC cannot proceed as long as there is still some $\tR_i$ that has not executed \iA. Once both \iA are done, \zA blocks by admissibility, waiting for \lhs{p} to become $0$. Due to both forms of blocking, \tC will not overwrite \lhs{cmnd} until both $\tR_i$ have read the previous command. 

  We use \lhs{p::PTSync} here for didactical purposes. We could equally well directly wrap \lhs{State} as a \lhs{PSMVar} with \lhs{read} and \lhs{write} methods, and make \lhs{read} take precedence over \lhs{write}. Also note that precedence policies keep our synchronisation independent of the number of \lhs{rower} threads and whether they execute \lhs{read cmnd} at all. This\onlylv{ depends on the synchronisation run-time of PSM and} is not possible with \lhs{TVars} and STM Haskell alone.  \qed
\end{example}

\begin{figure}[ht]
\begin{minipage}{\linewidth}
 \centering
\begin{lstlisting}[style=hs, label=lst:fork-join-pvar-2, numbers=none, numberstyle=\tiny, firstnumber=20]
cmnd :: State 
p :: PTSync 
cox, rower, main :: PSM ()
cox = zeroPTSync p >>> write cmnd >>> pause >>> cox
rower = incPTSync p >>> read cmnd
          decPTSync p >>> pause >>> rower
main = cox ||| rower ||| rower
\end{lstlisting}  
\end{minipage}
\caption{A Simple Concurrency Scenario.}
\label{fig:cox-rower}
\end{figure}

The essence of the thread scheduling mechanism, adapted for our purposes from~\cite{AguadoMPRH18}, is as follows:
\lvsv{In order to apply a policy we assume 
that every}{Every} active thread $t$ publishes a \emph{prediction} $\pi(t)$ which is a multiset of actions (represented via counters) that can potentially be executed by $t$ until it completes the current cycle.
This covers all potential control-flow paths until $t$ either terminates or pauses.
If a thread $t$ can execute an action $a$ and progress $t \lderives{}{} t'$, then $a$ is in its prediction, i.e., $\pi(t)(a) \geq 1$\footnote{$\pi(t)$ is a multiset that counts how often an action is contained in it.}. Furthermore, \onlylv{during each macro-step,} as the threads progress to execute their actions, the prediction reduces monotonically, i.e., $\pi(t) \geq \pi(t')$ under point-wise multiset inclusion. At the clock barrier, where $\pi(t') = \emptyset$, the predictions get refreshed for the next cycle.  
Predictions for concurrent compositions are aggregated by union $\Pi = \pi(t_1 \cpar t_2 \cpar \cdots \cpar t_n) = \sum_i \pi(t_i)$ of multisets.
Finally, an action $a$ is \emph{enabled} (for a thread $t$) in a state $s$ of the structure (under control) for a prediction $\Pi$ (of all threads concurrent to $t$) if (i) $a$ is admissible, $s \Vdash a$, and (ii) there is no precedence conflict in $\Pi$, i.e., for any action $b$, if $s \Vdash b \prec a$ then $\Pi(b) = \emptyset$. Hence, for enabling an action, the prediction $\Pi$ acts as a vector of barrier counters.

\begin{example}
 Let us apply these principles to Fig.~\ref{fig:cox-rower}.
 We want to schedule the parallel configuration $\tC \cpar \tR_1 \cpar \tR_2$ acting on the shared \lhs{PTSync} from its initial state $s = 0$.
 Without the policy control, there would be three scheduling steps\footnote{Since we assume actions are atomic we can use an interleaving model.} possible: $\tC \cpar \tR_1 \cpar \tR_2 \lderives{}{} \tC' \cpar \tR_1 \cpar \tR_2$ where $\tC \lderives{}{} \tC'$ is the \zA test; Or $\tC \cpar \tR_1 \cpar \tR_2 \lderives{}{} \tC \cpar \tR_1' \cpar \tR_2$ or $\tC \cpar \tR_1 \cpar \tR_2 \lderives{}{} \tC \cpar \tR_1 \cpar \tR_2'$ if one of $\tR_1$ or $\tR_2$, respectively, performs the \iA action. 
 Before the policy scheduler unblocks a thread, it needs to extract predictions. 
 For instance, initially, since $s = 0$, the \zA test is admissible, $s \Vdash \zA$.  
 This does not mean, however, \tC can execute \zA.
 This depends on the aggregated prediction $\pi(\tR_1 \cpar \tR_2) = \pi(\tR_1) + \pi(\tR_2)$ from $\tC$'s concurrent siblings.
 Their predictions are $\pi(\tC) = \{ \zA^1\}$, $\pi(\tR_1) = \{ \iA^1, \dA^1 \} = \pi(\tR_2)$. 
 But now $\pi(\tR_1 \cpar \tR_2)(\iA) = \{ \iA^2, \dA^2 \}(\iA) = 2 > 0$ and $s \Vdash \iA \prec \zA$. So, \zA is blocked by $s$ under $\pi(\tR_1 \cpar \tR_2)$. 
 
 On the other hand, the initial \iA actions of $\tR_1$ and $\tR_2$ are admissible and do not have any precedence conflicts in state $s = 0$.
 Therefore they are enabled. Either \lhs{rower} can take \iA as the first action. 
 Suppose $\tR_1$ executes \iA, leading to $\tR_1 \lderives{}{} \tR_1'$ and $s \lderives{}{} s'$ where $s' = +1$.
 In the new state $s'$ the \lhs{zero} test is not enabled, because it is not admissible, $s' \nVdash \zA$.
 The \lhs{cox} is still blocked.
 The actions \iA and \dA in $\tR_2$ and \dA in $\tR_1$, however, remain enabled.
 They can freely advance in any order.
 The \lhs{cox} remains being blocked, either by precedence or by admissibility, until $\tR_1$, $\tR_2$ have reached their \lhs{pause}, at which point $\tR_1$, $\tR_2$ block to wait for the clock. 
 Let $\tR_1^*$, $\tR_2^*$ be the \lhs{rower}s' threads at that moment.
 At this pausing position, we have $\pi(\tR_1^*) = \emptyset = \pi(\tR_2^*)$ since we only aggregate predictions within the current cycle of a thread. 
 The state of \lhs{p} has been incremented twice and decremented twice, so $s^* = 0$. Now the \zA test is finally enabled for $\tC$, since $s^* \Vdash \zA$ and $\pi(\tR_1^* \cpar \tR_2^*) = \emptyset$. 
 Then the \lhs{cox} advances $\tC \lderives{}{} \tC^*$ and arrives at the \lhs{pause}, too. 
 We have reached the configuration $\tC^* \cpar \tR_1^* \cpar \tR_2^*$.
 Now the \lhs{pause} method is executed synchronously in all waiting threads (and the counter object), orchestrated by the run-time scheduler. 
 This brings all threads back to their initial points and the scheduling restarts with configuration $\tC \cpar \tR_1 \cpar \tR_2$.
 \qed
\end{example} 

\onlylv{%

\begin{example} 
Let us apply these principles to our example.
We want to schedule the parallel configuration $\tF \cpar \tA \cpar \tB$ acting on the shared \lhs{tSync}\footnote{The shared ``counter'' \lhs{tSync} as the subject of control must not be confused with the prediction counters $\Pi$ as  the means for control.} from its initial state $s = 0$.
Without the policy control, there would be three scheduling steps\footnote{Since we assume actions are atomic we can use an interleaving model. 
} possible: $\tF \cpar \tA \cpar \tB \lderives{}{} \tF' \cpar \tA \cpar \tB$ where $\tF \lderives{}{} \tF'$ is the \zA test; Or $\tF \cpar \tA \cpar \tB \lderives{}{} \tF \cpar \tA' \cpar \tB$ or $\tF \cpar \tA \cpar \tB \lderives{}{} \tF \cpar \tA \cpar \tB'$ if one of \tA or \tB, respectively, performs the increment action \iA.
Before the policy scheduler unblocks a thread, it needs to extract predictions. 
For instance, initially, since $s = 0$, the \zA test is admissible, $s \Vdash \zA$.  
This does not mean, however, farmer \tF can execute \zA.
This depends on the aggregated prediction $\pi(\tA \cpar \tB) = \pi(\tA) + \pi(\tB)$ from the \farmer's concurrent siblings.
Their predictions are $\pi(\tF) = \{ \zA^1\}$, $\pi(\tA) = \{ \iA^1, \dA^1 \} = \pi(\tB)$ . 
But now $\pi(\tA \cpar \tB)(\iA) = \{ \iA^2, \dA^2 \}(\iA) = 2 > 0$ and $s \Vdash \iA \prec \zA$. So, \zA is blocked by $s$ under $\pi(\tA \cpar \tB)$. 
 
On the other hand, the initial increment actions \iA of the workers \tA and \tB are admissible and do not have any precedence conflicts in state $s = 0$.
Therefore they are enabled. Either workers can take \iA as the first action. 
Suppose \tA executes \iA, leading to $\tA \lderives{}{} \tA'$ and $s \lderives{}{} s'$ where $s' = +1$.
In the new policy state $s'$ the zero test is not enabled, because it is not admissible, $s' \nVdash \zA$.
The farmer is still blocked.
The actions \iA and \dA in \bWorker and \dA in \aWorker, however, remain enabled.
They can freely advance in any order.
The farmer remains being blocked, either by precedence or by admissibility, until \tA, \tB have reached their \lhs{pause}, at which point \tA, \tB block to wait for the clock. 
Let $\tA^*$, $\tB^*$ be the workers' threads at that moment, e.g., the control-flow point where \lhs{aWorker} executes the recursive call \lhs{cycle} in line~8 of Lst.~\ref{fig:workers-race}. 
At the \lhs{cycle} position, we have $\pi(\tA^*) = \emptyset = \pi(\tB^*)$ since we only aggregate predictions within the current cycle of a thread. 
The state of \lhs{tSync} has been incremented twice and decremented twice, so $s^* = 0$. Now the \zA test is finally enabled for the farmer, since $s^* \Vdash \zA$ and $\pi(\tA^* \cpar \tB^*) = \emptyset$. 
Then the farmer advances $\tF \lderives{}{} \tF^*$ and waits at the \lhs{pause}, too. 
We have reached the configuration $\tF^* \cpar \tA^* \cpar \tB^*$.
Now 
the \lhs{pause} method is executed synchronously in all waiting threads (and the counter object), orchestrated by the run-time scheduler. 
This brings all threads back to their initial points and the scheduling restarts with configuration $\tF \cpar \tA \cpar \tB$.
Note that it is important that the predictions $\pi(\tA)$ and $\pi(\tB)$ collect only the residual method calls up to the next \lhs{cycle} point.
Otherwise, we would get $\pi(\tA^*) = \{ \iA^\infty, \dA^\infty \} = \pi(\tB^*)$ and the \zA action in \tF would remain blocked forever. 
\qed
\onlylv{%
In other words, the precedence relation of the access policy is applied to method calls only within a macro-step, not across macro steps.} 

\onlylv{%
In sum we find that by wrapping the shared synchoniser \lhs{tSync} with a policy that adds precedence and a clock, fixes the synchronisation problems with the naive attempt in Fig.~\ref{fig:fork-join-dist} to distribute the farmer-worker. 
Another way to look at it is to say that the policy in Fig.~\ref{fig:psync-policy} encodes the centralised main-thread synchronisation pattern from Fig.~\ref{fig:fork-join-mvar} in a (statically) distributed fashion.
To stress the difference, let us call the policy-enriched synchroniser \lhs{pSync::PTSync} rather than \lhs{tSync::TSync}.
Its methods are typed as seen in Fig.~\ref{fig:psync-methods}.
Our new policy-revised farmer-worker example is given in Fig.~\ref{fig:pvar-farmer-dist} in \psm syntax which will be explained below.
}
\onlylv{%
The program in Fig.~\ref{fig:pvar-farmer-dist} not only fixes the scheduling problems with our naive model from Fig.~\ref{fig:fork-join-dist}. It is also determinate by construction.
To understand why this is so, it is useful to bring in the other side of the interface, that of the shared resource.
}
\onlylv{
The policy in Fig.~\ref{fig:psync-policy} acts as a contract that protects each shared \lhs{PTSync} instance from concurrent accesses that would jeopardise the determinacy of its memory state.
To stress the difference, let us call the policy-enriched synchroniser \lhs{pSync::PTSync} as opposed to \lhs{tSync::TSync} from Fig.~\ref{fig:counter-race}. 
}

\end{example}
} 

\onlylv{%

\subsection{Conformance} 
\label{sec:conformance}

\emph{Conformance} for the policy of
Fig.~\ref{fig:psync-policy} restricts the scheduling of a program's accesses in the way described above: First, assuming all calls to the actions \lhs{incPTSync}, \lhs{decPTSync} and \lhs{zeroPTSync}
are done atomically, only the interaction sequences specified by the automaton are
admissible. This is independent of whether the actions are performed by a single thread
or several threads. Thus, the policy of Fig.~\ref{fig:psync-policy} specifies
that \lhs{zeroPTSync} is only executable when \lhs{pSync} 
has policy state $0$. Otherwise, the policy will block the access. The other two actions \lhs{incPTSync} and \lhs{decPTSync}, in contrast,
are always admissible and can be executed in any order. Note that admissibility in a policy captures the synchronous blocking of \lhs{MVars} or the blocking in the \lhs{retry} of \stm. An entirely different form of controlling enabledness of methods arises from the precedence relation: The precedences at state $0$ express that the run-time environment must 
schedule the execution of \lhs{zeroPTSync} in a given thread strictly after any
execution of \lhs{incPTSync} or \lhs{decPTSync} in any other \emph{concurrent} thread. This is a proper extension in the scheduling infrastructure provided by \lhs{PSM} over and above \stm. Only when both these
scheduling constraints (admissibility and precedence) are fulfilled, we say the execution is \emph{policy-conformant}. 

In addition, policies use a clock as a synchronisation barrier to make threads cooperatively wait for each other to start the next computation round. 
The \lhs{PSM} context provides the action \lhs{pause::PSM()} to make a thread wait on \lhs{pause} at the end of each cycle. We can thus make the farmer and workers cooperate cyclically in lock-step, as intended by our naive distribution model in  Fig.~\ref{fig:fork-join-dist}.

}

Memory accesses in \psm Haskell 
are synchronised in \lhs{STM} using a generic enabling function
\begin{spec}
enabled :: PSMType a => Method a b c -> 
             PSMState a -> Prediction -> STM Bool
\end{spec}
\ignore{
\begin{code}
enabled :: PSMType a => Method a b c -> PSMState a -> Prediction -> Prediction -> STM Bool
enabled mtd pvar potCnt recCnt = do
  let totalPotential = addPred potCnt recCnt 
  maybeInhibs <- prec pvar (methodId mtd)
  case maybeInhibs of
    Nothing -> return False
    Just inhibs -> return $ -- $
      and (Prelude.map (\b -> findWithDefault 0 b totalPotential <= 0) inhibs)
\end{code}
}
which implements the policy-specific blocking for a given method\onlylv{ to ensure determinate behaviour of multi-threaded evaluations on type \lhs{a}}.
The type \lhs{Prediction} is the type of
\emph{predictions} 
collected in a map of counters
\begin{spec}
type Prediction = Map MethodId Int
\end{spec}
that provide conservative over-approximations of the potential future method calls against which to test enabledness.
If \hlst{$p\rrcol$ Prediction}, then
\hlst{$p$!$mId'\rrcol$ Int} stores the number of outstanding calls on each method index \hlst{$mId'$}, i.e.,
the calls that are potentially performed by the concurrent threads within the current macro-step. 
The 
test \hlst{enabled$\,m\,s\,p$} will return \lhs{True} if the policy relation \lhs{prec} indicates that (i) $m$ is admissible in state $s$ and (ii) there is no $m'$ with precedence over $m$ in state $s$ and \hlst{$p$!(methodId$\,m'$)} $> 0$.

\ignore{
\begin{code}
data PSMVarId = PSMVarId Int deriving (Show, Read, Eq, Ord)
\end{code}
}

\onlylv{%
Each prediction map \lhs{p::Prediction} records the accesses to a given \lhs{PSMVar} with unique index \lhs{PSMVarId}, made by a given thread indexed by \lhs{ThreadId}. 
} 
\onlylv{%
In the evaluation of an expression \lhs{e::PSM b}, in general, accesses to many different \lhs{PSMVar}s are made.}
\onlylv{%
The global prediction information about which \lhs{PSMVar}s are accessed in the concurrent environment by which threads is represented by maps
\begin{spec}
type ThrdPSMVarMap = Map PSMVarId Prediction
type PSMVarPotMap  = Map ThrdId Prediction
\end{spec}
}
\ignore{
\begin{code}
newtype CmplId = CmplId Int -- completion codes
  deriving (Eq, Ord)
instance Show CmplId where 
  show (CmplId 0) = "thr" -- through paths
  show (CmplId 1) = "snk" -- sink paths
  show (CmplId 2) = "rec" -- recurrent paths
  show (CmplId n) = "CmplId " ++ show n
type ThrdPSMVarMap = Map CmplId (Map PSMVarId Prediction)
  -- predictions indexed by completion code; if a completion is not in
  -- the keyset then the thread cannot exhibit the associated form of exit

-- potential
type PSMVarPotMap = Map ThrdId (Prediction, Prediction)

showThrPredMapx :: Map ThrdId PSMVarPred -> String
showThrPredMapx thrdPredMap = 
  snd(Map.foldlWithKey assemble (False, "[") thrdPredMap) ++ "]" where 
    assemble (comma, s) t p = (comma || True, 
      s ++ (if comma then ", \n   " else "") ++ show t ++ ":" ++ showPSMVarPred p)

showThrPredMapxx :: Map ThrdId (Prediction, Prediction) -> String
showThrPredMapxx thrdPredMap = 
  snd(Map.foldlWithKey assemble (False, "[") thrdPredMap) ++ "]" where 
    assemble (comma, s) t p = (comma || True, 
      s ++ (if comma then ", \n   " else "") ++ show t ++ ":" ++ show2Potential p)

showThrPredMap :: Map ThrdId Prediction -> String
showThrPredMap thrdPredMap = 
  snd(Map.foldlWithKey assemble (False, "[") thrdPredMap) ++ "]"where 
    assemble (comma, s) t p = (comma || True, 
      s ++ (if comma then ", \n   " else "") ++ show t ++ ":" ++ showPotential p)

\end{code}
}
\onlylv{%
Each running thread keeps a \lhs{ThrdPSMVarMap} to publish the set of \lhs{PSMVars} (i.e., the keys of the map)
that is has access to, and associating with each of these \lhs{PSMVars} a prediction of the method calls potentially still to be executed (i.e., the value for the key) during the current cycle.
Orthogonally, each allocated \lhs{PSMVar} maintains a \lhs{PSMVarPotMap} to record the set of threads accessing it and for each of these the predictions for method calls potentially arising from the thread during the current cycle.
}

\onlylv{%

\begin{figure*}[ht]
\begin{minipage}{\linewidth}
 \centering
\begin{minipage}[t]{.47\linewidth}
  \lstinputlisting[style=hs, numbers=left, numberstyle=\tiny, label=lst:fork-join-pvar-1]
       {fork-join-pvar-1a.lst}  
 \end{minipage}
 \hfill
 \begin{minipage}[t]{.47\linewidth}
  \lstinputlisting[style=hs, label=lst:fork-join-pvar-2, numbers=left, numberstyle=\tiny, firstnumber=20]
       {fork-join-pvar-2a.lst}  
\end{minipage}
\end{minipage}
\caption{The race-free distributed farmer-worker implemented in \lhs{PSM}.}
\label{fig:pvar-farmer-dist}
\end{figure*}
}

\onlylv{%

\begin{example}
Let the \lhs{farmer}, \lhs{aWorker} and \lhs{bWorker} threads in the \lhs{main} action of Fig.~\ref{fig:fork-join-dist} be called \tF, \tA and \tB, respectively.
For our discussion we ignore the threads' local memory and focus on their accesses to the shared \tSync.
We will use the terms `(memory) action' and `method call' synonymously.
During their first joint cycle, \tF executes the action \zero (\lhs{zeroTSync}), while each of \tA and \tB execute \inc (\lhs{incTSync}) followed by \dec (\lhs{decTSync}).
We want \tF to wait with \zA until none of \tA or \tB can execute \iA any more. This is blocking by \emph{precedence} where a method call in one thread takes priority over a method call in another thread. 
Once the \iA calls have been executed, \tF can call \zA to block on the counter's state until that has (again) reached $0$.  
This is blocking by \emph{admissibility}. 
Finally, to complete the cycle, \tA and \tB must wait for \tF to exchange states and then all start the next cycle synchronously together. 
This is blocking at a \emph{clock barrier}.
The clock barrier is essential also for precendence control: The method call \zA of \tF in the \emph{first} cycle need only give precedence to \iA calls in the \emph{first} cycle. It must ignore those arising from the \emph{second} and repeated rounds. Otherwise, no progress can be made. 
\qed 
\end{example}
}

\onlylv{%

The three forms of blocking together make up what has been called a \emph{concurrent access policy}~\cite{AguadoMPRH18}. 
The access policy for \tSync is seen in Fig.~\ref{fig:psync-policy}.
It is a labelled transition system where the states correspond to (an abstraction of) the states of the object under control. The transition labels are the actions through which the memory structure is manipulated. 
\onlylv{The control states track the states of the memory protected by the policy and so are abstractions of the memory.}
In our example, the states of the policy are identical to the states of the counter. 
The initial state is pointed to by the \new action which creates the
object instance. 
The three dimensions of blocking are reflected in the policy as follows:
}

\ignore{%

\begin{code}

newtype PTSync = PTSync(TVar Int)

stm2psm :: STM a -> PSM a 
stm2psm stm = evalSTM $ val $ pure stm 

newPTSync' :: IO PTSync
newPTSync' = atomically $ (fmap PTSync $ newTVar 0)

incPTSync :: PTSync -> PSM()
incPTSync (PTSync pSync) = stm2psm $ do
   cnt <- readTVar pSync
   writeTVar pSync (cnt + 1)

decPTSync :: PTSync -> PSM()
decPTSync (PTSync pSync) = stm2psm $ do
   cnt <- readTVar pSync
   writeTVar pSync (cnt - 1)
  
zeroPTSync :: PTSync -> PSM()
zeroPTSync (PTSync pSync) = stm2psm $ do
  cnt <- readTVar pSync
  if cnt == 0 then return () else retry

\end{code}

%
%

\begin{code} 

newPTSync :: (PTSync -> PSM a) -> PSM a
newPTSync p = p undefined -- dummy for testing

updateA = PSMVal id 
updateB = PSMVal id

combine :: PSMVal AState -> PSMVal BState 
  -> (PSMVal AState, PSMVal BState)
combine aState bState = (bState, aState)

combineA :: AState -> BState -> AState
combineA aState bState = bState

combineB :: AState -> BState -> BState
combineB aState bState = aState

exchange'' aState bState = 
  takePMVar aState >>>= \a ->
  takePMVar bState >>>= \b ->
  putPMVar aState (pure combineA <*> a <*> b) >>>
  putPMVar bState (pure combineB <*> a <*> b) 

initA = PSMVal () 
initB = PSMVal ()

loop :: (PSM() -> PSM()) -> PSM() 
loop f = p where p = f p

aWorker'' :: PMVar AState -> PTSync -> PSM()
aWorker'' aState pSync =
  let cycle =
        incPTSync pSync >>> 
        updPMVar aState updateA >>> 
        decPTSync pSync >>> 
        pause >>> 
        cycle
  in cycle 

bWorker'' :: PMVar BState -> PTSync -> PSM()
bWorker'' bState pSync =
  let cycle =
        incPTSync pSync >>> 
        updPMVar bState updateB >>> 
        decPTSync pSync >>> 
        pause >>> 
        cycle
  in cycle 

farmer'' :: PMVar AState -> PMVar BState -> PTSync -> PSM()
farmer'' aState bState pSync = let
  cycle =
    zeroPTSync pSync >>> 
    exchange'' aState bState >>> 
    pause >>>
    cycle
  in cycle

main'' :: PSM ThreadId
main'' =
  newPTSync $ \ pSync ->
  newPMVar  "aState" $ \ aState ->
  newPMVar  "bState" $ \ bState -> 
  putPMVar aState initA >>>
  putPMVar bState initB >>>  
  farmer''  aState bState pSync |||
  aWorker'' aState pSync ||| 
  bWorker'' bState pSync

\end{code}

}

\section{Some Basic PSMVars}
\label{sec:psmvars-assortment}

\ignore{

\subsection{Example: \lhs{MVars} as \lhs{PSMVars}}
\label{sec:mvars}

Let us see how the policy-wrapped version \lhs{PTMVar} of \lhs{MVars} from Sec.~\ref{sec:psmvars-assortment} is obtained in \psm.
Recall the method interface from Fig.~\ref{fig:mvar-methods} and the policy from Fig.~\ref{fig:mvar-policy}.
First, we define the core type \lhs{TMVar_ a} for our memory objects that we want to become an instance of the \lhs{PSMVar} class.
\begin{spec}
data PMVar_ a = PMVar_ (TVar (Maybe a)) 
\end{spec}
The contents are stored in a \lhs{TVar} which may be \lhs{Nothing} when the \lhs{PMVar} is empty or \lhs{Just v} if it is full holding value \lhs{v::a}.
Next we define the policy for our methods, assuming that \lhs{takePMVar} is indexed by \lhs{MethodId 1} and \lhs{putPMVar} by \lhs{MethodId 2}:
\begin{spec}
instance PSMType (PMVar_ a) where 
  prec (PMVar_ mustRef) (MethodId 1 _) = do    -- takePMVar = MethodId 1
    must <- readTVar mustRef
    case must of 
      Just _  -> return (Just [MethodId 1 "take"])  -- precedence loop
      Nothing -> return Nothing           -- not admissible

  prec (PMVar_ mustRef) (MethodId 2 "put") = do    -- putPMVar MethodId 2
    must <- readTVar mustRef
    case must of 
      Just _  -> return Nothing           -- not admissible 
      Nothing -> return (Just [MethodId 2 "put"])  -- precedence loop  
\end{spec}
Notice that the precedence relation \lhs{prec} introduces a precedence loop for \lhs{MethodId 1} when the \lhs{TMVar_} is full and for \lhs{MethodId 2} when it is empty. 
In the other cases, the methods are not admissible.
Now we construct the \lhs{PSMVar} class functions for creation, pausing and termination:
\lstset{numbers=left,numbersep=5pt}
\begin{spec}
instance PSMVar (PMVar_ a) where 
  type PSMState (PMVar_ a) = ()
  init _ = fmap PMVar_ $ atomically $ newTVar Nothing
  tick _ = return ()
  term _ = return ()
\end{spec}
\lhs{PMVars} are initialised as empty, so the static type from which \lhs{newPMVar} will instantiate is the trivial type \lhs{()} (line~2) and the \lhs{init} function creates an empty cell (line~3). The \lhs{tick} and \lhs{term} actions have no effect on a \lhs{PMVar}. 
The last step is to install the methods.  
This works analogously for both methods, so we only describe the definitions for \lhs{takePMVar}.
To anchor the method in the type system we first define singleton method type
\lstset{numbers=none}
\begin{spec}
data TakePMVar = TakePMVar
\end{spec}
Then we create a singleton family of methods indexed by \lhs{TakePMVar}, declare the method's parameter type as \lhs{()} (line~3), the return type as \lhs{a} (line~4):
\lstset{numbers=left,numbersep=5pt}
\begin{spec}
instance PSMVarMtd (PMVar_ a) TakePMVar where

  type PSMVarMtdIn (PMVar_ a) TakePMVar = ()
  type PSMVarMtdOut (PMVar_ a) TakePMVar = a

  methodId _ _ = MethodId 1 "take"

  method _ = runSTMMtd TakePMVar _must where
    _must _ (PMVar_ mVarRef) _ = do
      mvar <- readTVar mVarRef
      case mvar of 
        Nothing -> error "panic: read from empty PMVar"
        Just v -> return v
\end{spec}
and the generic function \lhs{method} that implements the action via \lhs{runSTMMtd} as described abovein Sec.~\ref{sec:methods-enabling}. The method is induced from the \stm action \lhs{_must::m->(PMVar_ a)->()->STM a} defined in lines~8--13.
It reads the contents of the memory which is called \lhs{mVarRef} in line~9 and returns the content if it is non-empty. The case that it is empty throws a (data corruption) error. This should never happen because the action \lhs{takePMVar} is not admissible in this case. 
Finally, we use the generic \lhs{method} function from the \lhs{PSMVarMtd} class to export the method as \lhs{takePMVar} 
\lstset{numbers=none}
\begin{spec}
type PMVar a = PSMVar (PMVar_ a)

takePMVar :: PMVar a -> PSM a
takePMVar mvar = method TakePMVar mvar (PSMVal ())
\end{spec} 

\begin{spec}
data PutPMVar = PutPMVar

instance PSMVarMtd (PMVar_ a) PutPMVar where

  type PSMVarMtdIn (PMVar_ a) PutPMVar = a
  type PSMVarMtdOut (PMVar_ a) PutPMVar = ()

  methodId _ _ = MethodId 1 "put"

  method _ = runSTMMtd PutPMVar _must where
    _must _ (PMVar_ mVarRef) v = do
      mvar <- readTVar mVarRef
      case mvar of 
        Nothing -> writeTVar mVarRef (Just v)
        Just _  -> error "panic: write into full PMVar"

putPMVar :: PMVar a -> a -> PSM ()
putPMVar mvar v = method PutPMVar mvar (PSMVal v)

newPMVar :: String -> (PMVar a -> PSM b) -> PSM b  
newPMVar name = newPSMVar name () 
\end{spec}

} 

\onlylv{ 

Consider the $\mathsf{PSMType}$ instance declaration in lines 963--973 for $\mathsf{PMVar}$. With this code, the programmer introduces two methods $\mathsf{takePMVar}$ ($\mathsf{MethodId 1}$) and $\mathsf{putPMVar}$ ($\mathsf{MethodId 2}$) and scheduling precedences $\lhs{prec}$ between them. This establishes the scheduling policy. It decrees that \lhs{takePMVar} takes precedence over itself when the variable is full (line 966), i.e., the state \lhs{must} of the PSMVar is of the form \lhs{Just _}. In the notation of Sec.~2.1 we have $\lhs{Just _} \Vdash \lhs{takePMVar} \prec \lhs{takePMVar}$.
This has the effect that the PSM scheduler will refuse to execute \lhs{takePMVar} concurrently with itself.  This is necessary since \lhs{takePMVar}, whose behaviour is defined in lines 1001--1004,  does not commute with itself: if we execute \lhs{takePMVar} on state $\lhs{Just _}$ it will take the value and empty the \lhs{PMVar}. Any other concurrent thread also executing \lhs{takePMVar} would now be blocked. If the programmer had left out the self-precedence in line 966, writing \lhs{Just _ -> return []}, then the method \lhs{takePMVar} would be concurrently independent from itself, i.e., $\lhs{Just _} \Vdash \lhs{takePMVar} \indep{} \lhs{takePMVar}$. In that case, the requirement of ``policy coherence'' would oblige the programmer to verify that any \emph{two} applications of \lhs{takePMVar} commute; Specifically, that after the first application of \lhs{takePMVar} another application remains admissible. The verification attempt to show that \lhs{takePMVar} (as implemented in lines 1001--1004) is isolated from itself would fail in this case. In synchronous programming, the compiler which is checking policy-constructiveness will refuse to compile. 

} 

The barrier synchronisers \lhs{PTSync} are a prime example of a \pvar that combines blocking by admissibility 
and precedence.  
As described in Sec.~\ref{sec:psm-idea}, they are universal for implementing the Cox-Rower Protocol which lies at the heart of PSM. 
Here we mention other useful (application-independent) primitive policy-controlled and race-free objects.

\begin{figure*}[ht]

 \begin{subfigure}[c]{0.5\linewidth}
 \begin{center}
 \includegraphics[scale=0.5]{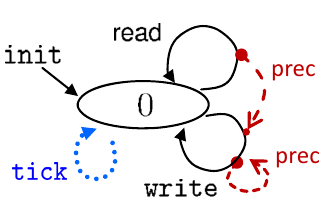}
 \end{center}
 \caption{Single-state \ptvar policy} 
 \label{fig:tvar-policy}
 \end{subfigure}
 \hfill
 \begin{subfigure}[c]{0.47\linewidth}
 \begin{center}
  \begin{spec}
   newPTVar   :: a->PSM (PTVar a)  -- init
   readPTVar  :: PTVar a->PSM a  -- read
   writePTVar :: PTVar a-> PSMVal a->PSM () -- write
  \end{spec}
 \end{center}
 \caption{\lhs{PTVar} method interface} 
  \label{fig:tvar-methods}
 \end{subfigure}

\onlylv{%
 \begin{subfigure}[l]{0.47\linewidth}
 \begin{center}
 \includegraphics[scale=0.5]{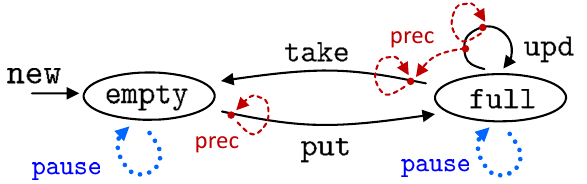}
 \end{center}
 \caption{Two-state \lhs{PMVar} policy.} 
  \label{fig:pmvar-policy}
 \end{subfigure}
\begin{subfigure}[l]{0.47\linewidth}
 \begin{center}
  \begin{spec}
   newPMVar  :: (PMVar a->PSM b)->PSM b
   updPMVar  :: PMVar a->PSMVal (a->a)->PSM ()
   putPMVar  :: PMVar a->PSMVal a->PSM ()
   takePMVar :: PMVar a->PSM a
  \end{spec}
 \end{center}
 \caption{\lhs{PMVar} method interface} 
  \label{fig:pmvar-methods}
 \end{subfigure}
} 

\begin{subfigure}[l]{0.4\linewidth}
 \begin{center}
 \includegraphics[scale=0.5]{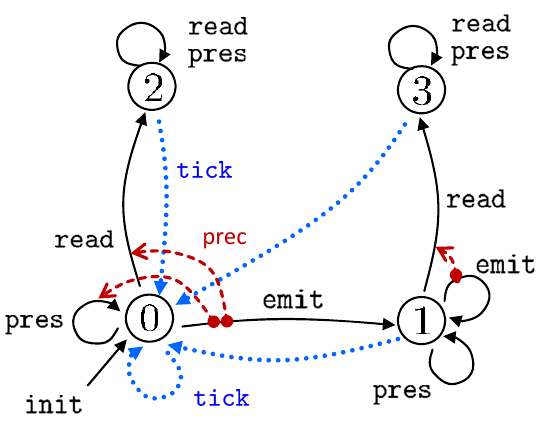}
 \end{center}
 \caption{Four-state \lhs{PTSig} policy.} 
  \label{fig:ptsig-policy}
 \end{subfigure}
 \begin{subfigure}[l]{0.52\linewidth}
 \begin{center}
  \begin{spec}
   newPTSig  :: a->(a->a->a)->(PTSig a->PSM b)->PSM b
   emitPTSig :: PTSig a->PSMVal a->PSM ()  -- emit
   presPTSig :: PTSig a->PSM Bool -- pres
   readPTSig :: PTSig a->PSM a  -- read
  \end{spec}
 \end{center}
 \caption{\lhs{PTSig} method interface} 
  \label{fig:ptsig-methods}
\end{subfigure}

\begin{subfigure}[c]{0.4\linewidth}
 \begin{center}
 \includegraphics[scale=0.5]{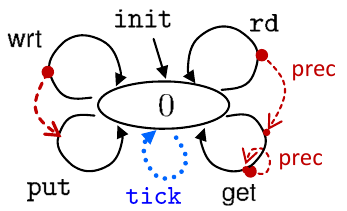}
 \end{center}
 \caption{State-less \lhs{PStrIO} policy.} 
  \label{fig:striovar-policy}
 \end{subfigure}
 \begin{subfigure}[c]{0.52\linewidth}
 \begin{center}
  \begin{spec}
   newPStrIO :: (PStrIO->PSM b)->PSM b
   wrStrIO   :: PStrIO->PSMVal String->PSM() -- wrt 
   rdStrIO   :: PStrIO->PSM String  -- rd
   getStrIO  :: PStrIO->PSM() -- get 
   putStrIO  :: PStrIO->PSM() -- put
  \end{spec}
 \end{center}
 \caption{\lhs{PStrIO} method interface} 
  \label{fig:striovar-methods}
\end{subfigure}

 \caption{Policy Interfaces for different instances of the \lhs{PSMType} class.} 

\label{fig:ptsig-interface}
\end{figure*}


\subsubsection*{Race-free TVars.} 

There are two canonical ways of protecting \lhs{TVars} from  data races created by accesses from concurrent \io threads.
One is the write-before-read protocol encoded in the policy seen in Fig.~\ref{fig:tvar-policy}. Under this policy, \lhs{writePTVar} has precedence over itself (eliminating w/w conflicts) and over \lhs{readPTVar} (eliminating r/w conflicts). We call this interface \lhs{PTVar}.
The alternative variant would be to reverse the causality and make reading take precedence over writing. Under this policy, the structure would behave like a \emph{register} in synchronous hardware: it is first read, before it is overwritten. We call this interface \lhs{PTReg}. 
\onlylv{It is the exactly as in Fig.~\ref{fig:tvar-policy} except renaming of the methods and reversing the ``prec'' edges in the policy.} 
Both \lhs{PTVar} and \lhs{PTReg} form the universal building blocks for (synchronous or asynchonous) data-flow programming.

\onlylv{%

\subsubsection*{Race-free MVars.} 

\lvsv{Although \lhs{MVars} implement synchronisation this is not enough for determinacy. 
We still need the single-reader and single-writer requirement.}
{\lhs{MVars} need the single-reader and single-writer requirement for determinacy. The update method, if permitted, creates a race with reading the \lhs{MVar} which must be ordered by precedence.} 
Adding the \psm policy interface seen in Figs.~\ref{fig:pmvar-policy} and~\ref{fig:pmvar-methods} generates a race-free version which we call a \lhs{PMVar}.
A \lhs{PMVar} has two states, \lhs{empty} and \lhs{full}. A fresh \lhs{x::PMVar} is initially empty. \lvsv{Whenever it is empty it}{It} can be filled with a value \lhs{v::a} via method \lhs{putPMVar x v}. When it is full, the value can be update with \lhs{updPMVar}, or extracted by \lhs{takePMVar x} which empties the variable again. Like for regular Haskell \lhs{MVars}, the admissibility of methods in the policy make sure that reading will block until a value is available and writing will block until \lhs{x} is empty. 
Obviously, a race occurs when two concurrent threads aim to put a value into an empty cell or 
both want to read from a full cell.  
\lvsv{To eliminate the races we can give \lhs{putPMVar} and \lhs{takePMVar} precedence over themselves.}{To eliminate write-write and read-read races we give all methods \lhs{putPMVar}, \lhs{takePMVar} and \lhs{updPMVar} precedence over themselves.} 
\onlylv{%
This 
protects the \lhs{PMVars} from write-write and read-read races.}
The precedences have the effect that concurrent accesses such as \lhs{putPMVar x v1 ||| putPMVar x v2} \onlylv{and \lhs{takePMVar x ||| takePMVar x}} will block each other from creating non-deterministic behaviour. 
In this fashion we replace non-determinism by blocking.  
\onlylv{When integrity is more important than availability, the forced deadlock is preferable than unpredictable behaviours. Avoiding deadlock and non-determinism requires static program analysis.}
The race between \lhs{updPMVar} and \lhs{takePMVar} is resolved by giving the former precedence over the latter.   
Hence in a parallel \lhs{takePMVar x ||| updPMVar x f}, the update will be executed first. 
Both these features of \lhs{PMVar}s properly extend the expressiveness over regular \lhs{MVar}s in \stm.  
\onlylv{
As an aside, note that Haskell's \lhs{IVars}, which can only be written once, are obtained if we make the \lhs{take} transition in Fig.~\ref{fig:mvar-policy} point back to state \lhs{full} and remove the ``prec'' self-loop on it.
}
}

\subsubsection*{Temporary Signals.}

A useful type of \lhs{PSMVar}, not available in the existing concurrency abstractions of Haskell, are Esterel temporary signals~\cite{Esterelv705}, 
\onlylv{They are useful in applications implementing \emph{synchronous control-flow} where concurrent processes cooperate in lock-step, like in the introductory farmer-worker scenario. Processes cycle through strictly sequentialised \emph{macro-steps}. \onlylv{This abstracts the computation principles of clock-synchonised sequential hardware.} Within a macro-step processes communicate on shared \emph{signals},
} 
here called \lhs{PTSig}.
A \emph{temporary} signal is analogous to a wire in a sequential circuit that carries a unique value at each macro-step. 
The value computed in a cycle is not memorised for the next cycle but instead re-initialised.
\onlylv{Since temporary signals do not need to be stored between ticks they behave like reusable (deterministic) \lhs{IVars}.} A combination function and synchronous policy are used to permit deterministic multi-writer, multi-reader accesses to the value.
\onlylv{This makes signals more powerful than Haskell's \lhs{IVars}.} 
The methods are given in Fig.~\ref{fig:ptsig-methods}.
A signal \hlst{$x\rrcol$ PTSig a} is created by calling \hlst{newPTSig$\,\mi{def}\,\mi{cmb}$} which sets the signal's \emph{default} value \hlst{$\mi{def}$} and \emph{combination} function \hlst{$\mi{cmb}$}.  
A value \hlst{$v\rrcol$ PSMVal a} is emitted on signal $x$ with \hlst{emitPTSig$\,x\,v$}. When more than one thread emits a value, then the combination function\footnote{The value type \lhs{a} with binary operation {\footnotesize\lhs{cmb}} and constant {\footnotesize\lhs{def}} must be a commutative monoid.} \lhs{cmb} is used to generate a unique value that is available to all the readers. 
This aggregated value can be \emph{read} with method \lhs{readPTSig} which will be blocked until all concurrent initialisations or emissions have taken place.
This is coded in the policy of Fig.~\ref{fig:ptsig-policy} by the precedences $s \Vdash \lhs{emitPTSig} \prec \lhs{readPTSig}$ for $s \in \{0,1\}$. Such a precedence is not needed in policy states $s \in \{2, 3\}$, because no thread is permitted to emit a signal more than once. 
Beside its value, a signal also has a status, \emph{present} or \emph{absent} which can be obtained by \lhs{presPTSig}. At the start of each cycle, the status is \emph{absent}, by default. As soon as the signal is emitted, its status is set to \emph{present}. To avoid a data race, the policy enforces \onlylv{the precedence} $0 \Vdash \lhs{emitPTSig} \prec \lhs{presPTSig}$. After a signal has been emitted first time, in policy states $s \in \{ 1, 3 \}$, the status is present and cannot change any more. In those states, emission and present test are independent, i.e., $s \Vdash \lhs{emitPTSig} \indep{} \lhs{presPTSig}$.
Note that the combination of default status with precedence permits an application to test for the \emph{absence} of a signal without the absence being explicitly communicated.
This is the key of Esterel-style synchronous event-based programming. Programs using only value-free signals of type \lhs{PTSig()} correspond to constructive circuits~\cite{ShipleBT96}. 
Reaction to absence (called ``quiescence'') is also a feature that has been exploited in the ReactiveAsync library~\cite{HallerGES16} implemented in Scala. The shared objects of ReactiveAsync, called \emph{cells}, can be seen as a 
generalisation of Esterel signals with persistent state and could be modelled as \lhs{PSMVar}s, too.

\subsubsection*{Race-free IO Ports.}

\onlylv{Communication with the external world does not take place in the \lhs{IO} monad, as for normal Haskell programs, but in the synchronised \lhs{PSM} context.}  
The external world can be connected through 
synchronised IO ports that prevents the environment from introducing data races. 
Prototypically, we provide \lhs{PStrIO} objects, specified in Fig.~\ref{fig:striovar-methods}, for \lhs{String} output and input, wrapping the system console. Each \lhs{ioPort::PStrIO} stores the current \emph{cycle count}, an \emph{inbuf} from which strings are synchonously read and an \emph{outbuf} to which characters are asynchonously appended.  
The method \lhs{wrStrIO} appends a string to outbuf and \lhs{putStrIO} asynchonously prints the current cycle count together with the content of outbuf to std-out. 
The method \lhs{getStrIO} sends a prompt \lhs{">> "} to std-out and then calls \lhs{getLine} to obtain a string from std-in. This input string is stored in the inbuf of the \lhs{ioPort}. The method \lhs{rdStrIO} is used to access the contents of inbuf. 
The \lhs{tick} function increments the cycle count stored with the \lhs{ioPort} and empties the outbuf, while it keeps the contents of inbuf for the next cycle. 
Observe that the policy (Fig.~\ref{fig:striovar-policy}) has only one state where all methods are admissible.
Since \lhs{wrStrIO}, \lhs{putStrIO} operate on outbuf and 
\lhs{getStrIO}, \lhs{rdStrIO} operate on inbuf, which are separated, no precedences between them are needed for coherence. However, we force \lhs{wrStrIO} to take precedence over \lhs{putStrIO} to make sure that the std-out print will contain all strings written by any thread during the current macro-step.
Note that neither \lhs{wrStrIO} nor \lhs{putStrIO} has precedence over itself, whence they may be called concurrently.  
Of course, the printout will reflect the scheduling order of the method calls.
This is ok, since std-out is off-limits and no \psm method depends on it.
On the input side we must block \lhs{getStrIO} from being called concurrently, for otherwise the content of inbuf is nondeterministic. 
\lvsv{Two calls of \lhs{getStrIO} do not commute\footnote{Here we mean calls to the same instance of a \lhs{PStrIO}. Different instances of io-ports are assumed to be independent and non-sharing. The displayed prompt to the user could disambiguate, so that any non-determinism arising through reading from different ports is a non-determinism in the external environment, not in the \psm program. If such is not desired, then we could use a policy for the instantiation function \lhs{newStrIO} to prevent concurrent threads from instantiating an io-port to the ``same user'' twice. This would just be a higher-level form of method call abstracted in \psm.} 
with each other. 
}
{Two calls of \lhs{getStrIO} do not commute with each other.} 
Hence the precedence loop on \lhs{getStrIO} in Fig.~\ref{fig:striovar-policy}. Finally, since all \lhs{PStrIO} readers must see the same content of inbuf, 
we give \lhs{getStrIO} precedence over \lhs{rdStrIO}. 
\onlylv{An example using IO-interaction is listed in Fig.~\ref{fig:iotest}.}

\onlylv{%

\begin{figure*}[ht]
\begin{minipage}{\linewidth}
 \centering
\begin{minipage}{.4\linewidth}
  \lstinputlisting[style=hs, label=lst:iotest-1]
       {iotest-1.lst}  
 \end{minipage}
 \hfill
 \begin{minipage}{.58\linewidth}
  \lstinputlisting[style=hs, label=lst:iotest-2]
       {iotest-2.lst}  
\end{minipage}
\end{minipage}
\caption{A sample \psm program with console IO interaction.
The action \lhs{ioLoop} connects an \lhs{io} port to the console by operating \lhs{getStrIO} and \lhs{putStrIO}. It iterates reading a string from std-in into the inbuf of \lhs{io} and writes back the outbuf at the end of each cycle. It also kills the \psm action when the user types the string ``\lhs{:q}'' to quit. 
The parallel action \lhs{ioTest} writes ``Hello'' and ``World'' into outbuf of \lhs{io} (\lhs{wrStrIO}) in the first two cycles. It reads (\lhs{rdStrIO}) the inbuf \lhs{io} as \lhs{l1} of in cycle 1 and as \lhs{l2} in cycle 2, immediately puts an output message containing both strings into outbuf.
Because of the synchronisation of \lhs{getStrIO} and \lhs{rdStrIO} in \lhs{io} the external event queue enters the \psm computations only at the beginning of a clock cycle.
Since all \psm threads run in lock-step with the clock, the evaluation remains deterministic. 
}
\label{fig:iotest}
\end{figure*}

}

\ignore{%

\begin{code}

--- RECORD
test14 = runStrIO $ \log ->
  wrStrIO log (pure "Hello") >>>
  pause >>>
  rdStrIO log >>>= \str ->
  wrStrIO log (pure "World") >>>
  pause >>>
  wrStrIO log (do s <- str; return $ "Here is what I read in tick 1: " ++ s ++ " ...") >>>
  rdStrIO log >>>= \str -> 
  wrStrIO log (do s <- str; return $ " ... and this is the input from this tick: " ++ s) >>>
  pause >>>
  wrStrIO log (pure "Bye")

{- 
*PSM13> test14
>> ASDF
Tick 0: Hello
>> QWER
Tick 1: World
>> VBNM
Tick 2: Here is what I read in tick 1: QWER ... ... and this is the input from this tick: VBNM
>> ZUIO
Tick 3: Bye
>>
Tick 4: 
>> :q
ThreadId 85
-}
\end{code}

}

The above examples are simple but universal. Many more complex application-specific data structure can be built
to instantiate the \hlst{PSMType} class.
We can lift arbitrary memory transactions in \stm to \psm actions and bundle them as a set of methods for a \lhs{PSMVar}.
\onlylv{
Admissibility and precedence must be chosen so that the memory transactions behave coherently on memory, in the sense discussed in Sec.~\ref{{sec:impl-PSMVars}}.
} 
In this way, we can generate data structures ranging from small blocks like synchronisers, buffers, channels and priority queues to larger units such as GUI or software components for embedded AI or robotics.

\section{The Semantics of PSM}
\label{sec:psm-semantics}

As PSM Haskell is built on top of STM, its semantics can be understood as an extension of the semantics presented in~\cite{HarrisMPJH:PPoPP05}. STM Haskell distinguishes two forms of reduction steps, viz.\ $\Rightarrow$ in STM and $\rightarrow$ in IO. Here, we take the latter to be evaluation in PSM, which subsumes (part of) IO but guarantees determinacy.  
A step~$\rightarrow$ may be an atomic method call that encapsulates an arbitrary sequence of STM-evaluations $\Rightarrow$ in a single thread.
The small-step semantics are defined in Fig.~\ref{fig:sos-semantics}, with concurrent PSM on the top, pure $\lambda$-reductions in the center and STM on the bottom.  
We focus on the PSM part, referring the reader to~\cite{HarrisMPJH:PPoPP05} for STM. 
\onlylv{The policy-based scheduling of PSM adds priorities to STM and implements the theory proposed in~\cite{AguadoMPRH18}. In the following we transcribe it as an extension of STM.}

\begin{figure}[ht]
\[ 
\begin{array}{rcl} 
  x, y &\in& \mathit{Variable} \\
  p, r, m &\in& \mathit{Name} \\[1ex]
  m \eqdef K &\in& \mathit{Method Declaration} \\[1ex]
  \text{Value}\; V &::=&
    p \mid \ttTrue \mid \ttFalse \mid \lambda x.\, M \mid \ttpause \\ 
    &\mid& \ttdone M \mid \ttkill \mid 
           \ttretry \mid M \ttbind N  \\
    &\mid& M \ttfork N \mid \ttif\ M\, \ttthen\ N\, \ttelse\, K \\ 
    &\mid& \lhs{newTVar} \mid \lhs{readTVar}\, r \mid 
     \lhs{writeTVar}\, r\, M \\
    &\mid& \lhs{newPSMVar}\, M\, N \mid 
     \ttmethod\, m\, p\, M \\[1ex] 
  \text{Term}\; M, N, K &::=& 
    x \mid V \mid M\, N \mid \ldots \\[1ex] 
\end{array} 
\] 
\caption{Values and terms.}
\label{fig:values-terms}
\end{figure}

\subsection{Syntax}
\label{sec:syntax}

Fig.~\ref{fig:values-terms} presents the core syntax of STM Haskell~\cite{HarrisMPJH:PPoPP05} extended by the operators for PSM. 
Variables $x, y$ refer to thunks on the stack where access is governed by the rules of the lazy $\lambda$-calculus. Names refer either to STM variables $r$ with single-threaded access, or PSM variables $p$ with shared access. Names $m$ refer to methods, each of which is assumed to have an associated method declaration $m \eqdef K$ where $K$ is an STM term. 

Values $V$ (cf.~\cite{HarrisMPJH:PPoPP05}) are those terms whose evaluation has side-effects and involves the scheduling environment. 
A special form of values are the \emph{pausing} values. These are $\ttpause$ or of form $V \ttbind M$ where $V$ is pausing. A value is \emph{completed} if it is pausing, of form $\ttkill$ or $\ttdone\, M$.
They represent fully evaluated terms at the end of a macro-step.

It will be convenient to consider terms up to  structural congruence induced by the identifications
\begin{eqnarray*}
  M \ttfork N &\equiv& N \ttfork M \\
  M \ttbind (N \ttbind K) &\equiv& (M \ttbind N) \ttbind K 
\end{eqnarray*}
which express commutativity of $\ttfork$ and associativity of $\ttbind$. 

In the untyped value syntax it is possible to overload the control operators \lhs{return} and \lhs{>>=} for STM with \lhs{done} and \lhs{>>>=} for PSM, respectively. Since PSM is the more general context, we use the latter\lvsv{. We rely on typing to disambiguate, in the sense that if $M \ttbind N$ has type \lhs{STM b} then $M \rrcol \lhs{STM a}$ and $N \rrcol \lhs{a -> STM b}$ so that the expression can be written $M \lhs{>>=} N$ in the actual Haskell code.}{ and rely on typing to disambiguate.}

In extension to STM, we now have general objects, consisting of compounds of TVars to implement arbitrary abstract data structures. Through these we also can access IO. Hence IO-actions such as \tcode{putChar} and \tcode{getChar}, modelled as exemplary cases in the STM semantics~\cite{HarrisMPJH:PPoPP05}, are subsumed by method calls $\lhs{method}\, m\, p\, M$.

When a method is defined $m \eqdef K$ then the execution of $m$ is an atomic evaluation of the method body $K$ in STM. Thus, our method calls subsume the \lhs{atomic} construct of STM.

For simplicity, we do not consider the general \lhs{throw}/\lhs{catch} mechanism at STM level, which could be added. Instead we provide the global exception \ttkill at PSM level.

\subsection{Operational Semantics}
\label{sec:sos}

\begin{figure}[t]
\[ 
\begin{array}{rrcl} 
 \text{Heap} & 
  \Theta &::=& \{r, p\} \hookrightarrow M \\ 
 \text{STM Context} & 
  \bbE &::=& [\cdot] \mid \bbE \ttbind M \\ 
 \text{PSM Context} & 
  \bbP &::=& 
     [\cdot] \mid 
    (\bbP \ttfork P) \mid 
    \bbP \ttbind M \\ 
    &&& \ttif\, \bbP\, \ttthen\, N\, \ttelse\, K \\ 
 \text{Action} & 
  a &::=& \sigma \mid \iota \mid \tau \mid \ldots 
\end{array} 
\] 
\caption{Program state and evaluation contexts.}
\label{fig:state-contexts}
\end{figure}

Fig.~\ref{fig:state-contexts} defines state and evaluation contexts. 
A \textit{program configuration} $M; \Theta$ consists of a term $M$ and a \emph{heap} $\Theta$ which is a finite function from names to terms.  
Like in STM Haskell, we have two \emph{evaluation contexts} $\bbP$ and $\bbE$. The former selects a node in the thread tree and is used to implement the PSM scheduler. The latter selects the current action in the active thread and controls the reduction in the STM monad.

\begin{figure*}[ht]

\fbox{PSM Reduction $M; \Theta \lderives{a}{} N; \Theta'$}

\[ 
\begin{array}{rclll}
  \bbP[\ttpause \ttbind M \ttfork \ttpause \ttbind N]; \Theta &
    \lderives{\tau}{} & \bbP[\ttpause \ttbind \lambda x.\, (M() \ttfork N())] ; \Theta
  & 
  & (\PAUSE) \\ 
  \bbP[\ttpause \ttbind M \ttfork \ttdone N]; \Theta &
    \lderives{\tau}{} & \bbP[\ttpause \ttbind 
      \lambda x.\, (M() \ttfork \ttdone N)]; \Theta 
  & 
  & (\PAUSE) \\
  \bbP[\ttdone M \ttfork \ttdone N]; \Theta &
    \lderives{\tau}{} & \bbP[\ttdone\, (M, N)]; \Theta 
  & 
  & (\DONE) \\
  \bbP[\ttkill \ttfork V]; \Theta &
    \lderives{\tau}{} & \bbP[\ttkill]; \Theta 
  & \text{$V$ complete}
  & (\KILL) \\
   \bbP[\ttkill \ttseq N]; \Theta & 
    \lderives{\tau}{} & \bbP[\ttkill]; \Theta 
  & 
  & (\KILL) \\
  \ttpause \ttbind M; \Theta 
  &\lderives{\sigma}{} & M(); \tick(\Theta) 
  & 
  & (\TICK) \\
  \bbP[\lhs{newPSMVar}\, M\, N]; \Theta 
  &\lderives{\iota}{}& 
    \bbP[N\, p]; \Theta[p \mapsto \init(M)] 
  & \text{where } p \not\in \domain(\Theta) 
  & (\NEW) 
\end{array}  
\] 
\begin{center} 
\[ \hfill
 \begin{Prooftree}
  \Bproof
     M \lderives{}{} N 
     \\[(\ADMIN)]
     \bbP[M]; \Theta \lderives{\tau}{} 
       \bbP[N]; \Theta 
  \EEproof 
  \qquad 
  \Bproof
    \Lproof
       K\, p\, M; \Theta \lDerives{\ast}{} \ttdone N; \Theta'
    \ANDproof 
       m \eqdef K
    \ANDproof 
       \forall m' \in \Pi_p(\bbP).\, 
         \Theta \Vdash m' \nprec m 
    \Rproof[(\METHOD)]
    \bbP[\lhs{method}\, m\, p\, M]; \Theta 
      \lderives{\tau}{} 
    \bbP[\ttdone N]; \Theta'
  \EEproof 
  \end{Prooftree}
\hfill
\]
\end{center}

\medskip 

\fbox{Administrative $M \derives{}{} N$}

\[ 
\begin{array}{rclll}
  M & \lderives{}{} & V 
  &\text{if ${\mathcal V}\sem{M} = V$ and $M \not\equiv V$}
  & (\EVAL) \\ 
  \ttdone N \ttbind M & \lderives{}{} & M\, N  
  & 
  & (\BIND) \\ 
  \ttretry \ttbind M & \lderives{}{} & \ttretry 
  & 
  & (\RETRY) \\
  \ttif\, V\, \ttthen\, M_{\ttTrue}\, \ttelse\, M_{\ttFalse} 
    & \lderives{}{} & M_V 
  & \text{if $V \in \{\ttTrue, \ttFalse \}$}
  & (\COND) 
\end{array} 
\] 

\medskip 

\fbox{STM Reduction $M; \Theta \Derives{}{} N; \Theta'$}

\[ 
\begin{array}{rclll}
  \bbE[\lhs{readTVar r}]; \Theta
  &\lDerives{}{}& 
    \bbE[\ttdone \Theta(r)]; \Theta
  & \text{if } r \in \domain(\Theta)
  & (\READ) \\ 
  \bbE[\lhs{writeTVar r M}]; \Theta 
  &\lDerives{}{}& 
    \bbE[\ttdone\, ()]; \Theta[r \mapsto M]
  & \text{if } r \in \domain(\Theta)
  & (\WRITE) \\ 
  \bbE[\lhs{newTVar M}]; \Theta 
  &\lDerives{}{}& 
    \bbE[\ttdone r]; \Theta[r \mapsto M]
  & \text{where } r \not\in \domain(\Theta) 
  & (\ANEW) 
\end{array}  
\]
\[ 
 \begin{Prooftree}
   \Bproof
     M \lderives{}{} N 
     \\[(\AADMIN)]
     \bbE[M]; \Theta \lDerives{}{} 
       \bbE[N]; \Theta 
   \Eproof 
  \end{Prooftree}
\]
\caption{Abstract Operational Semantics of PSM Haskell.}
\label{fig:sos-semantics}
\end{figure*}

The operational rules are seen in Fig.~\ref{fig:sos-semantics}. The core are the administrative steps $M \rightarrow N$ that capture the reduction semantics of the pure (call-by-name) $\lambda$-calculus. In this part we do not need any heap 
information. 
The rule (\EVAL) reduces a non-value term $M$ to a value ${\mathcal V}\sem{M}$ in a single step, with the help of an external evaluation function.  
Rule (BIND) models sequential composition for both PSM and STM, while (\RETRY) implements the \ttretry exception of STM.
This administrative part is as in~\cite{HarrisMPJH:PPoPP05} and entirely standard. 

The semantics of STM is expressed as a transition\footnote{The transitions of~\cite{HarrisMPJH:PPoPP05} also carry allocation effects $\Delta$ that are redundant here since we do not treat general exceptions.} 
\[ 
  M; \Theta \Rightarrow N; \Theta'
\] 
where $M$, $N$ are STM terms and $\Theta$, $\Theta'$ heaps. 
All administrative rules can be applied in an STM term via rule (\AADMIN).
The rules (\READ), (\WRITE), (\ANEW) which handle \lhs{TVar} references in STM are standard~\cite{HarrisMPJH:PPoPP05}.  

While the administrative and STM steps are standard, we now come to discuss the \bbP context and associated reductions 
\[ 
  M; \Theta \lderives{a}{} N; \Theta'
\] 
where the extension of STM Haskell is made. First notice that as in STM Haskell, rule (\ADMIN) lifts reductions in the pure $\lambda$-calculus into the $\bbP$ context. These are \emph{silent} steps labelled $\tau$.
The three forms of thread completion 
are propagated up the tree by the rules (\PAUSE), (\DONE), (\KILL), adopting the semantics of Esterel. Note that the \ttkill completion waits for all threads to complete the current macro-step before it kills the evaluation context. It permits all threads a ``last wish''. This is called a \emph{weak preemption} in Esterel. For contrast, a rule like 
\[ 
  \bbP[\ttkill]; \Theta
    \lderives{\tau}{} \ttkill; \Theta 
\]
would preempt the $\bbP$ context at any moment, without control about how far its active threads have advanced. This would result in a non-deterministic semantics. 
Esterel's deterministic trap mechanims could be easily added to $\bbP$. 

A macro-step of a term $M$ consists of a sequence of PSM reductions with label $a \in \{\tau,\iota\}$, abbreviated $M; \Theta \rightarrow^\ast V; \Theta'$, such that $V$ is complete. If $V \equiv \ttpause \ttbind M$ is pausing, the rule (\TICK) permits the clock to tick, genrating a $\sigma$-step. This activates the next macro cycle with the continuation term $M$ in an updated heap $\tick(\Theta')$, obtained by executing the \emph{tick} function on each PSM variable. If $V$ is $\ttkill$ or $\ttdone M$, then no reduction is possible anymore. We have reached a normal form for the $\bbP$ context.
Notice that the $\sigma$-step can only be executed top-level at the root, not anywhere inside the thread tree, such as with a rule of the form 
\[ 
 \bbP[\ttpause \ttbind M]; \Theta 
  \lderives{\sigma}{} \bbP[M()]; \tick(\Theta). 
\] 
We do not generalise (\TICK) in this way in order to force all threads to synchronise on the clock. 
Note that any IO with the world happens at the clock tick ($\sigma$-step), 
whose effect on the heap is modelled by the associated $\tick$ function. In this way the $\sigma$-actions subsume the IO-actions with labels $!c$ and $?c$ of STM Haskell~\cite{HarrisMPJH:PPoPP05}. Other IO with side-effects on a PSM variable may occur when its memory is initialised and at the end of its life-time ($\init$, $\term$ functions). 

Rule (\NEW) initialises a thread-local instance of a \lhs{PSMVar} on the heap with a fresh name $p$ and passes this reference to the continuation term. We indicate that this may involve IO-actions by adding the action label $\iota$ to this step.

The excution of a method call \hlst{method$\,m\,p\,M$} is handled by rule (\METHOD). It evaluates the method body $K$ by passing the reference $p$ of the variable and the method parameter $M$. The evaluation proceeds as a non-interruptible reduction sequence in STM that runs to completion where it must reach termination. 
When this is not possible, because of a \ttretry exception, the transaction $K\, p\, M$ is abandoned. 
Rule (\METHOD) corresponds to the rule (\ARET) of STM Haskell, except that we execute an STM transaction indirectly as a call $K\, p\, M$ with a \lhs{PSMVar} reference $p$ and argument $M$, rather than a fixed $\ttatomic K$. 
It is the reference $p$ that gives us a handle for policy-based scheduling.  
The novelty of PSM Haskell over STM Haskell lies in the scheduling side condition 
$$
  \forall m' \in \Pi_p(\bbP).\, 
      \Theta \Vdash_p m' \nprec m
$$
of rule (\METHOD) which does not exist in (\ARET).
It expresses that in the current heap $\Theta$, the prediction $\Pi_p(\bbP)$ for $p$ in the concurrent context does not block method $m$ under the policy of $p$.  
This is implemented in STM (cf. Sec.~\ref{sec:psm-scheduling}) by
\[ 
  \lhs{enabled}\, m\, (\Theta(p))\, (\Pi_p(\bbP)); \Theta
     \lDerives{\ast}{} \ttreturn\, \ttTrue; \Theta'.  
\] 
$\Pi_p(\bbP)$ is a multiset of methods on variable $p$ that can be computed from $\bbP$. In our Haskell implementation this is done dynamically by the distributed Cox-Rowers protocol via prediction counters \hlst{pred$\,p\,m$}, as explained in Sec.~\ref{sec:psm-idea}.
\medskip 

Let $\mi{Mtd}(p)$ denote the methods defined for a PSM variable $p$. Coherence requires that all concurrently independent method calls from $\mi{Mtd}(p)$ must be confluent. 

\begin{definition}
 A PSM variable $p$ is \emph{locally coherent} in store $\Theta$ if for all $m_1, m_2 \in \mathit{Mtd}(p)$ with $m_i \eqdef M_i$ and $\Theta \Vdash m_1 \indep{} m_2$ the following holds: For any $N_1$, $N_2$, if $$M_i\, p\, N_i; \Theta \Derives{\ast}{} \ttreturn K_i; \Theta_i$$ then $\Theta_i \Vdash m_{3-i}$ and there exists $\Theta'$ such that $$M_{3-i}\, p\, N_{3-i}; \Theta_{i} \Derives{\ast}{} \ttreturn K_{3-i}; \Theta'.$$
 A PSM variable $p$ is \emph{coherent} if it is locally coherent in all (well-formed reachable) stores. \qed
\label{def:coherence}
\end{definition}

\begin{theorem}
  If all PSM variables occurring in a term $M$ are coherent, then the reduction relation $\derives{\ast}{}$ is confluent. Moreover, if $M \derives{\ast}{} \ttpause \ttbind N_1$ and $M \derives{\ast}{} \ttpause \ttbind N_2$ then $N_1 \equiv N_2$. \qed
\label{thm:determinacy}
\end{theorem}

\begin{proof}
  The reduction patterns in $\bbP$ are non-overlapping and the administrative evaluation of $\lambda$-terms is referentially transparent and thus Church-Rosser. Thus, any conflicting reductions must come from the order of method calls that change the heap. However, since all successful STM method calls (accessing the heap) are coherent, the global confluence for method calls in contexts $\bbP$ follows from the results in \cite{AguadoMPRH18}.
\end{proof}

\section{Conclusion and Related Work}

In this paper we have introduced the new type context \psm for safe concurrent programming that provides determinacy out-of-the-box within GHC's Haskell static typing discipline.
The \psm library is extensible with arbitrary pre-defined shared data structures as instances of the \lhs{PSMType} type class.
These structures must export policy interfaces and be verified to be policy-coherent (race-free) with respect to their interface. 
\onlylv{We have motivated the design of \psm, explained the principles on which it is built and presented several examples of primitive \lhs{PSMVar} structures that can be used to implement a mix of Kahn-style data flow or Berry-style control flow, including IO interaction. 
}

\emph{Policy-controlled CSM types}~\cite{AguadoMPRH18} were proposed originally as a theoretical concept aiming to generalise the model of computation that lies behind synchronous programming languages and extended to abstract data structures (ADS) and multi-threaded shared memory. 
PSM is inspired by that work, from where we have adopted terminology and some notation.
It is the first practical implementation of the scheduling model of CSM. 
Specifically, the function \lhs{runPSM} executes a policy-controlled scheduler implemented in line with the CSM theory~\cite{AguadoMPRH18}.
In fact, our implementation is a simplification  in which the predictions via barrier counters are a conservative over-approximation of the policy-control automata of CSM~\cite{AguadoMPRH18}. Hence, the scheduling of \lhs{runPSM} is conservatively restricting the schedules considered in CSM. 

\onlylv{ 
\onlylv{Our Haskell implementation is a simplification of the scheduling model of~\cite{AguadoMPRH18}.}
The \onlylv{predictions and} enabling used in \psm (barrier counters) are a conservative over-approximation of the \lvsv{concurrent access policies (CAP)}{CAP} of~\cite{AguadoMPRH18}. Those are compositions of policy control automata with predictions consisting of trace sets. These are more expressive, because they preserve the order of actions, where we only record how often an action can occur (Parikh image). Since \psm is conservatively restricting the schedules considered in~\cite{AguadoMPRH18}, the \onlylv{Diamond Property and Macro Step} Determinacy Theorems of~\cite{AguadoMPRH18} apply. 
}

The genericity of PSM, and key advance over CSM, comes from the binding operator \lhs{newPSMVar} for thread-local \lhs{PSMVar_a} variables. 
The \lhs{PSMVar} wrapper defines for each such type \lhs{a} an abstract data structure (ADS)
which may be shared by concurrent threads, protected under the control of a scheduling policy consisting of admissibility, precedence and a clock barrier.
 
Besides providing a prototype implementation, we make two main technical advances over the earlier theoretical developments.
Firstly, in the theory of CSM all object types are fixed in a small imperative kernel language, called DCoL, which is not extensible.
In \psm can encapsulate any Haskell programmable data structure \lhs{a} in the \lhs{PSMType} class by marshalling it with a scheduling policy for a choice of methods. 
Of course, every library of such \lhs{PSMVar}s must be verified to be coherent for the policy interfaces that are exported. 
\green{Since the interface methods of the ADS wrapped by a \lhs{PSMVar} must be controlled by scheduling policies,
the code of the \lhs{PSMVars} library must be curated. The library programmer is responsible for checking coherence, in the same way that programmers have an obligation to verify the monad laws for any application-specific monad instance.}
This is methodologically analogous to the obligation that Haskell programmers have to verify that their code satisfies the monad laws, if they want to use the monadic facilities of Haskell.
Secondly, in CSM all objects are global and statically allocated. A PSM program, for contrast, can dynamically instantiate an arbitray number of lexically-scoped, thread-local \pvar{s} into a thread. 
These extensions are necessary to leverage the theory of CSM for main-stream (concurrent) functional programming, outside the narrow field of synchronous languages. 

The \psm context is only ``monad-like'', not a proper monad, like \stm and \io. 
\onlylv{This is seen, e.g., in the \lhs{exchange} function of Fig.~\ref{fig:pvar-farmer-dist} (lines 18--19) where use the applicative structure (\lhs{pure}, \lhs{(<*>)}) of \lhs{PSMVal} to lift value functions into the \psm context.}
\onlylv{%
This is further seen, e.g., in the \lhs{ioLoop} example of Fig.~\ref{fig:iotest}.
The conditional expression of the \lhs{ifte} is a \lhs{PSVal} action \lhs{do s <- l; return s == ":q"} and not a plain Boolean test \lhs{l == ":q"}.
The reason is that the string input is of type \lhs{l::PSVal String} rather than \lhs{l:String}.
}
This is not an artefact of our implementation. It is 
an intrinsic consequence of the Cox-Rowers scheduling protocol which separates two (run-time) phases, one for predictions (recovery) and one for values (drive), where the former must not depend on the latter.
Consider the function \hlst{f x = if x then$\,f_1\,$else$\,f_2$} of type \hlst{$f\rrcol$ a->PSM b} that selects one of two actions \hlst{$f_1, f_2 \rrcol$ PSM b} based on an input of type \lhs{a}.
This means that the prediction of an application \hlst{$f v \rrcol$ PSM b} depends on the value of $v$. 
Suppose we obtain this value in a monadic bind \hlst{g >>= f} from an upstream action \hlst{$g \rrcol$ PSM a}. Then, the return value of $g$ is not available in the prediction phase. However, if we instead compose $g$ with a function of type \hlst{$f \rrcol$ PSMVal a -> PSM b}, then we are guaranteed that the prediction of the action \hlst{$f v$} can be computed without knowing the value $v$. The reason is that \lhs{PSMVal a} shields the input value so it cannot be accessed in the prediction phase. 
However, since \psm a \emph{relative monad}~\cite{AltenkirchCU:FoSSaCS10} (relative to \lhs{PSMVal}), programmers can use the Qualified do-notation (GHC9) to assemble PSM in convenient monadic style.

As \psm is a type context for reactive programming, it should not surprise that it is not a monad. Functional embeddings, specifically on (stateful) stream processing, typically do not satisfy monad properties. For these, the weaker properties of \emph{arrows}~\cite{hughes:SCP00,Paterson:ICFP01} have been found useful.
Indeed, the type operator \lhs{SF} defined as \lstinline[style=hs]{SF a b = PSMVal a -> PSM b} defines an arrow with input \lhs{a} and output \lhs{b}.
However, the arrow syntax based on the standard operators 
\begin{tabbing} 
  \=\lstinline[style=hs]{arr :: (a -> b) -> SF a b} \\
  \>\lstinline[style=hs]{>>> :: SF b c -> SF c d -> SF b d} \\
  \>\lstinline[style=hs]{first :: SF b c -> SF (b, d) (c, d)}
\end{tabbing} 
would only capture the sequential composition of \psm which is already present from the monadic syntax of \lhs{PSMVal} and the binder \lhs{>>>=} of \psm.  
Arrows would only explain a small part of \psm, namely the control flow arising from termination, which is merely one form of completion. 
The information flow related to clocks (synchronous pausing) and the concurrent structure of \psm arising from \lhs{(|||)} would not be part of this. 
The category-theoretic structure of shared-memory concurrency is highly non-trivial. 
In particular, the notion of causal commutative arrows~\cite{LiuCH:ICFP2009}, which are meant to capture non-interference in arrow structures will not be sufficient. It does not adequately capture the difference between sequential \lhs{(>>>=)} and concurrent \lhs{(|||)} composition of PSM. 
\onlylv{%
Note that the arrow operator \lhs{(***)}, which is derivable from \lhs{arr}, \lhs{>>>} and \lhs{first}, is sometimes referred to as ``parallel''~\cite{PerezBN:Haskell16}. It does not capture true concurrency, however, but a form of sequential composition where the hidden side effects (e.g. here, memory accesses) of the arrow are sequentialised. The operator \lhs{(***)} does not permit ping-pong communication like our \lhs{(|||)}.
}

\onlylv{%
An interesting case in point is the \emph{imperative streams} model~\cite{Scholz:ICFP98}, which happens to be an exception from the rule that FRP on streams is not monadic. 
A Haskell monad \lhs{St a} is defined that represents (possibly finite) episodes of synchronous streams producing values of type \lhs{a} during every cycle they are active.
Haskell's monadic binder \lhs{f >>= \ x -> g} passes the value produced by stream \lhs{f} to the stream \lhs{g} (which may depend on \lhs{x}) at \emph{every} cycle.
This is a \emph{parallel} composition of \lhs{f} and \lhs{g}, communicating with each other synchronously at every macro-step, but a \emph{sequential} composition for the computation of values during a macro-step. 
In particular, the information flow in \lhs{f >>= \ x -> g} is unidirectional.
In \psm, the bind \lhs{f >>>= \ x -> g} is unidirectional, too, but it passes values when \lhs{f} terminates as opposed to when it pauses. 
The parallel communication between \lhs{f} and \lhs{g} in the sense of~\cite{Scholz:ICFP98} is obtained via the independent operator \lhs{(|||)} permitting explicit access to shared \lhs{PSMVar} memory. 
Since \psm is truly concurrent, \lhs{f} and \lhs{g} can then communicate bi-directionally (ping-pong) during a macro-step, which is not possible for imperative streams.  
Note that the monad \lhs{St a} lives outside of \io, whence it is trivially deterministic.
}

\onlylv{%
As pointed out before, the \psm context lies between \stm and \io.
The \lhs{PSMType} class currently does not force data structures to be implemented in \stm, though this is the recommended way of extending the \psm library. \stm forces atomicity and isolation of transactions and makes it easier to maintain coherence at different levels of abstraction. 
The use of \io inside \pvar instances should be restricted to dedicated IO objects interfacing with the external world that must be carefully crafted.
Similar to the monadic extension of Yampa~\cite{PerezBN:Haskell16}, the responsibility is on the \lhs{PSMVar} library designer not the user of the \psm module. 
}

\onlylv{%
The implementation is a simplification of the scheduling model of~\cite{AguadoMPRH18}.
The predictions in PSM (barrier counters) are a conservative over-approximation of the policy-control automata. Hence, the scheduling in our PSM library is conservatively restricting the schedules considered in~\cite{AguadoMPRH18}. Therefore, the determinacy argument still applies.
The generalisations in PSM concern thread-local object instantiation which does not break the determinacy argument. The extension such as to permit full Haskell data types and $\lambda$-calculus under lazy evaluation, do not impinge on the basic combinatorial argument. The PSM programming combinators are essentially those of~\cite{AguadoMPRH18}:
The \lhs{var} combinator implements the method calls \lhs{let x = c.m(e) in P} of~\cite{AguadoMPRH18} while \lhs{newPSMVar} is our extension to instantiate objects \lhs{c} from a library of \lhs{PSMVar}s. The statements \lhs{nothing} and \lhs{pause} are \lhs{skip} and \lhs{pause} of~\cite{AguadoMPRH18}.

We have added a \lhs{kill} statement that preempts the main thread. It permits all running threads to complete their current clock cycle (deterministically according to the theory) and then kills all threads in one sweep across the thread tree. Parallel and sequential composition 
\lhs{(|||)}, \lhs{(>>>=)}, \lhs{(>>>)} are parallel and sequential composition
 $\cpar$ and $\cseq$ of~\cite{AguadoMPRH18}; \lhs{ifte} and \lhs{switch} correspond to branching \lhs{if e then P else P} of~\cite{AguadoMPRH18}. The function \lhs{evalSTM} which embeds a deterministic computation of an STM transaction as a pure value into \psm that executes this transaction atomically. The same applies to \lhs{val} which embeds a side-effect free computation of a Haskell value atomically into PSM.
}

Our interest is on the determinacy of concurrent (in-place, non-monotonic) memory updates. 
Note that \lvars and \ivars provide deterministic concurrency in the more general understanding of  ``concurrency''~\cite{Turon:2013} where "Parallelism always involves concurrency". The point in relation to \lvars and \ivars is that PSM achieves determinacy from commutation (confluence) of actions, which is an operational concept ($\lambda$-calculus), not via monotonicity in abstract lattice structures (domain theory). The former permits general destructive updates, the latter only monotonic updates.
Having said that, we do not exclude the possibility that the synchronisation mechanism of PSM might be encodable into LVars by monotonic threshold operations on abstract lattices. In fact, we believe this is the case. However, such ``deep'' encoding would be a purely theoretical exercise rather than 
practically useful.

\onlylv{
Our work aims to reconcile determinacy and concurrency without un-necessarily restricting expressiveness.
In this sense, it differs from functional reactive programming (FRP) which restricts expressiveness for good reasons.  
The well-known Yampa library~\cite{NilssonCP:Haskell02} is based on stream abstractions in pure Haskell and lives outside of the \io monad.
All user-interaction is packaged in an external ``GUI/reactimate'' function that is composed with the application code to form a closed system.
Since the code is executed in a sequential fashion, determinacy is trivially satisfied.
However, real-time interaction is extremely difficult to control from within the ``reactive'' program.
Also, this cannot leverage shared memory concurrency, which can be important for space-time efficiency. 
\psm can express sharing of memory resources and concurrent decomposition, gives full control over the external reactivity of the program, and still ensures determinate behaviour.
Recent work~\cite{PerezBN:Haskell16} on Yampa, called ``monadic FRP'' makes a move to increase expressiveness this direction. It refactors the restricted language of Yampa stream functions by a monad.
In this way, the programmer can introduce \io into the reaction function at each step of the stream production.
In \psm this would correspond to the use of a concurrent interface process such as the \lhs{ioLoop} in Fig.~\ref{fig:iotest}.
Of course, the \io monad in Yampa breaks the determinism guarantee again.
It would be instructive to refactor FRP into PSM and compare.
}
{
The well-known Yampa library~\cite{NilssonCP:Haskell02}, based on stream abstractions in pure Haskell, lives outside of the \io monad.
All user-interaction is packaged in an external ``GUI/reactimate'' function that is composed with the application code to form a closed system.
Since the code is executed in a sequential fashion, determinacy is trivially satisfied.
Recent work on Yampa, called \emph{Monadic FRP}, introduces monadic \io into the reaction function at each step of the stream production~\cite{PerezBN:Haskell16}. This breaks the determinism guarantee.
It would be instructive to refactor Monadic FRP into PSM, where we can express sharing of memory resources and concurrent decomposition, with full control over the external reactivity of the program and still ensure determinate behaviour.
}

\lvsv{
Languages like concurrent revisions~\cite{BurckhardtFLS12},
VHDL~\cite{KloosB95} or ForeC~\cite{YipRBG13} use special forms 
of shared memory in which (essentially) concurrent writes are
permitted, but reads are delayed by a clock cycle.
However, these approaches are not integrated into a typed functional language and also  
lack support for more expressive shared
ADTs essential for programming complex systems.
We believe that basic mechanism of concurrent revisions can be captured in \psm for general data structures through access policies. 
Since reads are delayed, no precedence between read and write need to be scheduled and coherence is trivial.
}
{
Deterministic parallel programming like concurrent revisions~\cite{BurckhardtFLS12}, VHDL~\cite{KloosB95} or ForeC~\cite{YipRBG13} use special forms of shared memory in which (essentially) concurrent writes are
permitted but reads are delayed by a clock cycle.
Since reads are delayed, no precedence between read and write need to be scheduled and coherence is trivial.
We believe that basic mechanism of concurrent revisions can be captured in PSM for general data structures through access policies. 
}

\lvsv{%
The constructs of \psm constitute a coordination language for assembling race-free concurrent applications from an extensible library of primitive data structures.  
Our focus is on determinacy of concurrent (in-place) memory updates.
We do not address space-time leaks arising from dynamic memory allocation as studied, e.g., by~\cite{Krishnaswami:ICFP13}, for functional stream processing in the sequential rather than concurrent $\lambda$-calculus.
Assuming shared data structures (\lhs{newPSMVar}) are statically allocated, \psm users can exploit in-place memory updates to avoid space leaks by hand. 
In our future theoretical work we plan to investigate extension of typing disciplines such as~\cite{Nakano:LICS00,Guatto:LICS18} for \psm in order to verify the absence of time-leaks (causality deadlock) that arise from the access policies of \psm. 
}
{
We do not address space-time leaks arising from dynamic memory allocation as studied, e.g., by~\cite{Krishnaswami:ICFP13}. In our future theoretical work we plan to investigate extension of typing disciplines such as~\cite{Nakano:LICS00,Guatto:LICS18} for PSM in order to verify the absence of time-leaks (causality deadlock) that arise from the access policies of PSM.
}

\lvsv{
We deliberately focus on concurrency abstractions (as opposed to parallel programming) where performance is not the main concern. Still, we agree that performance tests regarding the overhead introduced by our Cox-Rowers scheduling mechanism would be useful. One might hope to arrange things so that applications using only TVars when embedded into \psm will not be punished, so that PSM is computationally conservative over \stm. However, the envisaged applications of PSM will be where the clock-synchronised scheduling is necessary and non-trivial. The measurements would then show the performance of STM as an implementation of clocked synchronisation. We would then have to compare with other APIs that provide the same functionality and also in Haskell, which do not exist. since this is the first implementation of a new ide, it will likely be improved by later versions. Since our code is a first prototype, performance tests were not our priority at this stage. 
}
{
We deliberately focussed on concurrency abstractions (as opposed to parallel programming) where performance is not the main concern. Performance tests regarding the overhead introduced by our enriched \psm scheduling mechanism would certainly be useful. Since our code is a first functional prototype, performance optimisations are not our priority, at this stage.
}

We found \stm a most natural abstraction for implementing policy-controlled scheduling which essentially reduces precedence of transactions to blocking on \lhs{retry} plus clock synchronisation.
For better performance, it may be possible to integrate precedence blocking 
directly as an extension of STM.

Concerning the existing implementation, there is certainly potential for optimisations. 
Parts of the core scheduling engine could be refactored with \lhs{MVars} or barrier locking in \lhs{IORefs}.
Also, currently we maintain a single thread tree through which all threads and all \lhs{PSMVars} are linked with each other. 
Where this dependency matrix is sparse, better run-time structures would preserve a loose coupling. 
Specifically, \lhs{PSMVar} structures (such as data-flow buffers and queues) that are coherent for precedence-free policies (only blocking on state) do not need to be synchronised by the clock. 
Clusters of threads that are coupled only through precedence-free objects could run asynchronously with independent local clocks. 
This leads to an obvious refinement of PSM to globally asynchronous, locally synchronous scheduling with multiple clocks.
We leave this to future work.


\bibliography{../bib/sp-references}

\newpage 
\pagebreak 

\appendix
\onecolumn

\begin{center}
\Large\bf Appendix 
\end{center}

The purpose of this appendix is to provide a commented description of the implementation of \psm. 
The appendix has three parts, 
\ref{appendix:partA}--\ref{appendix:partC}.
Sec.~\ref{appendix:partA} covers the back-end infrastructure of \psm for policy-based thread scheduling extending the \stm context of Haskell. Sec.~\ref{appendix:partB} describes the main combinators exported as the application programming interface of the \psm context. The combinators give shared multi-threaded access to memory structures abstracted as instances of the \lhs{PSMType} type class. A few examples of primitive \psm memory abstractions are provided in Sec.~\ref{appendix:partC}. These \pvar{s} generalise the \mvar, \tvar, \ivar, \lvar of Haskell.  
These may be extended by other, more complex, data structures by library designers. 
The library designer is responsible for ensuring the coherence of the implementation of an abstract data structure relative to the published policy interface. The application programmer is responsible for ensuring policy-constructiveness of their programs on the other side of the interface.  
The appendix also contains an example of a \psm program in Sec.~\ref{sec:tmp-sig}. It demonstrates the use of Esterel-style temporary signals coded as \pvar{s}. 

\smallskip 

Note that the 
implementation presented here is slightly extended over the submitted research paper. It also implements a combinator \lhs{repUntil} for unguarded repeat loops, i.e. loops which can run through an unbounded number of iterations without pausing. In the research paper we assume that loops are expressed via ordinary Haskell recursion. This, however, requires that each iteration of the loop is broken by a \lhs{pause} synchonisation, like in standard synchronous languages (Esterel, Lustre etc). With \lhs{repUntil} this restriction is now lifted. An arbitrary amount of iterated memory accesses can be exercised within one synchronous tick.

\medskip 

This extended code documentation together with the source code will be made available online and as a github (or similar public) repository. 

\medskip

{\small Smallprint: This file is produced by LaTeX from a Literate Haskell file, i.e. a file that contains both the Haskell source code as well as its LaTeX documentation. In this way, we can guarantee that the available Haskell code is complete and type correct. However, for conciseness, we do not always expose the full Haskell code in our LaTeX description, but a slightly cleaned-up version. In particular, all instrumentation, such as for debugging, has been removed to reduce clutter. The disadvantage is that the following text may contain some typos that were introduced inadvertently as a result the code duplication.    
}

\section{The PSM Scheduling Context}
\label{appendix:partA}

\lhs{PSMType}s when wrapped under \lhs{PSMVar} are abstract and cannot be
used directly. They can be
accessed only within the \lhs{PSM} scheduling context, just like
|MVar|s can be read and written only inside the \lhs{STM} monad
context. Although \lhs{PSM} is not a monad
it behaves almost like a monad.
To export this structure into the Haskell context we provide three
type operators \lhs{PSMVar}, \lhs{PSMVal} and \lhs{PSM} which together implement
the required scheduling control.

\subsection{The \lhs{PSMVar} and \lhs{PSMVal} type Operators}
\label{sec:PSMVarVal}

Each \lhs{PSMVar} that is to be passed to a function needs to be
protected by a wrapper that adds prediction information from
the context (that has access to the \lhs{PSMVar}) outside of and concurrent 
to the thread evaluating the function. Likewise, the
expected accesses arising from the function application must
be communicated to the concurrent environment. This is done by
encapsulating a variable inside a type operator \lhs{PSMVar} which
adds essentially two things: A unique identifier of type
\lhs{PSMVarId} for the \lhs{PSMVar} to identify it to the scheduler and
a prediction table of type \lhs{PSMVarPotMap} that contains all
potential method accesses to the given \lhs{PSMVar} by all threads
that have access to it. 

\ignore{%

\begin{code}
data PSMVar a = Idle String | Active PSMVarId String (PSMState a) (TVar PSMVarPotMap)

instance Show (PSMVar pa) where
  show (Active vId _ _ _) = show vId ++ "(...)"
  show (Idle sId) = "_" ++ show sId
\end{code}

} 

\begin{spec}
data PSMVar a = Idle | Active PSMVarId (PSMState a)  (TVar PSMVarPotMap) 

PSMVarPotMap  = Map ThreadId (Prediction, Prediction)
data PSMVarId = PSMVarId Int deriving (Show, Read, Eq, Ord)
type Prediction = Map PSMVarId Int
\end{spec}

\noindent
The type \lhs{PSMVar} has two constructors \lhs{Idle} and 
\lhs{Active}. They represent the two phases of evaluation for a
\lhs{PSMVar}. Only in the \lhs{Active} phase the \lhs{PSMVar} is allocated
in memory and accessible for evaluation. It is here 
where the \lhs{TVar PSMVarPotMap} stores the prediction for
scheduling control. The prediction over-approximates the potential memory accesses still to be expected by all running threads during the current macro-step. We also call this the \emph{(capability) potential} in analogy with the terminology introduced by Boussinot 
for synchronous programming. 
The potential is contained in a map \lhs{Map ThreadId (Prediction, Prediction)} that associates two method counters \lhs{Prediction} for each \lhs{ThreadId}. 
Suppose the prediction of an active \lhs{vId::PSMVarId} is \lhs{pred::PSMVarPotMap}.
For a running thread \lhs{tId::ThreadId} and method \lhs{mId::MethodId}, the first of these \lhs{fst(pred!tId)!mId::Int} tells us how many times \lhs{tId} may  still execute \lhs{mId} on any instantaneous control-flow path, until completion, and without repeating any program loop (\lhs{repUntil}). 
Each time \lhs{tId} executes \lhs{mId}, this counter is decremented by the thread. The second counter \lhs{snd(pred!tId)!mId::Int} is for keeping track of how many backward loops might still be taken by the thread, before completion, which would reintroduce a potential for excuting \lhs{mId}.
If all loops are guarded by pauses, i.e., there is no instantaneous control-flow loop, then this second counter is always $0$ and can be ignored in the scheduling. This is what we assume in the accompanying research paper.

The potential \lhs{pred::PSMVarPotMap} in the \pvar gets updated every time (i) a new thread with access to the \lhs{PSMVar} is spawned, (ii) a method is executed on
the \lhs{PSMVar}, (iii) a conditional branching is evaluated, (iv) a loop is repeated or exited or (v) a new synchronous macro-step is started. 
\lhs{Idle} is used for \lhs{PSMVars} which are already in the prediction of a thread but not yet allocated and thus not yet active.
The reference gets destroyed when the thread which allocated the variable leaves the local scope of the variable or dies.

\smallskip

In contrast to \lhs{PSMVar} which is wrapping volatile shared memory in \psm, the type operator \lhs{PSMVal} is wrapping thread-local pure values.  
The \lhs{PSMVal} type operator is duplicating the \lhs{Maybe} monad by virtue of the following declarations:

\begin{code}
data PSMVal a = Bot | PSMVal a

instance Functor PSMVal where
  fmap f Bot = Bot
  fmap f (PSMVal v) = PSMVal (f v)

instance Applicative PSMVal where 
  pure = PSMVal 
  Bot <*> _ = Bot 
  _   <*> Bot = Bot
  (PSMVal f) <*> (PSMVal v) = PSMVal (f v)

instance Monad PSMVal where
  Bot >>= f = Bot
  (PSMVal v) >>= f = f v

\end{code}

\ignore{
\begin{code}

instance Show a => Show (PSMVal a) where
  show Bot = "undef"
  show (PSMVal v) = show v

-- unsafe, not to be exported
eval :: PSMVal a -> a
eval Bot = error "eval: value undefined"
eval (PSMVal v) = v

\end{code}
}

\noindent 
The monad structure of \lhs{PSMVal} induces natural syntax to embed ordinary Haskell evaluations inside the \lhs{PSM} context. For instance, 
suppose we have a value \lhs{val::PSMVal Int} returned from a method call and want to test if this value is equal to \lhs{59::Int}, returning a \lhs{PSMVal Bool} that can be used to control the evaluation of a conditional control-flow in a \lhs{PSM} process. With the applicative structure we can write this test predicate 
as 
\begin{spec}
   pure (==) <*> val <*> pure 59
\end{spec}
or in monad syntax
\begin{spec}
   do x <- val; return (x == 59).
\end{spec}
Any Haskell function of type 
\lstset{mathescape=true}
\begin{spec}
f :: $a_1$ -> $a_2$ -> $\cdots$ -> $a_n$ ->$b$
\end{spec}
thus lifts to a function of type 
\begin{spec}
lift$_n$ f :: PSMVal $a_1$ -> PSMVal $a_2$ -> $\cdots$ -> PSMVal $a_n$
lift$_n$ f =   $\lambda\,$ $xv_1$ $xv_2$ $\cdots$ $xv_n$ -> do $x_1$ <- $xv_1$; $x_2$ <- $xv_2$; $\ldots$; $x_n$ <- $xv_n$; return f $x_1$ $x_2$ $\cdots$ $x_n$
\end{spec}
\lstset{mathescape=false}

\subsection{The \lhs{PSM} Type Operator}

\ignore{%

\begin{code}
data PSM a = 
  PSM (ThrdPSMVarMap, Int,
         TVar String -> TVar ThrdEnv -> ThrdPSMVarMap -> IO a)
\end{code}

}


The type operator \lhs{PSM} provides the synchronisation 
context for the evaluation of a thread in a concurrent environment. 
A \psm process \lhs{proc::PSM a}, where 
\begin{spec}
data PSM a = PSM (ThrdPSMVarMap, TVar ThrdEnv -> ThrdPSMVarMap -> IO a)
\end{spec}
when it is installed and executed will produce a value of type \lhs{IO a} upon termination. 
A process value \lhs{proc::PSM a} is a pair 
\lhs{proc = PSM(pred, act)}
where the first component \lhs{pred::ThrdPSMVarMap} is a \emph{prediction} of type
\begin{spec}
type ThrdPSMVarMap = Map CmplId (Map PSMVarId Prediction)
newtype CmplId = CmplId Int
type Prediction = Map MethodId Int
\end{spec}
The process' \emph{prediction} \lhs{pred} summarises all the \lhs{PSMVar} memory accesses potentially made in the evaluation of the process' \emph{action} \lhs{act} during the first macro step if and when \lhs{act} is run. 
The prediction is relevant at the moment
when (i) \lhs{act} is started in a freshly forked thread, (ii) at the start of a new tick of a pausing thread, or (iii) when a loop is rentered through a backward control-flow edge.
The prediction \lhs{pred} is used by the main thread to initialise the barrier counters (i.e., the potential) of the active \pvar{s}.
This prevents the concurrent environment from premature accesses that conflict with \lhs{proc} according to the scheduling policy.

\smallskip 

A \lhs{ThrdPSMVarMap} is a map containing a method counter \lhs{Prediction} for each way in which the action may complete, indexed by \lhs{CmplId}, and each \pvar indexed by \lhs{PSMVarId} (see below). 
A prediction \lhs{pred::ThrdPSMVarMap}, in particular, provides a prediction of the completion codes for the action. 
We use the three completion codes, viz.\ $0$ for instantaeous \emph{termination}, $1$ for \emph{pausing} and $2$ for \emph{looping}.
If code $0$ is present as a key in \lhs{pred}, i.e., if \lhs{member (CmplId 0) pred == True} then the action has an instantaneous control-flow path ending in termination. We call these paths \emph{through paths}. If code $1$ is present, the action contains a \emph{sink path}. This is a control-flow path that ends in a pause instruction. Finally, the presence of completion code $2$ indicates that the action can potentially execute an instantaneous loop.  
The completion information is needed to compute the method predictions for sequential composition of actions. 

\smallskip 

The second component \lhs{act::TVar ThrdEnv -> ThrdPSMVarMap -> IO a} of a process is the evaluation \emph{action}. 
For compositionality, the action \lhs{act} is generic and can be instantiated to a particular execution context, consisting of a reference to a thread context \lhs{TVar ThrdEnv} and a prediction map \lhs{ThrdPSMVarMap}.
These capture the context of threads \emph{concurrent} with \lhs{act} and
the \emph{sequential} context of the action.
In an application \lhs{act ctxRef cntPred :: IO a} the former is captured 
by a given thread environment stored in \lhs{ctxRef::TVar ThrdEnv} and the latter by a thread prediction \lhs{cntPred::ThrdPSMVarMap}.  
The thread context \lhs{ctxRef} contains references to the current thread tree and to a global map listing all the \lhs{PSMVars} owned by each thread.
The structure of \lhs{ThrdEnv} is described below in section Sec.~\ref{sec:threads}.
The second argument \lhs{cntPred} of type \lhs{ThrdPredMap} represents the predictions for
the processes sequentially downstream from \lhs{act}. These must be passed to \lhs{act} so it can set the potentials for the next macro-step when the control flow pauses inside \lhs{act}.     

\smallskip

The separation into prediction and action applies generally to all
\lhs{PSM} type operators. The parameters \lhs{PSMVar} and \lhs{PSMVal} of a generic \lhs{PSM} process
\begin{spec}
  proc :: PSMVar a => PSMVar a -> PSMVal b -> PSM c
\end{spec}
also have predictions and actions, except that these are degenerated in different ways. 
The prediction of \lhs{PSMVar a} is the identifier of the reference and the
action is the memory structure \lhs{a}. The prediction of a \lhs{PSMVal b} is 
the unit type \lhs{()} and the action of a \lhs{PSMVal b} is a value of type 
\lhs{b}. Then, \lhs{proc} is a \emph{layered} (2-stage) function such that the prediction 
of \lhs{proc var val} only depends on the predictions of
\lhs{var} and \lhs{val} while the action of \lhs{proc var val}
may depend on both the predictions and the action of \lhs{var} and \lhs{val}.
In effect, this means that the prediction of the process \lhs{proc var val}
only depends on the identifier of the reference \lhs{var} but not on the memory
value of \lhs{var} or the value of \lhs{val}. In other words, we can extract the
prediction of \lhs{proc var val} without knowing the values of \lhs{var} or \lhs{val}.
This is explained further in the next section~\ref{sec:run-method}.

\smallskip

The tracking of predictions and potentials manipulates method counters \lhs{Prediction}. These naturally form a commutative group, with neutral element, addition and subtraction:
\begin{code}
zeroPred :: Prediction
zeroPred = empty
addPred :: Prediction -> Prediction -> Prediction
addPred = unionWith (+)
subPred :: Prediction -> Prediction -> Prediction
subPred = unionWith (-)
\end{code}
The binary operation \lhs{addPred} is used to 
combine predictions from concurrent and sequential threads during the prediction phase, or to recharge potentials upon loop reiteration.
Subtraction \lhs{subPred} is used to reduce potentials during the action phase, as the control flow moves forward, e.g., when method calls are executed or loops are exited. 
The constant \lhs{zeroPred} is the neutral element for \lhs{addPred}. 
Furthermore, we need maximum operation and difference for control-flow branching:
\begin{code}
maxPred :: Prediction -> Prediction -> Prediction
maxPred = unionWith max
diffBranchPred :: Prediction -> Prediction -> Prediction
diffBranchPred pred1 pred2 = 
  let apred2 = union pred2 pred1 in
    unionWith (\x y -> if x >= y then 0 else y-x) pred1 apred2
\end{code}
The maximum \lhs{maxPred} is used to over-approximate the predictions of a conditional statement, or generally where control flow branches depend on computed values during the action phase. 
The ``difference'' \lhs{diffBranchPred pred1 pred2} computes for each method the
value by which the over-approximation \lhs{maxPred pred1 pred2}
must be reduced if the prediction \lhs{pred2} can be excluded when a conditional branch is taken, thereby eliminating \lhs{pred2} from the current potential.

\smallskip

Note that the function \lhs{unionWith} (from module \lhs{Data.Map}) acts neutrally for all keys that are contained in only
one of the two maps. However, for methods occurring in \lhs{pred1} but not in
\lhs{pred2} the ``difference'' should be $0$. Therefore, we must use 
\lstinline[style=hs]{apred2 = union pred2 pred1} instead of \lhs{pred2}.

\subsection{Running a \lhs{PSM} Method (\lhs{callVal} and \lhs{bindVar})}
\label{sec:run-method}

\ignore{
\begin{code}
callVal :: PSM(a -> b) -> PSMVal a -> PSM b
callVal (PSM (pred, trp, fact)) (PSMVal v) = PSM (pred, trp, act) where
  act ioRef ctxRef cntPred = do 
    f <- fact ioRef ctxRef cntPred
    return $ f v -- $
callVal (PSM(pred, trp, fact)) Bot = PSM (pred, trp, act) where
  act _ _ _ = error "callVal: undefined value parameter"
\end{code}
}

\noindent
Suppose we have a process \lhs{PSM (a -> b)} computing a function inside the \psm context and receive a value \lhs{PSMVal a} from a method call.
To apply the function to the value we use the \lhs{callVal} combinator:
\begin{spec}
callVal :: PSM (a -> b) -> PSMVal a -> PSM b
callVal (PSM (fPred, fAct)) (PSMVal v) = PSM (fPred, act) where
  act ctxRef cntPred = do 
    f <- fAct ctxRef cntPred
    return $ f v -- $
callVal (PSM (fPred, fAct)) Bot = PSM (fPred, act) where
  act _ _ = error "callVal: undefined value parameter"
\end{spec}
The application \lhs{callVal fProc (PSMVal v)} with \lhs{fProc = PSM (fPred, fAct)} has the same prediction for memory accesses as the process \lhs{fProc} that computes the function.
This prediction \lhs{fPred} is passed out directly. 
The action \lhs{act} of the application first executes the action \lhs{fAct} that obtains the function and then applies the computed function \lhs{f} to the value \lhs{v}.
As long as the value is undefined, the application \lhs{callVal fproc Bot} has a prediction, too, which is the prediction of the function process \lhs{fProc}. The action, however, is undefined in this case and will never be executed.  We raise an error if this happens for diagnostics.   

\smallskip

All \stm transactions can be embedded as atomic actions inside \psm, wrapping the \lhs{atomically} operator of \stm: 
\begin{spec}
evalSTM :: PSM (STM a) -> PSM a
evalSTM (PSM (pred, act)) = PSM (pred, act') where
  act' ctxRef cntPred = do
    stm <- act ctxRef cntPred
    atomically stm
\end{spec}
Using \lhs{evalSTM} we can make \stm actions interact with each other concurrently in \io, under the protection of the \psm context.
Memory access methods in \psm are generated with \lhs{evalSTM} from \stm transactions that depend on external parameters.

\smallskip 

To construct a method \lhs{m} for a \lhs{PSMVar a} such that \lhs{b = PSMVarMtdIn a m} and \lhs{d = PSMVarMtdOut a m} we use the operator
\begin{spec}
runSTMMtd :: PSMType p => Method p a b -> 
  (PSMState p -> a -> STM b) -> 
      PSMVar p -> PSMVal a -> PSM b

runSTMMtd mtd f p b = 
  bindVar mtd g p where 
    g = \pp -> evalSTM $ callVal (callVal (pure m) pp) b -- $
\end{spec}
The function \lhs{runSTMMtd} renders a functional \stm transaction \lhs{f mtd::a->b->STM d} programmed in the 
ambient Haskell context for a \lhs{PSMVar} \lhs{p} and method argument \lhs{b} for execution in the \lhs{PSM} context. 
The wrapped \psm method \lhs{runSTMMtd mtd f} has type \lhs{PSMVar a -> PSMVal b -> PSM d}. This is a method that can be called on a \lhs{PSMVar a} with a method value parameter \lhs{PSMVal b} and it produces a return value of type \lhs{d} in the protected \psm context.

\smallskip 

\ignore{

\begin{code}

evalSTM :: PSM (STM a) -> PSM a
evalSTM (PSM(pred, trp, act)) = PSM(pred, trp, act') where
  act' ioRef ctxRef cntPred = do
    stm <- act ioRef ctxRef cntPred
    atomically stm

\end{code}

\begin{code}

-- unsafe, not to be exported
unsafeEvalIO :: PSM (IO a) -> PSM a
unsafeEvalIO (PSM(pred, trp, act)) = PSM(pred, trp, act') where
  act' ioRef ctxRef cntPred = do
    stm <- act ioRef ctxRef cntPred
    stm

unsafeIO2PSM :: IO a -> PSM a 
unsafeIO2PSM io = unsafeEvalIO $ val (pure io)


runIOMtd :: PSMType p => 
  Method p a b -> (Method p a b -> PSMState p -> a -> IO b) -> 
     PSMVar p -> PSMVal a -> PSM b

runIOMtd mtd f p b =
  bindVar mtd g p where 
    g = \pp -> unsafeEvalIO $ callVal (callVal (pure (f mtd)) pp) b
  
runSTMMtd :: PSMType p => Method p a b -> 
  (PSMState p -> a -> STM b) -> 
      PSMVar p -> PSMVal a -> PSM b

runSTMMtd mtd m p b = 
  bindVar mtd g p where 
    g = \ pp -> evalSTM $ callVal (callVal (pure m) pp) b -- $

\end{code}

} 

The main code for accessing a \lhs{PSMVar} resides in the \lhs{bindVar} function. It extracts from a variable \lhs{PSMVar a} a value \lhs{PSMVal a}. This access can be associated with a method index \lhs{MethodId} and is scheduled, with this index, under the policy of the \lhs{PSMType a}.

\lstset{numbers=left,numberstyle=\tiny}
\begin{spec}
bindVar :: PSMType a => Method a d c -> (PSMVal (PSMState a) -> PSM b) -> PSMVar a -> PSM b

bindVar mtd f (Active vId _ pvar predMapRef) = PSM(pred', act') where
  PSM(pred, _)  = f Bot
  pred' = Map.map (\p -> unionWith addPred 
            (Map.singleton vId (fromList [(mtd, 1)])) p) pred
  act' ctxRef cntPred = do
    let PSM(_, act) = f (PSMVal pvar)
    -- check enabledness
    thrdPath <- atomically $  waitForEnabled mtd vId pvar predMapRef ctxRef
    -- execute function call
    x <- act ctxRef cntPred
    -- adjust prediction   
    atomically $ do -- $ 
      predMap <- readTVar predMapRef
      let canDn (potCnt, recCnt) = Just $ 
          -- reduce the transient count, keep recurrent count
            (adjust (\x -> x - 1) (methodId mtd) potCnt, recCnt) 
      writeTVar predMapRefx $ Prelude.foldr (update canDn) predMap thrdPath
    -- return method result
    return x

bindVar _ f (Idle sId) = PSM (pred', act') where
  PSM(pred', _)  = f Bot
  act' _ _ = error $ "bindVar: cannot execute on idle pvar" ++ sId

\end{spec}
\lstset{numbers=none}

\ignore{

\begin{code}
bindVar :: PSMType a => Method a d c -> (PSMVal (PSMState a) -> PSM b) -> PSMVar a -> PSM b


bindVar mtd f (Active vId _ pvar predMapRef) = 
  PSM(pred', trp, act') where
    PSM(pred, trp, _)  = f Bot
    pred' = Map.map (\p -> unionWith addPred 
              (Map.singleton vId (fromList [(methodId mtd, 1)])) p) pred
    act' ioRef ctxRef cntPred = do
      let PSM(_, _, act) = f (PSMVal pvar)
      -- check enabledness
      thrdPath <- atomically $ waitForEnabled mtd vId pvar predMapRef ctxRef
      -- execute function call
      x <- act ioRef ctxRef cntPred
      -- adjust prediction   
      atomically $ do -- $ 
        predMap <- readTVar predMapRef
        let canDn (potCnt, recCnt) = Just $ 
            -- reduce the transient count, keep recurrent count
              (adjust (\x -> x - 1) (methodId mtd) potCnt, recCnt) 
            -- (adjust (\x -> x - 1) mId potCnt, recCnt)
        writeTVar predMapRef $ Prelude.foldr (update canDn) predMap thrdPath
      -- return method result
      return x

bindVar mId f (Idle sId) = PSM (pred', 0, act') where
  PSM(pred', _, _)  = f Bot
  act' _ _ _ = error $ "bindVar: cannot execute on idle pvar" ++ sId

\end{code}
}

The application \lhs{bindVar mtd f pvar} generates a process \lhs{PSM(pred',act')} to execute the method body \lhs{f} with method index \lhs{mId} on the \lhs{pvar}, which may be \lhs{Active} (lines 3--21) or \lhs{Idle} (lines 23--25). 
The prediction \lhs{pred'} for this access on an active/allocated 
\pvar takes the prediction \lhs{pred} of the method body \lhs{f}  (line 4) and increments the counter for \lhs{mtd} by $1$ (line 5--6). 
For an \lhs{Idle} and thus inactive or non-allocated \pvar, the prediction of all other \pvar{s} in the method body is simply passed back unchanged (lines 23--24). 

\smallskip 

Let us look at the memory actions. First note that we never execute a method \lhs{f} on a \pvar which is \lhs{Idle} or in the process of being allocated, hence the error message in line~25. 
The \psm action \lhs{act'} for the \lhs{Active} case is seen in lines~7--21. 
It performs the following steps:
\begin{itemize}

\item Line 8--10: We extract the action of the method \lhs{f} on the \pvar and wait for enabledness of the method call on \pvar with identifier \lhs{vId}.

\item Line 12: When the method is enabled, then we execute 
the action \lhs{act} of \lhs{f} which returns a value \lhs{x}.

\item Lines 14--19: We adjust the potential for the
executing thread since the method just executed now is no longer
part of the prediction. For this, we decrement the method count
with \lhs{canDn} for the given thread and all its ancestors
(obtained as \lhs{thrdPath} in line~10) in the potential \lhs{predMap} of the \lhs{PSMVar}. We reduce the potentials for the thread's ancestors since each method call of a thread counts also
as a method call of its ancestors. In this way, the enabling test for
a thread only needs to visit the \emph{direct} children of the thread's
ancestors, rather than \emph{all descendants} of every ancestor.

The function \lhs{canDn} (lines~16--18) implements the change of potential
that happens when the method \lhs{mId} is executed. More precisely, this is decrementing the barrier counter \lhs{potCnt} for the transient calls to \lhs{mtd} but keeps the barrier counter \lhs{recCnt} for predicted recurrent calls.

\item Line 21: Finally, the return value is passed back.
\end{itemize}

\noindent
The blocking of \lhs{bindVar} is achieved in the \lhs{STM} monad using
the auxiliary function \lhs{waitForEnabled} in line~10. This function is
defined thus: 

\lstset{numbers=left,numberstyle=\tiny}
\begin{code}

waitForEnabled :: PSMType p =>  
  Method p a b -> PSMVarId -> PSMState p -> TVar PSMVarPotMap -> TVar ThrdEnv -> STM [ThrdId]
waitForEnabled mtd vId pvar predMapRef ctxRef = do
  ctx <- readTVar ctxRef
  let (TTreeNode (_, _, _, _, pVarGlobMapRef, _) _ _ _ _) = ctx
  pVarGlobMap <- readTVar pVarGlobMapRef
  if not (member vId pVarGlobMap)
    then error "waitForEnabled: PSMVar not known" 
    else do
      pVarxPredMap <- readTVar predMapRef
      (concThrds, thrdPath) <- collectSpine (\ (x, _, _, _, _, _) -> x) ctxRef 
      let (potCnt, recCnt) = collectPSMVarPot concThrds pVarxPredMap in do
        c <- enabled mtd pvar potCnt recCnt
        if c then return thrdPath else retry
\end{code}
\lstset{numbers=none}

\begin{itemize}

 \item In lines 5--10 we check that the \lhs{PSMVar} with identifier
 \lhs{vId} is actually accessible and fully instantiated in memory.
 This is for code robustness only, as in a well-formed program obtained 
 through the exported abstract interface this should never fail.

 \item Lines 11--13 collects the potential on the \lhs{PSMVar} currently 
 recorded by all concurrent threads. For this we need to traverse 
 the thread tree in the environment. We use the function
 \lhs{collectSpine} (from the \lhs{TTree} package)\footnote{The \lhs{TTree} module is not decribed here. It provides triply-linked trees to store and navigate in the thread tree. It links each node to its parent, children and siblings.} to get hold of all
 concurrent threads and then the function \lhs{collectPSMVarPot} to
 aggregate all their predictions, both transient and recurrent:
\begin{code}
collectPSMVarPot :: [ThrdId] -> PSMVarPotMap -> (Prediction, Prediction)
collectPSMVarPot tIds pMap = 
  Prelude.foldr add2Potential (zeroPred, zeroPred) $ 
    Prelude.map thrdToPSMVarPot tIds where 
      add2Potential (x1, y1) (x2, y2) = (addPred x1 x2, addPred y1 y2)
      thrdToPSMVarPot tId = findWithDefault (zeroPred, zeroPred) tId $ pMap
\end{code}

 \item The \lhs{enabled} test for the method is performed in line 14--15. 
 When it succeeds, we return the thread identifiers of all ancestors
 in the return value \lhs{thrdPath}. Otherwise, we use the \lhs{retry} combinator
 of the \lhs{STM} monad to abandon the evaluation and retry until it
 succeeds. This may never happen, though, if the program is not policy-constructive (this can be checked by static program analysis in an approximate way, see e.g.~\cite{AguadoMPRH18}). We assume that all \lhs{PSM} 
 programs have been statically verified to be policy-constructive.
 In the current prototype implementation we leave
 the check for deadlocks to the \lhs{STM} scheduler. In an advanced version we
 plan to catch this in the \lhs{PSM} context. 
A natural solution would be to use Template Haskell to implement static schedulability analysis on the \psm structure, similar to constructiveness analyses in synchronous programming.
It will further be interesting to see if Haskell's template mechanism can be leveraged to make PSM look like a monadic extension, essentially hiding the use of \lhs{PSMVal}. 

\end{itemize}

The central \lhs{enabled} predicate to check if a method identifier \lhs{mId} on a \lhs{PSMType a} can proceed under at a potential \lhs{potCnt} and \lhs{recCnt} is defined thus:
\lstset{numbers=left,numberstyle=\tiny}
\begin{spec}
enabled :: PSMType a => Method a b c -> PSMState a -> Prediction -> Prediction -> STM Bool
enabled mtd pvar potCnt recCnt = do
  let totalPotential = addPred potCnt recCnt 
  maybeInhibs <- prec pvar (methodId mtd)
  case maybeInhibs of
    Nothing -> return False
    Just inhibs -> return $ -- $
      and (Prelude.map (\b -> findWithDefault 0 b totalPotential <= 0) inhibs)
\end{spec}
\lstset{numbers=none}
It first sums up the transient count and recurrent count of methods (line~3), looks up the method indices in the policy of the \lhs{PSMType a} that take precedence over \lhs{mId} (line~4) and then checks that if there are any inhibitors, the total method count in the potential for these inhibitors is zero (lines~5--8).  

\smallskip 

The special case of \lhs{bindVar} above in which the method \lhs{f} to receive the value from the \pvar is the ``identity method'' \lhs{val:: PSMVal a -> PSM a} yields the \lhs{var} function  
\begin{code}
var :: PSMType a => Method a b c -> PSMVar a -> PSM (PSMState a)
var mtd = bindVar mtd val
\end{code} 
This function represents the direct access to a \pvar inside the \psm context. 
The \lhs{bindVar} function can be obtained from \lhs{var} and action sequencing \lhs{(>>>=)} that will be introduced below in Sec.~\ref{sec:sequencing}.

\subsection{Threads}
\label{sec:threads}

All active threads are linked with each other on a (binary) tree of nodes,
the \emph{thread environment} \lhs{ThrdEnv}:
\begin{spec} 
type ThrdEnv = TTree ThrdTTreeNode
\end{spec}
Each node in this tree
has type \lhs{ThrdTTreeNode} and stores the information to identify the process
thread running in it:
\begin{spec} 
type ThrdTTreeNode = (ThrdId, ThrdStatus, TVar PSMVarGlobMap, TVar PSMVarLocMap)
\end{spec}

\ignore{
\begin{code} 

data ThrdStatus =
  Init ThrdPSMVarMap | Start | Running | Terminated | Pausing ThrdPSMVarMap | Kill | Exiting
   deriving (Show)

type ThrdTTreeNode =
  (ThrdId, -- thread Id 
   ThrdStatus, -- status of thread 
   TVar Int, -- current counter of PSMVars (for this thread?!)
   TVar Int, -- current counter of trap points 
   TVar PSMVarGlobMap, -- reference to all PSMVars
   TVar PSMVarLocMap -- reference to list of thread local PSMVars
  )

type ThrdEnv = TTree ThrdTTreeNode

\end{code}
}

\noindent 
More specifically, each thread node \lhs{ctx::ThrdTTreeNode} has the components
\begin{spec}
ctx = (tId, tSt, pVarGlobMap, pVarLocMap)
\end{spec}
where \lhs{tId::ThrdId} is the thread's id, \lhs{tSt::ThrdStatus} its thread status,  
\lhs{pVarGlobMap::TVar PSMVarGlobMap} a reference to the accessible \lhs{PSMVar}s and \lhs{pVarLocMap::TVar PSMVarLocMap} 
the map of \lhs{PSMVar}s locally created by the given thread at hand.
To access the thread status there are getter and setter
functions such as
\begin{spec}
getStatus    :: ThrdEnv -> ThrdStatus
setStatus    :: ThrdEnv -> ThrdStatus -> ThrdEnv
isStatus     :: ThrdEnv -> ThrdStatus -> Bool
getThreadId  :: ThrdEnv -> ThreadId
\end{spec}

\ignore{

\begin{code}

-- status
getStatus :: ThrdEnv -> ThrdStatus
getStatus ctx =
  case ctx of 
    (TTreeNode (_, tSt, _, _, _, _) _ _ _ _) -> tSt

-- id
getThreadId :: ThrdEnv -> ThreadId
getThreadId (TTreeNode (ThrdId tId _, _, _, _, _, _) _ _ _ _) = 
  tId

getPSMVarMax :: ThrdEnv -> STM Int
getPSMVarMax (TTreeNode (_, _, numRef, _, _, _) _ _ _ _) = 
  readTVar numRef

grabPSMVarInd :: ThrdEnv -> STM Int
grabPSMVarInd (TTreeNode (_, _, numRef, _, _, _) _ _ _ _) = do
  num <- readTVar numRef
  writeTVar numRef (num + 1)
  return num    

-- trapCntRef
getTrapMax :: ThrdEnv -> STM Int
getTrapMax (TTreeNode (_, _, _, numRef, _, _) _ _ _ _) = 
  readTVar numRef

grabTrapInd :: ThrdEnv -> STM Int
grabTrapInd (TTreeNode (_, _, _, numRef, _, _) _ _ _ _) = do
  num <- readTVar numRef
  writeTVar numRef (num + 1)
  return num    

isStatus :: ThrdEnv -> ThrdStatus -> Bool
isStatus (TTreeNode (_, stat, _, _, _, _) _ _ _ _) s = compare s stat where 
  compare (Init _) (Init _) = True
  compare Start Start = True
  compare Running Running = True
  compare Terminated Terminated = True
  compare (Pausing _) (Pausing _) = True
  compare Kill Kill = True
  compare Exiting Exiting = True
  otherwise = False

isTerminated :: ThrdEnv -> Bool
isTerminated ctx =
  case ctx of 
    TTreeNode (_, Terminated, _, _, _, _) _ _ _ _ -> True
    otherwise -> False

isPausing :: ThrdEnv -> Bool
isPausing ctx =
  case ctx of
    TTreeNode (_, Pausing _, _, _, _, _) _ _ _ _ -> True
    otherwise -> False

isStart :: ThrdEnv -> Bool
isStart ctx =
  case ctx of
    TTreeNode (_, Start, _, _, _, _) _ _ _ _ -> True
    otherwise -> False

isKill :: ThrdEnv -> Bool
isKill ctx =
  case ctx of
    TTreeNode (_, Kill, _, _, _, _) _ _ _ _ -> True
    otherwise -> False

setStatus :: ThrdEnv -> ThrdStatus -> ThrdEnv
setStatus ctx st =
  case ctx of 
    (TTreeNode (tId, _, numRef, trapCntRef, pVarGlobMapRef, pVarLocMapRef) tpar tsib tlft trgt) 
      -> TTreeNode (tId, st, numRef, trapCntRef, pVarGlobMapRef, pVarLocMapRef) tpar tsib tlft trgt

\end{code}

}

The derived getter and setter actions for the status on mutable references are called 
\begin{spec}
getStatusRef :: TVar ThrdEnv -> STM ThrdStatus 
setStatusRef :: TVar ThrdEnv -> ThrdStatus -> STM ()
\end{spec}

\ignore{

\begin{code}
getStatusRef :: TVar ThrdEnv -> STM ThrdStatus 
getStatusRef ctxRef = do 
  ctx <- readTVar ctxRef
  return $ getStatus ctx 
\end{code}

\begin{code}
setStatusRef :: TVar ThrdEnv -> ThrdStatus -> STM ()
setStatusRef ctxRef stat = do 
  ctx <- readTVar ctxRef -- $
  writeTVar ctxRef $ setStatus ctx $ stat
\end{code}

}

\noindent
The node tree \lhs{ThrdEnv} stores the current \lhs{PSM} execution environment at any moment. 
Each node in the tree corresponds to an active thread. Only the leaves of the 
thread tree are the currently active \lhs{PSM} processes. The inner nodes of the tree 
are parent threads executing a \emph{synchroniser} (named \lhs{syncTChildren}, see below) to wait for completion of their children. They also connect to the \lhs{PSM} continuation processes that get activated when the forked children terminate and join.

\smallskip

Each process thread keeps a global reference map \lhs{PSMVarGlobMap} of those
\lhs{PSMVars} that are known and accessibly in this thread
and a-fortiori, by transitive scoping, by its descendants.
The \lhs{PSMVar}s accessible in a thread are those which have been 
instantiated by this thread or any of its ancestors. We use
existential types to abstract from the particular
\lhs{PSMVar} type. Every \lhs{PSMVar} can be embedded into the type of all
\lhs{PSMVar}s via the \lhs{AnyPSMVar} constructor. The global map is then
given as type \lhs{PSMVarGlobMap} 
In addition to the global map \lhs{PSMVarGlobMap} we also keep a separate
record of the local \lhs{PSMVar}s that were allocated by this process thread
and thus are accessible exclusively by the thread
itself and its descendants. These \lhs{PSMVar}s are listed in a map
with type \lhs{PSMVarLocMap}. However, since they are also contained in 
\lhs{PSMVarGlobMap}, the local map only stores the \lhs{PSMVarId}s as keys
but not any values.

\begin{code}
data AnyPSMVar = forall p. PSMType p => AnyPSMVar (PSMVar p)

type PSMVarGlobMap = Map PSMVarId (TVar AnyPSMVar)
type PSMVarLocMap  = Map PSMVarId ()
\end{code}

\begin{figure}[ht]
\begin{center}
\includegraphics[scale=0.6]{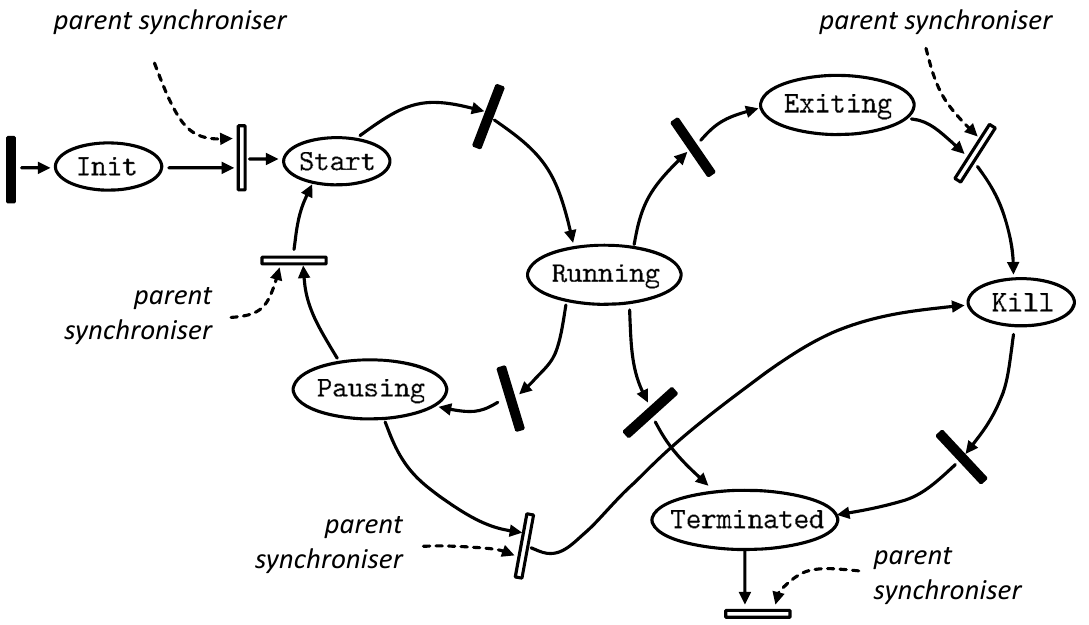}
\end{center}
\caption{The life cycle of an active thread as a Petri net. The places correspond to the statuses of the thread node. The solid transitions are the thread's synchronisation actions, the unfilled transitions are from the thread's synchronising parent.}
\label{fig:thrd-statuses}
\end{figure}

\noindent 
Besides the references to accessible \lhs{PSMVar}s, each process thread has a 
status \lhs{ThrdStatus}
\begin{spec}
data ThrdStatus =
  Init ThrdPSMVarMap | Start | Running | Terminated | Pausing ThrdPSMVarMap | Kill | Exiting
\end{spec}
which is used to synchronise the thread with its concurrent environment,
using its ancestors as proxies.  This synchronisation controls forking and
joining of threads, including the deletion of all \lhs{PSMVars} that a terminating
thread may have allocated during its lifetime. Also, the \lhs{ThrdStatus} helps
to implement the synchronous lock-step execution of all active threads along
sequences of macro-steps. 
The life cycle of a thread is depicted in Fig.~\ref{fig:thrd-statuses}. 
Each active thread begins its life with status \lhs{Init}
where it waits to be started. With the \lhs{ThrdPSMVarMap} parameter of the \lhs{Init} status, the new 
thread communicates to all other concurrent threads (any active thread that is not a descendant or an ancestor) a prediction of all methods calls it is potentially performing during the imminent macro-step. When the predictions have been installed in the potentials of the \pvar{s} by
the thread's parent synchroniser, the parent will set the thread's status to 
\lhs{Start}. The child thread then advances its status to \lhs{Running}. 
In this status it 
executes its action and eventually completes with one of the three statuses \lhs{Terminated}, 
\lhs{Pausing} or \lhs{Exiting}. When \lhs{Pausing}, the thread registers a new prediction for the \lhs{PSMVar}
accesses this thread is potentially going to perform during the next macro-step. 
When the thread is \lhs{Terminated}, its parent synchroniser will remove it from the thread environment and all the potentials registered in the \pvar{s} of the \lhs{PSMVarGlobMap}. 
Finally, a thread may also complete by \lhs{Exiting} when it wants the main program to stop. This is the only and most extreme case of an exeption trap currently implemented in the \psm library.  
When a child thread sets its status to \lhs{Exiting}, the parent synchroniser distributes the exit signal all the way up the thread tree. The root synchoniser eventually transforms the stop signal into a \lhs{Kill} status that is propagated back down the thread tree. 
Every thread that is completed with a \lhs{Pausing} or \lhs{Exiting} status will detect the \lhs{Kill} signal from the parent. If this is received, it immediately abandons its residual computation (using Haskell trap mechanism and a \lhs{PSMTrap} exception) and sets its own status to \lhs{Terminated}.
In this way, any \lhs{Exiting} signal by any active leaf thread will eventually end up forcing all threads to finish their computations as in the regular \psm termination procedure. 
Likewise, the \lhs{Pausing} completion is passed up the thread tree and reflected back down by the root synchroniser as a \lhs{Start} signal to the (remaining non-terminated) leaf threads. These will then progress into their next macro-cycle, setting their statuses to \lhs{Running}, again.
Note the lock-step synchronisation: When the status is \lhs{Init}, \lhs{Pausing} or \lhs{Exiting}, then the change of status is made by the parent, while the child is waiting. 
When the status is \lhs{Start}, \lhs{Running} or \lhs{Kill} the change is made by the child and the parent is waiting. 
The waiting mechanism for status change is implemented with the \stm \lhs{retry} abstraction.

\subsection{Thread Environment}
\label{sec:thrd-environment}

As indicated above, the \lhs{ThrdEnv} has two purposes, (i) as a central hub
for collecting the concurrent predictions in order to control the access to
shared \lhs{PSMVar}s and (ii) to organise the synchronisation of concurrent
threads for joining and pausing.
The following auxiliary functions are central for thread synchronisation.
\begin{spec}
setPSMVarPot  :: TVar ThrdEnv -> ThrdPSMVarMap -> STM ()
remThrdPot  :: TVar ThrdEnv -> STM ()
addPSMVarPot  :: TVar ThrdEnv -> ThrdPSMVarMap -> STM ()
subPSMVarPot  :: TVar ThrdEnv -> ThrdPSMVarMap -> STM ()
exitPSMVarPot :: TVar ThrdEnv -> ThrdPSMVarMap -> STM ()
\end{spec}
The first, \lhs{setPSMVarPot} publishes the given \lhs{ThrdPSMVarMap} as a fresh \pvar potential for the executing thread in the current \lhs{ThrdEnv}.
If a \lhs{PSMVar} mentioned in the \lhs{ThrdPSMVarMap} cannot be found, then a run-time error is generated.
The function \lhs{remThrdPot} is the opposite.
It removes the current thread from all \lhs{PSMVar}s in the \lhs{ThrdEnv}.
It is called by a thread just before it terminates and removes itself from \lhs{ThrdEnv}.
The function \lhs{addPSMVarPot} recharges the predictions of a loop body into the potentials of the current \pvar context when loop body is reiterated. 
In this way, we ensure that the potential always corresponds to the worst-case count of methods that is remaining for the current execution of the loop body. 
While a \lhs{repUntil} loop is running and not exited, the potential keeps all method calls recurrent in the loop. 
The function \lhs{subPSMVarPot} subtracts predictions from the potential, e.g., when the control flow branches and thereby narrows down the possible outstanding method calls of the continuation code. 
The function \lhs{exitPSMVarPot} is called when a loop is exited, in order to discharge (subtract) the loop's recurrent prediction from the current pvar potential. 
In this fashion, the \lhs{repUntil} loop acts as a consumer of the recurrent counters (named \lhs{recCnt} in the code) just like each method call in the \lhs{var}/\lhs{bindVar} consumes from the transient counters (named \lhs{potCnt}). 
The code for these accounting functions is presented in the following.

\begin{spec}
setPSMVarPot = alterPSMVarPot predToPot

predToPot :: Map CmplId Prediction -> Maybe (Prediction, Prediction) -> Maybe (Prediction, Prediction)

predToPot pVarPred _ = (Just(potCnt', recCnt'))  where
  potCnt' = Map.foldr addPred empty (delete (CmplId 2) pVarPred)
  recCnt'   = findWithDefault empty (CmplId 2) pVarPred  
\end{spec}

\begin{spec}
remThrdPot :: TVar ThrdEnv -> STM ()
remThrdPot ctxRef = do
  ctx <- readTVar ctxRef
  let (TTreeNode (tId, _, pVarGlobMapRef, _) _ _ _ _) = ctx 
  pVarGlobMap <- readTVar pVarGlobMapRef
  mapM_ (deleteThrd tId) pVarGlobMap where
    deleteThrd tId pVarRef = do
      AnyPSMVar pVar <- readTVar pVarRef
      let (Active _ _ _ predMapRefx) = pVar 
      predMapx <- readTVar predMapRefx
      writeTVar predMapRefx $ delete tId predMapx
\end{spec}

\begin{spec}
addPSMVarPot = alterPSMVarPotTrans addPredFn

addPredFn :: Map CmplId Prediction -> Maybe (Prediction, Prediction) -> Maybe (Prediction, Prediction)
addPredFn pVarPred Nothing = Nothing 
addPredFn pVarPred (Just(potCnt, recCnt)) = Just (potCnt', recCnt') where
  potCnt' = addPred potCnt (Map.foldr addPred empty (delete (CmplId 2) pVarPred))
  recCnt' = addPred recCnt $ findWithDefault empty (CmplId 2) pVarPred
\end{spec}

\begin{spec}
exitPSMVarPot = alterPSMVarPotTrans exitPredFn

exitPredFn :: Map CmplId Prediction -> Maybe (Prediction, Prediction) -> Maybe (Prediction, Prediction)
exitPredFn pVarPred Nothing = Nothing 
exitPredFn pVarPred (Just(potCnt, recCnt)) = Just(potCnt, recCnt') where  
  recCnt' = subPred recCnt $ findWithDefault empty (CmplId 0) pVarPred
\end{spec}

\begin{spec}
alterPSMVarPot :: (Map CmplId Prediction -> Maybe (Prediction, Prediction) 
    -> Maybe (Prediction, Prediction)) -> TVar ThrdEnv -> ThrdPSMVarMap -> STM ()
alterPSMVarPot alterFn ctxRef pred = do
  ctx <- readTVar ctxRef
  -- get map of all PSMVars registered in the node
  let TTreeNode (tId, _, pVarGlobMapRef, _) _ _ _ _ = ctx
  pVarGlobMap <- readTVar pVarGlobMapRef
  mapM_ (alterPred pVarGlobMap tId) pVars where
    -- collect all pvars in the prediction
    pVars = Map.foldr Data.union Data.empty (Map.map keysSet pred)
    alterPred pVarGlobMap tId vId = do 
      let pVarPred = Map.map (findWithDefault empty vId) pred 
          pVarRef = findWithDefault (error message) vId pVarGlobMap
          message = "alterPSMVarPot: unknown PSMVar access on " ++ show vId ++ " by thread " ++ show tId
      -- get current predMap from PSMVar
      AnyPSMVar pVar <- readTVar pVarRef
      let Active _ _ _ predMapRef = pVar
      -- current registered predictions
      predMap <- readTVar predMapRef
      -- insert thread and its prediction to PSMVar
      writeTVar predMapRef $ alter (alterFn pVarPred) tId predMap
\end{spec}

\begin{spec}
alterPSMVarPotTrans :: 
  (Map CmplId Prediction -> Maybe (Prediction, Prediction) 
    -> Maybe (Prediction, Prediction)) -> String -> TVar String 
     -> TVar ThrdEnv -> ThrdPSMVarMap -> STM ()

alterPSMVarPotTrans alterFn s ioRef ctxRef pred  = do
  ctx <- readTVar ctxRef
  let TTreeNode (tId, _, pVarGlobMapRef, _) _ _ _ _ = ctx
  pVarGlobMap <- readTVar pVarGlobMapRef
  mapM_ (alterPred pVarGlobMap tId) pVars where
    pVars = Map.foldr Data.union Data.empty (Map.map keysSet pred)
    alterPred pVarGlobMap tId vId = do 
      let
        pVarPred = Map.map (findWithDefault empty vId) pred 
        pVarRef = findWithDefault (error message) vId pVarGlobMap
        message = "alterPSMVarPotTrans: unknown PSMVar access on " ++ show vId
      (_, thrdPath) <- collectSpine (\ (x, _, _, _) -> x) ctxRef -- $
      AnyPSMVar pVar <- readTVar pVarRef
      let Active _ _ _ predMapRef = pVar
          adjustThrdPred tId = do
            predMap <- readTVar predMapRef
            writeTVar predMapRef $ alter (alterFn pVarPred) tId predMap
      mapM_ adjustThrdPred thrdPath
\end{spec}

\ignore{

\begin{code}

-- simulates setPSMVarPot
setPSMVarPot = alterPSMVarPot predToPot
predToPot :: Map CmplId Prediction -> 
                Maybe (Prediction, Prediction) -> Maybe (Prediction, Prediction)
predToPot pVarPred _ = (Just(potCnt', recCnt'))  where
  potCnt' = Map.foldr addPred empty (delete (CmplId 2) pVarPred)
  recCnt'   = findWithDefault empty (CmplId 2) pVarPred  


alterPSMVarPot :: (Map CmplId Prediction -> Maybe (Prediction, Prediction) 
    -> Maybe (Prediction, Prediction)) -> String -> TVar String 
     -> TVar ThrdEnv -> ThrdPSMVarMap -> STM ()
alterPSMVarPot alterFn s ioRef ctxRef pred = do
  ctx <- readTVar ctxRef
  -- get map of all PSMVars registered in the node
  let TTreeNode (tId, _, _, _, pVarGlobMapRef, _) _ _ _ _ = ctx
  pVarGlobMap <- readTVar pVarGlobMapRef
  -- debug logging
  logStr ioRef $ s ++ ": " 
  mapM_ (alterPred pVarGlobMap tId) pVars where
    -- collect all pvars in the prediction
    pVars = Map.foldr Data.union Data.empty (Map.map keysSet pred)
    alterPred pVarGlobMap tId vId = do 
      let pVarPred = Map.map (findWithDefault empty vId) pred 
          pVarRef = findWithDefault (error message) vId pVarGlobMap
          message = "alterPSMVarPot: unknown PSMVar access on " ++ show vId ++ " by thread " ++ show tId
      -- get current predMap from PSMVar
      AnyPSMVar pVar <- readTVar pVarRef
      let Active _ _ _ predMapRef = pVar
      -- debug logging
      logStr ioRef $ "alter pVarPred for " ++ show vId ++ " and " ++ show tId ++ " by " ++ showPSMVarPred pVarPred
      -- current registered predictions
      predMap <- readTVar predMapRef
      -- debug logging 
      logStr ioRef $ "Current predMap for " ++ show vId ++
                " = \n  " ++ showThrPredMapxx predMap
      -- insert thread and its prediction to PSMVar
      writeTVar predMapRef $ alter (alterFn pVarPred) tId predMap
      -- debug logging 
      predMapx' <- readTVar predMapRef
      logStr ioRef $ "New predMap for " ++ show vId ++
                " = \n  " ++ showThrPredMapxx predMapx' ++ "\n"

\end{code}

\begin{code}

remThrdPot :: TVar ThrdEnv -> STM ()
remThrdPot ctxRef = do
  ctx <- readTVar ctxRef
  let (TTreeNode (tId, _, _, _, pVarGlobMapRef, _) _ _ _ _) = ctx 
  pVarGlobMap <- readTVar pVarGlobMapRef
  mapM_ (deleteThrd tId) pVarGlobMap where
    deleteThrd tId pVarRef = do
      AnyPSMVar pVar <- readTVar pVarRef
      let (Active _ _ _ predMapRefx) = pVar 
      -- complete for Idle pvars
      predMapx <- readTVar predMapRefx
      writeTVar predMapRefx $ delete tId predMapx

\end{code}

\begin{code}

addPSMVarPot = alterPSMVarPotTrans addPredFn

addPredFn :: Map CmplId Prediction -> Maybe (Prediction, Prediction) -> Maybe (Prediction, Prediction)
addPredFn pVarPred Nothing = Nothing 
addPredFn pVarPred (Just(potCnt, recCnt)) = Just (potCnt', recCnt') where
  potCnt' = addPred potCnt (Map.foldr addPred empty (delete (CmplId 2) pVarPred))
  recCnt' = addPred recCnt $ findWithDefault empty (CmplId 2) pVarPred

subPSMVarPot :: String -> TVar String -> TVar ThrdEnv -> ThrdPSMVarMap -> STM ()
subPSMVarPot = alterPSMVarPotTrans subPredFn

subPredFn :: Map CmplId Prediction -> Maybe (Prediction, Prediction) -> Maybe (Prediction, Prediction)
subPredFn pVarPred Nothing = Nothing
subPredFn pVarPred (Just(potCnt, recCnt)) = Just(potCnt', recCnt') where
  potCnt' = subPred potCnt $ Map.foldr addPred empty (delete (CmplId 2) pVarPred)
  recCnt' = subPred recCnt $ findWithDefault empty (CmplId 2) pVarPred

exitPSMVarPot = alterPSMVarPotTrans exitPredFn

exitPredFn :: Map CmplId Prediction -> 
                Maybe (Prediction, Prediction) -> Maybe (Prediction, Prediction)
exitPredFn pVarPred Nothing = Nothing 
exitPredFn pVarPred (Just(potCnt, recCnt)) = Just(potCnt, recCnt') where  
  recCnt' = subPred recCnt $ findWithDefault empty (CmplId 0) pVarPred


alterPSMVarPotTrans :: 
  (Map CmplId Prediction -> Maybe (Prediction, Prediction) 
    -> Maybe (Prediction, Prediction)) -> String -> TVar String 
     -> TVar ThrdEnv -> ThrdPSMVarMap -> STM ()

alterPSMVarPotTrans alterFn s ioRef ctxRef pred  = do
  ctx <- readTVar ctxRef
  let TTreeNode (tId, _, _, _, pVarGlobMapRef, _) _ _ _ _ = ctx
  pVarGlobMap <- readTVar pVarGlobMapRef
  -- debug logging
  logStr ioRef $ s ++ ": " 
  mapM_ (alterPred pVarGlobMap tId) pVars where
    pVars = Map.foldr Data.union Data.empty (Map.map keysSet pred)
    alterPred pVarGlobMap tId vId = do 
      let
        pVarPred = Map.map (findWithDefault empty vId) pred 
        pVarRef = findWithDefault (error message) vId pVarGlobMap
        message = "alterPSMVarPotTrans: unknown PSMVar access on " ++ show vId
      (_, thrdPath) <- collectSpine (\ (x, _, _, _, _, _) -> x) ctxRef -- $
      AnyPSMVar pVar <- readTVar pVarRef
      let Active _ _ _ predMapRef = pVar
          adjustThrdPred tId = do
            -- debug logging
            -- logStr ioRef $ s ++ ": " ++ "for " ++ show vId ++ " by " ++ show tId ++ ": " ++ showPSMVarPred predMapx
            predMap <- readTVar predMapRef
            -- debug logging
            -- logStr ioRef $ "Current predMapx for " ++ show vId ++
            --   " = \n   " ++ showThrPredMapx (Map.map snd predMapx)
            writeTVar predMapRef $ alter (alterFn pVarPred) tId predMap
            -- debug logging 
            -- predMapx' <- readTVar predMapRef
            -- logStr ioRef $ "New predMapx for " ++ show vId ++
            --  " = \n   " ++ showThrPredMapxx predMapx' ++ "\n"
      -- debug logging 
      logStr ioRef $ "alter pVarPred for " ++ show vId ++ " and " ++ show thrdPath ++ " by " ++ showPSMVarPred pVarPred
      -- debug logging 
      predMapx' <- readTVar predMapRef
      logStr ioRef $ "Current predMap for " ++ show vId ++
                " = \n  " ++ showThrPredMapxx predMapx'
      mapM_ adjustThrdPred thrdPath
      -- debug logging 
      predMapx' <- readTVar predMapRef
      logStr ioRef $ "New predMap for " ++ show vId ++
                " = \n  " ++ showThrPredMapxx predMapx' ++ "\n"

\end{code}

} 

\smallskip 

The synchronisation of threads with their parent synchoniser (or ``join thread'') is managed via the functions {waitForInit}, \lhs{waitForStart}, \lhs{waitForKill}, \lhs{waitForStartOrKill} and 
\lhs{WaitForCompletion}. These are described next. 

\begin{code}
waitForInit :: TVar ThrdEnv -> STM (ThrdPSMVarMap, ThrdPSMVarMap)
waitForInit ctxRef = do
  ctx <- readTVar ctxRef
  case ctx of
    TTreeNode _ _ _ (Just tleft) (Just tright) -> do
      left  <- readTVar tleft
      right <- readTVar tright
      case (getStatus left, getStatus right) of
        (Init predl, Init predr) -> return (predl, predr)
    orelse -> retry
\end{code}

The \lhs{STM} transaction \lhs{waitForInit} is called by a synchoniser (in \lhs{syncTChildren}, see below) immediately after it has been started in order to wait for its children to become installed in the thread tree.
When this happens, the transaction unblocks and returns the predictions of the children. 
These predictions are then entered into the potentials of the \pvar{s} by the synchroniser, before the children can begin. 

\begin{code}
waitForStart :: TVar ThrdEnv -> STM ()
waitForStart ctxRef = do
  ctx <- readTVar ctxRef
  case getStatus ctx of
    Start  -> return ()
    orelse -> retry
\end{code}

The \lhs{STM} transaction \lhs{waitForStart} is called by a \psm action immediately after it has been forked (in \lhs{forkChild}, see below) and entered the stage with \lhs{Init} status. Using the \lhs{waitForStart} transaction, the new action blocks until the parent synchroniser sends the \lhs{Start} signal. 
Then it can start its action for the current macro-step.

\begin{code}
waitForKill :: TVar ThrdEnv -> STM ()
waitForKill ctxRef = do
  ctx <- readTVar ctxRef
  case getStatus ctx of
    Kill   -> return () 
    orelse -> retry
\end{code}

The \lhs{STM} transaction \lhs{waitForKill} is called by a thread after it has gone into \lhs{Exiting} mode. It blocks until the parent synchroniser sends the \lhs{Kill} signal.
The blocked thread may itself be a synchoniser or an active leaf thread. 
In the former case, the unblocked synchoniser will forward the \lhs{Kill} signal to its children. In the latter case, if it is a leaf thread (executing the \psm action \lhs{kill}, see below), it will terminate prematurely and get itself removed from the \lhs{ThrdEnv} (in the \lhs{forkFinally} of the \lhs{forkChild}, see below).

\begin{code}
waitForStartOrKill :: TVar ThrdEnv -> STM Bool
waitForStartOrKill ctxRef = do
  ctx <- readTVar ctxRef
  case getStatus ctx of
    Start  -> return True
    Kill   -> return False
    orelse -> retry
\end{code}

The function \lhs{waitForStartOrKill} is an \lhs{STM} action which is entered by a thread when it is in the \lhs{Pausing} status. It blocks until the status is set to either \lhs{Start} or \lhs{Kill} in the \lhs{ThrdEnv}.
Again, this is done by the parent thread which is responsible for the synchronisation of the child thread with its siblings.
When the child's thread status is thus set to \lhs{Start}
by the parent synchroniser, the function \lhs{waitForStartOrKill} returns \lhs{True}. If the status is set to \lhs{Kill} it returns \lhs{False}. 
What happens then depends on the waiting thread. 
If the waiting thread is a synchroniser, it will pass the \lhs{Start} or \lhs{Kill} signal, respectively, to its children.
If the waiting thread is a leaf (executing the \psm action \lhs{pause} or \lhs{wait}, see below) and the signal is \lhs{Start}, then the leaf will move to \lhs{Running} and commence the action of the next macro-step . 
Otherwise, if the signal is \lhs{Kill}, the leaf thread will  terminate prematurely and get removed from the \lhs{ThrdEnv}.

\begin{code}
waitForCompletion :: TVar ThrdEnv -> TVar ThrdEnv -> STM (ThrdStatus, ThrdStatus)
waitForCompletion tleft tright = do
  left  <- readTVar tleft
  right <- readTVar tright
  case (getStatus left, getStatus right) of
    (statl@(Pausing _), statr@(Pausing _)) -> return (statl, statr)
    (statl@(Pausing _), Terminated) -> return (statl, Terminated)
    (Terminated, statr@(Pausing _)) -> return (Terminated, statr)
    (Terminated, Terminated) -> return (Terminated, Terminated)
    (Exiting, statr) -> return (Exiting, statr)
    (statl, Exiting) -> return (statl, Exiting)
    orelse -> retry
\end{code}

The transaction \lhs{waitForCompletion} is evaluated by a synchroniser in the \lhs{Running} status to wait for its children to complete the current macro-step. 
When this happens, the transaction returns the completion statuses of the children. The synchroniser will then set its own status accordingly and wait for its parent synchroniser to signal an appropriate continuation.

\subsection{Thread Forking}
\label{sec:thrd-fork}

The operator to fork processes is called \lhs{forkChild}. It takes a process
\lhs{proc = PSM(pred, act)} and runs it as a child thread of a given parent node 
pointed to by reference \lhs{ctxRef :: TVar ThrdEnv}. The result is a concurrent computation
$$\lhs{forkChild ctxRef proc} :: \lhs{IO ThreadId}$$
 in the \lhs{IO} context returning a \lhs{ThreadId}.

\lstset{numbers=left,numberstyle=\tiny}
\begin{spec}
forkChild :: TVar ThrdEnv -> PSM a -> IO ThreadId
forkChild ctxRef (PSM(pred, act)) =
  forkFinally (do 
    tId <- myThreadId
    -- we copy the PSMVar references from the parent to the child thread
    -- the child can impose predictions only on these
    ctx <- atomically $ readTVar ctxRef
    let (TTreeNode (_, _, pVarGlobMapRef, _) _ _ _ _) = ctx 
    pVarGlobMap <- atomically $ readTVar pVarGlobMapRef
    pVarGlobMapCpRef <- atomically $ newTVar pVarGlobMap
    -- we start with the empty local pvars map
    pVarLocMapRef <- atomically $ newTVar empty
    -- create and install child node
    current <- atomically $ 
      addChild ctxRef (tId, Init pred, pVarGlobMapCpRef, pVarLocMapRef)
    -- wait for synchroniser to indicate that computation phase begins
    atomically $ waitForStart current
    atomically $ setStatusRef current Running 
    -- execute action, catch PSMTrap exceptions
    res <- try $ act current (fromList [(CmplId 0, empty)])
    case res of
      Left (PSMTrap _) -> return current
      Right x -> return current)
    (\ret -> do
      case ret of
        Left e -> putStr $ show e -- error "forkChild: uncaught thread exception"
        Right childRef -> do
          -- set status to Terminated
          child <- atomically $ readTVar childRef 
          atomically $ writeTVar childRef (setStatus child Terminated)  
          atomically $ remThrdPot childRef)
\end{spec}
\lstset{numbers=none}

The child thread is created by Haskell's \lhs{forkFinally} function. It takes two
arguments, the actual child code (lines 4--23) and the code to be executed
after the child thread has terminated or has raised an exception (lines 25--31).
The \lhs{PSM} process \lhs{proc} is turned into a child thread as follows:
\begin{itemize}

 \item Line 4: The child first retrieves its own thread identifier as \lhs{tId}.

 \item Lines 7--10: Then, we create a copy of the global map of \lhs{PSMVar}s, which is called  \lhs{pVarGlobMapCpRef}.

 \item Line 12: Since the child thread is fresh, it has no local \lhs{PSMVar}s of its own, yet.
 Thus, we create a fresh empty map of local \lhs{PSMVar}s, referred to by
 \lhs{pVarLocMapRef}.   

 \item Lines 14--15: We create and allocate a new thread node, called \lhs{current}, in the execution
 environment with the thread identifier \lhs{tId}. It carries the initial thread status \lhs{Init pred} with the prediction \lhs{pred} of the action \lhs{proc} to be run on this child node. 
 It also receives 
 the global and local \lhs{PSMVar} reference
 maps. The auxiliary function \lhs{addChild} (from the \lhs{TTree} module) is responsible
 for adding a new child node below the current environment reference node \lhs{ctxRef}.
 We note that since the \lhs{TTree} module currently only supports binary trees,
 \lhs{forkChild} can be called at most two times from a given thread node \lhs{ctxRef} before a join synchronisation must be done.
If the ctx thread tries to fork a third thread a run-time error is generated.
 
 \item Line 17--18: Thereafter, the new child thread waits in the transaction \lhs{waitForStart} to be started by
 the synchroniser (\lhs{synchTChildren}) in the parent node. 
 This makes sure that the predictions of all concurrent siblings of \lhs{current}
 are installed in the shared \lhs{PSMVar}s before the child begins its action in line~20. Before it does so, it sets its own status to \lhs{Running} in line~18.

 \item Lines 21--23: When the child action \lhs{act} returns, we receive the result \lhs{res}. If a \lhs{PSMTrap} exception occurs the actions has executed a \lhs{kill} statement. Otherwise, it returns a value \lhs{x::a}. In both cases the \lhs{forkChild} returns the child node's thread node \lhs{current::TVar TrdEnv}.
 In the present implementation we do not use the return value \lhs{x::a}. 

\end{itemize} 

The second argument of the \lhs{forkFinally}, in lines~25--31, is the clean-up where the
child thread executing the child action sets the status of the
child in the thread environment to \lhs{Terminated} and removes itself from the potentials in all \lhs{PSMVar}s, by calling the function \lhs{remThrdPot} discussed
above. Nodes are removed by setting the node to ``Nothing'' by the synchronising parent thread.

Notice the communication between computation in the first argument of
\lhs{forkFinally} (lines~4--23) and the second argument (lines~25--31): The \lhs{childRef} node reference in lines 29 and 30 is the same as the value \lhs{current} returned in lines~14--15. 

\ignore{

\begin{code}

forkChild :: TVar String -> String -> TVar ThrdEnv -> PSM a -> IO ThreadId
forkChild ioRef tName ctxRef (PSM(predx, trp, act)) =
  forkFinally ( do
       tId <- myThreadId
       ctx <- atomically $ readTVar ctxRef
       let (TTreeNode (_, _, numRef, trapCntRef, pVarGlobMapRef, _) _ _ _ _) = ctx
       -- we copy the PSMVar references from the parent to the child thread
       -- the child can impose predictions only on these
       pVarGlobMap <- atomically $ readTVar pVarGlobMapRef
       pVarGlobMapCpRef <- atomically $ newTVar pVarGlobMap
       -- we start with the empty local pvars map
       pVarLocMapRef <- atomically $ newTVar empty
       -- create and install child node
       current <- atomically $ 
         addChild ctxRef (ThrdId tId tName, Init predx, numRef, trapCntRef, pVarGlobMapCpRef, pVarLocMapRef)
       -- wait for synchoniser to indicate computation phase begins
       atomically $ waitForStartOrKill current
       atomically $ setStatusRef current Running 
       -- execute action, mask all exceptions (we should only catch PSMTrap)
       res <- try (act ioRef current (fromList [(CmplId 0, empty)]))
       case res of -- here we could print diagnostics into stderr
         Left (PSMTrap _) -> return current
         Right _          -> return current)
    (\ret -> do
        case ret of
          Left e -> putStr $ show e -- error "forkChild: uncaught thread exception"
          Right childRef -> do
            -- set status to Terminated
            child <- atomically $ readTVar childRef 
            atomically $ writeTVar childRef (setStatus child Terminated)  
            -- print $ "TERMINATE" ++ tName 
            -- child thread removes itself from all pVars it is registered with
            atomically $ remThrdPot childRef)

\end{code}

}

\subsection{Thread Synchronisation}
\label{sec:thrd-sync}

Once a process has been forked, it synchronises with its parent as indicated in Fig.~\ref{fig:thrd-statuses}. 
The parent runs a joining synchroniser, called \lhs{syncTChildren}, which becomes active in the parent node sequentially after the forking has taken place.
The thread node (in the thread tree) of the parent synchroniser is referred to as the \emph{synchronising node} in the following.
All active nodes in the thread tree of the current \lhs{ThrdEnv} are either 
synchonising nodes, waiting for their children to terminate, or \emph{leaf nodes} corresponding to the \psm actions executing concurrently with each other.
While they are running, these leaf nodes may fork new child actions and turn into synchronising nodes.    
We assume that at most two children are forked when the synchroniser
is started. 
The purpose of the synchroniser is first to start the children and then to convergecast and broadcast status information between the root of the thread tree and the running leaves, eventually joining the terminated children threads and continue control in the sequentially down-stream actions of the parent. 
The code of the \lhs{syncTChildren} synchroniser is given below. In the following description we will refer to the code by line numbers.

\begin{itemize}

\item At the beginning, \lhs{syncTChildren} is responsible for initialising the thread environment and then starting off the children's actions.
This initialisation phase comprises the code lines 3--12 (see below). 
The problem is that the spawning of the children by \lhs{forkChild} is concurrent with the starting of \lhs{syncTChildren} in the parent thread.
Hence, a child node may not be active yet. Its node in the thread tree may still remain \lhs{Nothing} for a while, when the evaluation of \lhs{syncTChildren} has already started.
To account for this transient behaviour, \lhs{syncTChildren} first waits in the \lhs{STM} transaction \lhs{waitForInit} for both children to become installed and initialised. This happens when the function \lhs{waitForInit}
in line~3 unblocks and returns the initial predictions \lhs{predl} and \lhs{predr} of the children.
At this point both children are active and their
entries in the thread tree match the patterns \lhs{Just tleft} and \lhs{Just tright} (line~5) where \lhs{tleft} and \lhs{tright} are the references to the children's nodes. 
The inital code in lines~5--7 also grabs the identifiers \lhs{locKeys} of the local \pvar{s} for this node. These do not change during the life-time of \lhs{syncTChildren}, because new \pvar{s} can only be created by the thread after the synchroniser has returned (join).

\item Lines~8--10: Since both children have \lhs{Init} status their 
nodes then are extracted as \lhs{left} and \lhs{right} in lines~9--10.
The children's predictions \lhs{predl} and \lhs{predr} are published in the potentials of the \lhs{PSMVars} using the function \lhs{setPSMVarPot} explained above.

\item Lines~11--12: The children's actions are now released by setting their status to \lhs{Start} using the auxiliary function 
\lhs{setStatusRef} which overwrites the status of a node. 

\end{itemize}

It is important that the children threads begin their actions only after all predictions are installed in the potentials of the \pvar{s}.
To enforce this\footnote{A closer look reveals that this atomicity is in fact not necessary. The only way in which the predictions can change is via the children threads. But these have completed and are waiting for the \lhs{Start} signal which is given by the synchroniser thread only sequentially after the evaluation of the \lhs{setPredVars} transactions.}, the prediction setting happens in a single atomic transaction in lines~8--10. For contrast, the starting of the children can be interleaved and implemented in separate transactions in lines~11 and~12. 
The initialisation phase of lines~8--12 could be enriched to implement
``surface entry actions'' for a parallel region 
which must execute strictly before any action of a child process at the first time they become active.

\smallskip

After initialistion, the synchoniser enters its stationary behaviour in a big \lhs{repeat} loop, which is defined in lines~18--213. The repeat loop is started in line~219. 
The loop implements the necessary synchronisation logic by ``convergecasting'' and ``broadcasting'' the completion statuses of nodes across the thread tree in several waves. 
In doing so, the synchroniser's behaviour depends on the status of the synchronising node and the status of its children nodes.
The repeat loop first obtains the status of the synchronising node (line~18) with \lhs{getStatusRef} .
When the synchroniser \lhs{syncTChildren} is started by an active \psm thread (with \lhs{forkChild}), then the status of the synchronising node is always \lhs{Running}. 
The code in lines~26--136 covers the status of the synchronising node when it is \lhs{Running}. Then, \lhs{syncTChildren} is in the ``upwards'' phase. It is waiting for its children to complete and then to propagate the completion information from the leaves up towards the root node. 
After it has been started, a child can complete in three ways, by setting its status to one of \lhs{Pausing}, \lhs{Terminated} or \lhs{Exiting} (see Fig.~\ref{fig:thrd-statuses}).
In lines~26--27 the synchroniser uses the transaction \lhs{waitForCompletion} to wait for the children to complete. 
When they do, we receive their statuses as \lhs{stat1} and \lhs{stat2}.

\begin{itemize} 

  \item If both children are \lhs{Terminated} the synchroniser removes the children from the thread tree with the \lhs{removeChildren} transaction (line~30).  
  The function \lhs{removeChildren} 
  \begin{spec}
  removeChildren :: TVar ThrdEnv -> STM ()
  removeChildren ctxRef = do
    TTreeNode (tId, stat, pVarGlobMapRef, pVarLocMapRef) 
            mpar msib _ _  <- readTVar ctxRef  
    writeTVar ctxRef $ -- $
      TTreeNode (tId, stat, pVarGlobMapRef, pVarLocMapRef) mpar msib Nothing Nothing
  \end{spec}
  deletes the child nodes of a thread by setting them constant \lhs{Nothing}. With the return of \lhs{syncTChildren} no thread has access to the original memory locations any more so they can be garbage collected.
  With the execution of \lhs{removeChildren}, the synchoniser returns control to the sequentially down-stream code of the parent node.
  The parent action that has called the \lhs{fork} now continues and becomes an active leaf node (with \lhs{Running} status) that eventually completes with \lhs{Terminated}, \lhs{Pausing} or \lhs{Exiting}.

  \item If at least one of its children is \lhs{Exiting}, then \lhs{syncTChildren} sets the status of its own synchronising node to \lhs{Exiting}, if it is not the root thread (lines~115--117,~131--133).
  In this fashion the synchroniser switches itself into ``downward'' mode where it blocks in the transaction \lhs{waitForKill} to wait for the \lhs{Kill} status signal to arrive from the root. When this eventually happens (lines~202-206) it passes the \lhs{Kill} signal to those of its children that have not terminated yet. This is done by the \lhs{trySetStatusKill} transaction. Also, in that case, the synchronising node goes back into the \lhs{Running} status (line~205), again following an ``upward'' information flow, because it now needs to wait for the (remaining) children to become \lhs{Terminated}.

  \smallskip 

  If, however, the synchronising node (in the \lhs{Pausing} mode) is the root and at least one of its children is \lhs{Exiting} and (lines~109--113,~125--129), then its behaviour switches to ``downward'' directly. It sends the \lhs{Kill} signal back to its children and goes into \lhs{Running} status, where it waits for them to become \lhs{Terminated}. 
  
  \item Suppose the synchroniser is in ``upward'' mode, i.e. \lhs{stat == Pausing}, and one of its children is \lhs{Pausing} while none is \lhs{Exiting}, i.e. both are \lhs{Pausing} or \lhs{Terminated}. 
  In each of these these situations, which are handled in lines~34,~62 and~83, 
  the current macro-step is finished and a new macro-step must be initiated.
  If the synchoniser is not the root (an inner node) it passes the \lhs{Pausing} status up the tree, by setting its own status to \lhs{Pausing} (lines~58,~79 and~100). The predictions for the synchronising node, representing the concurrent composition of its children, are aggregated for the local \pvar{s} in the parameter of the \lhs{Pausing} status constructor. 
  The predictions are computed in lines~51--57 when both children pause, in lines~74--78 when the right child is pausing and the left is terminated, and symmetrically in lines~95--100 when only the left child is pausing. 

  \smallskip 

  By setting its own status \lhs{Pausing}, the synchroniser 
  goes into ``downward'' mode (line~140) where it waits for a \lhs{Start} or a \lhs{Kill} signal from the root.
  The blocking is done in the \lhs{waitForStartOrKill} transaction, line~148. The ``downward'' signal is indicated by the boolean flag returned by \lhs{waitForStartOrKill}, which is \lhs{True} if the signal is \lhs{Start} and \lhs{False} if it is \lhs{Kill}. 
  When one of the signals arrive, it is forwarded by the synchroniser to its non-terminated children in downward direction. 
  Also, in each case, the synchronising node goes into \lhs{Running} mode (lines~155, 161, 170, 186, 195) to handle the forthcoming terminiation or pausing of its children in ``upward'' direction. 
  
  \smallskip 

  At each node in the path of a ``downward'' broadcast of a \lhs{Start} signal 
  (lines~158--162,~173-177 and~189--196) the \lhs{tick} method is executed on the locally defined \pvar{s} using the \lhs{tickPSMVars} function: 
  \begin{spec}
  tickPSMVars :: ThrdEnv -> IO ()
  \end{spec}
  For each node \lhs{ctx :: ThrdEnv} the application \lhs{tickPSMVars ctx} calls the \lhs{tick} method on each \lhs{PSMVar} locally instantiated in the
  thread context \lhs{ctx}. In this way, the synchronisers ensure, that
  in the transition from one to the next macro-step the \lhs{tick} method is executed on all \lhs{PSMVar}s (by a single thread) before all the threads enter into the next macro-step. 
  \begin{spec}
  tickPSMVars ctx = do
    let (TTreeNode (tId, _, pVarGlobMapRef, pVarLocMapRef) _ _ _ _) = ctx
  pVarLocMap  <- atomically $ readTVar pVarLocMapRef
  pVarGlobMap <- atomically $ readTVar pVarGlobMapRef
  mapM_ (tickPSMVar pVarGlobMap) (keys pVarLocMap) where
    tickPSMVar pVarGlobMap vId = let
      pVarRef = findWithDefault (error message) vId pVarGlobMap
      message = "tickPSMVars: unknown PSMVar access" in do
        AnyPSMVar (Active vId _ state _) <- atomically $ readTVar pVarRef -- $
        tick state
  \end{spec}
  After all \pvar{s} have been updated for the next macro-step, the synchroniser will install the new prediction into the potentials of all \pvar{s} using \lhs{setPSMVarPred} in lines~159,~174 and~190--192.  
  The \lhs{Start} signal will make the pausing children at the leaves execute their next macro-step, the \lhs{Kill} will make them terminate prematurely. The synchronising node itself is set to \lhs{Running} in order to wait for the children to complete, accordingly.
  
\end{itemize}

\ignore{

\begin{code}

removeChildren :: TVar ThrdEnv -> STM ()
removeChildren ctxRef = do
  TTreeNode (tId, stat, num, traps, pVarGlobMapRef, pVarLocMapRef) 
          mpar msib _ _  <- readTVar ctxRef  
  writeTVar ctxRef $ -- $
    TTreeNode (tId, stat, num, traps, pVarGlobMapRef, pVarLocMapRef) mpar msib Nothing Nothing

\end{code}

\begin{code}
tickPSMVars :: TVar String -> ThrdEnv -> IO ()
tickPSMVars ioRef ctx = do
  let (TTreeNode (tId, _, _, _, pVarGlobMapRef, pVarLocMapRef) _ _ _ _) = ctx
  pVarLocMap  <- atomically $ readTVar pVarLocMapRef
  pVarGlobMap <- atomically $ readTVar pVarGlobMapRef
  mapM_ (tickPSMVar pVarGlobMap) (keys pVarLocMap) where
    tickPSMVar pVarGlobMap vId = let
      pVarRef = findWithDefault (error message) vId pVarGlobMap
      message = "tickPSMVars: unknown PSMVar access" in do
        AnyPSMVar (Active vId _ state _) <- atomically $ readTVar pVarRef
        tick state

termPSMVars :: TVar String -> ThrdEnv -> IO ()
termPSMVars ioRef ctx = do
  let (TTreeNode (tId, _, _, _, pVarGlobMapRef, pVarLocMapRef) _ _ _ _) = ctx 
  pVarLocMap  <- atomically $ readTVar pVarLocMapRef
  pVarGlobMap <- atomically $ readTVar pVarGlobMapRef
  mapM_ (termPSMVar pVarGlobMap) (keys pVarLocMap) where
    termPSMVar pVarGlobMap vId = let
      pVarRef = findWithDefault (error message) vId pVarGlobMap
      message = "termPSMVars: unknown PSMVar access" in do
        AnyPSMVar (Active vId _ state _) <- atomically $ readTVar pVarRef
        term state

\end{code}

} 

A complication arises by the need to put the new predictions in place
before each pausing leaf thread receives the starting signal for the next macro-step. 
These predictions must neither be installed too early nor too late. In order
to avoid being too early the predictions cannot be set by a local process at the moment
when it pauses. There may still be other active processes, still finishing the current
cycle. Indeed, we cannot replace the potentials until the global clock
synchronisation is detected which is when the status of the root node switches from
\lhs{Running} to \lhs{Pausing} (and then immediately back to \lhs{Running}).
On the other hand, we can neither have the predictions set by
a leaf process when the downwards wave sets its status from \lhs{Pausing} to \lhs{Start}, as
this would be too late. It is too late, because some leave nodes may receive their 
\lhs{Start} signal earlier than others. Hence, when a leaf process sets its prediction,
another may have already started the new macro-step.


The solution is to set the predictions in the downwards phase \emph{as soon as possible}.  
For this the predictions must be passed up the tree during the upwards phase. 
This means that we need to store predictions with the \lhs{Pausing} status 
in the upwards direction.
However, the prediction of a 
given \lhs{PSMVar} is pertinent, and thus can travel, only within the scope of the \lhs{PSMVar}, 
i.e., up to the thread node that has allocated it in its \lhs{PSMVarLocMap}.

\smallskip

\lstset{numbers=left,numberstyle=\tiny}
\begin{spec}
syncTChildren :: TVar ThrdEnv -> ThrdPSMVarMap -> IO ()
syncTChildren ctxRef cntPred = do 
  (predl, predr) <- atomically $ waitForInit ctxRef
  -- get references to children, etc (must now be defined!)
  ctx@(TTreeNode (tId,_,_,pVarLocMapRef) mpar _ (Just tleft) (Just tright)) <- atomically $ readTVar ctxRef
  pVarLocMap <- atomically $ readTVar pVarLocMapRef -- $  
  let locKeys = keysSet pVarLocMap       
  atomically $ do -- $
    setPSMVarPot tleft predl
    setPSMVarPot tright predr
  atomically $ setStatusRef tleft Start 
  atomically $ setStatusRef tright Start

  -----------------
  -- main loop 
  -----------------

  let repeat = do
      stat <- getStatusRef ctxRef 
      case stat of

        ----------------- UP PASS -----------
        Running -> do
        -------------------------------------

          (stat1, stat2) <- atomically $ waitForCompletion tleft tright -- $
          case (stat1, stat2) of 

            ---------------------------
            (Terminated, Terminated) -> atomically $ removeChildren ctxRef
            ---------------------------

            -------------------------------------------
            (Pausing upPred1, Pausing upPred2) -> do
            -------------------------------------------

              left  <- atomically $ readTVar tleft
              right <- atomically $ readTVar tright

              if not (isJust mpar) then do -- this is the root thread
                tickPSMVars ctx
                atomically $ do -- $
                  setPSMVarPot tleft upPred1
                  setPSMVarPot tright upPred2
                atomically $ writeTVar tleft  $ setStatus left Start
                atomically $ writeTVar tright $ setStatus right Start
                repeat

              else do -- this is an inner node, keep going up
              -- combine predictions of children and pass up the tree to parent 
                let upPred1' = Map.map (\p -> withoutKeys p locKeys) upPred1 
                    upPred2' = Map.map (\p -> withoutKeys p locKeys) upPred2 
                    cntPred' = Map.map (\p -> withoutKeys p locKeys) cntPred 
                    parPred' = concPred upPred1' upPred2'
                    thr'  = Map.member (CmplId 0) parPred'
                    pred' = if thr' then addSeqPred parPred' cntPred' 
                            else parPred'
                atomically $ writeTVar ctxRef $ setStatus ctx $ Pausing pred'
                repeat

            --------------------------------------------
            (Terminated, Pausing upPred) -> do
            -------------------------------------------- 
              
              right <- atomically $ readTVar tright

              if not (isJust mpar) then do -- we are root thread, switch to down-pass
                tickPSMVars ctx
                atomically $ setPSMVarPot tright upPred
                atomically $ writeTVar tright $ setStatus right Start
                repeat

              else do -- we are inner node, propagate non-local predictions of child upwards
                let upPred'  = Map.map (\p -> withoutKeys p locKeys) upPred 
                    cntPred' = Map.map (\p -> withoutKeys p locKeys) cntPred 
                    thr'  = Map.member (CmplId 0) upPred'
                    pred' = if thr' then addSeqPred upPred' cntPred'   
                            else upPred'
                atomically $ writeTVar ctxRef $ setStatus ctx $ Pausing pred'
                repeat

            -------------------------------------------
            (Pausing upPred, Terminated) -> do
            -------------------------------------------

              left  <- atomically $ readTVar tleft

              if not (isJust mpar) then do -- we are root thread, switching to down pass
                tickPSMVars ctx
                atomically $ setPSMVarPot tleft upPred -- $
                atomically $ writeTVar tleft $ setStatus left Start
                repeat

              else do -- we are not the root thread, keep going up
                let upPred'  = Map.map (\p -> withoutKeys p locKeys) upPred 
                    cntPred' = Map.map (\p -> withoutKeys p locKeys) cntPred 
                    thr'  = Map.member (CmplId 0) upPred'
                    pred' = if thr' then addSeqPred upPred' cntPred'   
                            else upPred'
                atomically $ writeTVar ctxRef $ setStatus ctx $ Pausing pred'
                repeat

            ---------------------------------
            (Exiting, _) -> do
            ---------------------------------

              left  <- atomically $ readTVar tleft

              if not (isJust mpar) then do -- we are root node, switch to down pass
                atomically $ writeTVar tleft $ setStatus left Kill
                atomically $ trySetStatusKill tright
                atomically $ writeTVar ctxRef $ setStatus ctx Running 
                repeat

              else do -- we are inner node, pass Exiting upwards  
                atomically $ writeTVar ctxRef $ setStatus ctx Exiting 
                repeat

            ---------------------------------
            (_, Exiting) -> do
            ---------------------------------

              right <- atomically $ readTVar tright

              if not (isJust mpar) then do -- we are root node, switch to down pass
                atomically $ writeTVar tright $ setStatus right Kill
                atomically $ trySetStatusKill tleft
                atomically $ writeTVar ctxRef $ setStatus ctx Running 
                repeat
              
              else do -- we are inner node, pass Exiting upwards
                atomically $ writeTVar ctxRef $ setStatus ctx Exiting
                repeat

            orelse -> error 
              "syncChildren: -panic- children have completed but status has glitched"


        ----------------- DOWN PASS -----------
        Pausing _ -> do - cannot be the root; we must be an inner node
        ---------------------------------------

          start <- atomically $ waitForStartOrKill ctxRef -- $
          left  <- atomically $ readTVar tleft
          right <- atomically $ readTVar tright 

          case (getStatus left, getStatus right) of

            --------------------------------------------
            (Pausing upPred, Terminated) -> do
            -------------------------------------------- 

              if (not start) then do -- Kill
                 atomically $ writeTVar tleft $ setStatus left Kill
                 atomically $ writeTVar ctxRef $ setStatus ctx Running
                 repeat
              else do -- Start next tick
                 tickPSMVars ctx
                 atomically $ setPSMVarPot tleft upPred -- $
                 atomically $ writeTVar tleft $ setStatus left Start
                 atomically $ writeTVar ctxRef $ setStatus ctx Running
                 repeat

            -------------------------------------------
            (Terminated, Pausing upPred) -> do
            -------------------------------------------

              if (not start) then do -- Kill 
                 atomically $ writeTVar tright $ setStatus right Kill
                 atomically $ writeTVar ctxRef $ setStatus ctx Running
                 repeat
              else do -- Start next tick
                 tickPSMVars ioRef ctx
                 atomically $ setPSMVarPot tright upPred -- $
                 atomically $ writeTVar tright $ setStatus right Start 
                 atomically $ writeTVar ctxRef $ setStatus ctx Running
                 repeat

            -----------------------------------------------------------
            (Pausing upPred1, Pausing upPred2) -> do
            -----------------------------------------------------------

              if (not start) then do -- Kill 
                 atomically $ writeTVar tleft  $ setStatus left Kill
                 atomically $ writeTVar tright $ setStatus right Kill 
                 atomically $ writeTVar ctxRef $ setStatus ctx Running
                 repeat cnt
              else do -- more ticks
                 tickPSMVars ctx
                 atomically $ do 
                   setPSMVarPot tleft  upPred1
                   setPSMVarPot tright upPred2
                 atomically $ writeTVar tleft  $ setStatus left  Start
                 atomically $ writeTVar tright $ setStatus right Start  
                 atomically $ writeTVar ctxRef $ setStatus ctx Running
                 repeat

        ----------------- DOWN PASS -----------
        Exiting -> do -- cannot be the root; we must be an inner node
        ---------------------------------------

          atomically $ waitForKill ctxRef
          atomically $ trySetStatusKill tleft
          atomically $ trySetStatusKill tright
          atomically $ writeTVar ctxRef $ setStatus ctx $ Running
          repeat

        -------------------------------------------------
        Kill -> do -- should not occur
        -------------------------------------------------
          error "syncTChildren: ThrdStatus Kill, should not occur"

        orelse -> repeat
  
  -----------------------------------
  -- now start synchronisation loop
  -----------------------------------

  repeat
\end{spec}
\lstset{numbers=none}

\ignore{

\begin{code}
trySetStatusKill :: TVar ThrdEnv -> STM ()
trySetStatusKill ctxRef = do 
  ctx  <- readTVar ctxRef
  case getStatus ctx of
    Terminated -> return ()
    _ -> writeTVar ctxRef $ setStatus ctx Kill

\end{code}

\begin{code}

syncTChildren :: (Int -> Bool) -> TVar String -> TVar ThrdEnv -> ThrdPSMVarMap -> IO ()
syncTChildren maxf ioRef ctxRef cntPred = do 
  -- wait for both children to Init
  (predl, predr) <- atomically $ waitForInit ctxRef
  -- get references to children, etc (must now be defined!)
  ctx@(TTreeNode (tId,_,_,_,_,pVarLocMapRef) mpar _ (Just tleft) (Just tright)) <- atomically $ readTVar ctxRef
  pVarLocMap <- atomically $ readTVar pVarLocMapRef -- $  
  let locKeys = keysSet pVarLocMap
  -- install predictions for children, and start them
  atomically $ do -- could be separated, since sequentialised by program order; but keep bundled for didactic reasons
    setPSMVarPot "INIT" ioRef tleft predl
    setPSMVarPot "INIT" ioRef tright predr
  atomically $ setStatusRef tleft Start 
  atomically $ setStatusRef tright Start 

  -----------------------------------
  -- now start synchronisation loop
  -----------------------------------
  
  let repeat cnt = do

       stat <- atomically $ getStatusRef ctxRef

       case stat of

        ----------------- UP PASS -----------
        Running -> do
        -------------------------------------

          (stat1, stat2) <- atomically $ waitForCompletion tleft tright -- $
          case (stat1, stat2) of 

            ---------------------------------
            (Terminated, Terminated) -> do
            ---------------------------------

              atomically $ removeChildren ctxRef -- $

            -------------------------------------------
            (Pausing upPred1, Pausing upPred2) -> do
            -------------------------------------------

              -- we obtain child nodes, they will not change 
              left  <- atomically $ readTVar tleft
              right <- atomically $ readTVar tright

              if not (isJust mpar) then do 
              -----------------------------
              -- this is the root thread, switch to down pass
              -----------------------------
                if not (maxf cnt) then do 
                  ---------------
                  -- more ticks
                  ---------------
                  tickPSMVars ioRef ctx
                  atomically $ do -- $
                    setPSMVarPot "PAUSE" ioRef tleft upPred1
                    setPSMVarPot "PAUSE" ioRef tright upPred2
                  atomically $ writeTVar tleft  $ setStatus left Start
                  atomically $ writeTVar tright $ setStatus right Start
                  repeat (cnt + 1)
                else do 
                  -----------------
                  -- no more ticks
                  ------------------
                  -- break, send Kill signal to children
                  atomically $ writeTVar tleft  $ setStatus left Kill 
                  atomically $ writeTVar tright $ setStatus right Kill
                  -- continue and wait for children to terminate
                  repeat cnt
              else do 
              ---------------------------
              -- this is an inner node, keep going up
              ---------------------------
              -- if ctx is not the root of the clock, set ctx "Pausing". 
              -- combine predictions of children and pass up the tree to parent 
              -- If both children have through paths we must add sequential downstream predictions from cntPred (we compute the prediction of a fork)
              -- We must filter and remove predictions for the local variables
                let upPred1' = Map.map (\p -> withoutKeys p locKeys) upPred1 
                    upPred2' = Map.map (\p -> withoutKeys p locKeys) upPred2 
                    cntPred' = Map.map (\p -> withoutKeys p locKeys) cntPred 
                    -- parPred'  =  unionWith addPred upPred1' upPred2'
                    parPred' =  concPred upPred1' upPred2'
                    -- does continuation of parallel have a thr path
                    thr' = Map.member (CmplId 0) parPred'
                    pred' = if thr' then addSeqPred parPred' cntPred'   
                            else parPred'
                atomically $ writeTVar ctxRef $ setStatus ctx $ Pausing pred'
                repeat cnt

            --------------------------------------------
            (Terminated, Pausing upPred) -> do
            -------------------------------------------- 
              
              right <- atomically $ readTVar tright

              if not (isJust mpar) then do 
              -----------------------
              -- we are root thread and switch to down-pass
              -----------------------
                if not (maxf cnt) then do
                  -------------------
                  -- more ticks
                  -------------------
                  tickPSMVars ioRef ctx
                  atomically $ setPSMVarPot "PAUSE" ioRef tright upPred -- $
                  atomically $ writeTVar tright $ setStatus right Start
                  repeat (cnt + 1)
                else do
                  --------------------
                  -- no more ticks
                  --------------------
                  atomically $ writeTVar tright $ setStatus right Kill
                  -- repeat, wait for child to terminate
                  repeat cnt
              else do 
              ------------------------
              -- we are not the root, we propagate non-local predictions of child upwards
              ------------------------
                let upPred' = Map.map (`withoutKeys` locKeys) upPred 
                    cntPred' = Map.map (`withoutKeys` locKeys) cntPred 
                    -- does continuation of parallel have a thr path
                    thr' = Map.member (CmplId 0) upPred'
                    pred' = if thr' then addSeqPred upPred' cntPred'   
                            else upPred'
                atomically $ writeTVar ctxRef $ setStatus ctx $ Pausing pred'
                repeat cnt

            -------------------------------------------
            (Pausing upPred, Terminated) -> do
            -------------------------------------------

              left  <- atomically $ readTVar tleft

              if not (isJust mpar) then do 
              -----------------------
              -- we are root thread, switching to down pass
              -----------------------
                if not (maxf cnt) then do
                  --------------------
                  -- more ticks
                  --------------------
                  tickPSMVars ioRef ctx
                  atomically $ setPSMVarPot "PAUSE" ioRef tleft upPred -- $
                  atomically $ writeTVar tleft $ setStatus left Start
                  repeat (cnt + 1)
                else do
                  ------------------
                  -- no more ticks
                  ------------------
                  atomically $ writeTVar tleft $ setStatus left Kill
                  -- repeat wait for child to terminate
                  repeat cnt
              else do 
              -------------------------------
              -- we are not the root thread, keep going up
              -------------------------------
                -- pVarLocMap  <- atomically $ readTVar pVarLocMapRef -- $
                let -- locKeys  = keysSet pVarLocMap
                    -- upPred'  = withoutKeys upPred locKeys 
                    upPred' = Map.map (\p -> withoutKeys p locKeys) upPred 
                    -- cntPred'  = withoutKeys cntPred locKeys 
                    cntPred' = Map.map (\p -> withoutKeys p locKeys) cntPred 
                    -- does continuation of parallel have a thr path
                    thr' = Map.member (CmplId 0) upPred'
                    -- pred  = if thr' then unionWith addPred upPred' cntPred'
                    --         else upPred'
                    pred' = if thr' then addSeqPred upPred' cntPred'   
                            else upPred'
                -- putStrLn "\n Up PT"
                --print $ "Prediction left: " ++ show (restrictKeys upPred locKeys)
                atomically $ writeTVar ctxRef $ setStatus ctx $ Pausing pred'
                repeat cnt

            ---------------------------------
            (Exiting, _) -> do
            ---------------------------------

              left  <- atomically $ readTVar tleft

              if not (isJust mpar) then do 
              ---------------------
              -- we are root node, switch to down pass
              ---------------------
              --  set both Children to Kill, if not Terminated already and continue 
                -- print "DONE" 
                atomically $ logStr ioRef "KILL\n" 
                atomically $ writeTVar tleft $ setStatus left Kill
                atomically $ trySetStatusKill tright
                -- repeat wait for child to terminate
                atomically $ writeTVar ctxRef $ setStatus ctx $ Running
                repeat cnt

              else do 
              ----------------------
              -- we are inner node, pass Exiting upwards              ----------------------
                -- pass up Exiting and continue (wait for termination)
                -- termination status must be set by forkChild
                atomically $ logStr ioRef "KILL\n" 
                atomically $ writeTVar ctxRef $ setStatus ctx $ Exiting
                repeat cnt

            ---------------------------------
            (_, Exiting) -> do
            ---------------------------------

              right <- atomically $ readTVar tright

              if not (isJust mpar) then do 
              ---------------------
              -- we are root node, switch to down pass
              ---------------------
              --  set both Children to Kill if not Terminated, and continue
              --  if child is terminated it will not respond to the Kill signal
                atomically $ logStr ioRef "KILL\n" 
                atomically $ writeTVar tright $ setStatus right Kill
                atomically $ trySetStatusKill tleft
                -- repeat wait for children to terminate
                atomically $ writeTVar ctxRef $ setStatus ctx $ Running
                repeat cnt 
              
              else do 
              ----------------------
              -- we are inner node, pass Exiting upwards
              ----------------------
                -- pass up Exiting and continue (wait for termination)
                -- termination status must be set by forkChild
                atomically $ logStr ioRef "KILL\n" 
                atomically $ writeTVar ctxRef $ setStatus ctx $ Exiting
                repeat cnt

            orelse -> error "syncChildren: -panic- children have completed but status has glitched"

        ----------------- DOWN PASS -----------
        Pausing _ -> do -- cannot be the root; we must be an inner node
        ---------------------------------------

          -- wait for ctx status to become Start, set status to Running
          -- in addition, if we get kill signal, then propagate it downwards 
          start <- atomically $ waitForStartOrKill ctxRef -- $
          -- do we need to grab nodes again, since children cannot have changed?
          left  <- atomically $ readTVar tleft
          right <- atomically $ readTVar tright 

          case (getStatus left, getStatus right) of

            --------------------------------------------
            (Pausing upPred, Terminated) -> do
            -------------------------------------------- 

              if (not start) then do -- must Kill
              ----------------
              -- no more ticks
              ----------------
                 atomically $ writeTVar tleft $ setStatus left Kill
                 -- wait for child to terminate
                 atomically $ writeTVar ctxRef $ setStatus ctx Running
                 repeat cnt
              else do
              ------------------
              -- more ticks
              ------------------
                 tickPSMVars ioRef ctx
                 atomically $ setPSMVarPot "PAUSE" ioRef tleft upPred -- $
                 atomically $ writeTVar tleft $ setStatus left Start
                 atomically $ writeTVar ctxRef $ setStatus ctx Running
                 repeat (cnt + 1)

            -------------------------------------------
            (Terminated, Pausing upPred) -> do
            -------------------------------------------

              if (not start) then do 
              ------------------
              -- no more ticks
              ------------------
                 atomically $ writeTVar tright $ setStatus right Kill
                 -- wait for child to terminate
                 atomically $ writeTVar ctxRef $ setStatus ctx Running
                 repeat cnt
              else do
              ---------------
              -- more ticks
              ---------------
                 -- tick local pvars
                 tickPSMVars ioRef ctx
                 atomically $ setPSMVarPot "PAUSE" ioRef tright upPred -- $
                 atomically $ writeTVar tright $ setStatus right Start  
                 atomically $ writeTVar ctxRef $ setStatus ctx Running 
                 repeat (cnt + 1)

            -----------------------------------------------------------
            (Pausing upPred1, Pausing upPred2) -> do
            -----------------------------------------------------------

              if (not start) then do
              -------------------
              -- no more ticks
              -------------------
                 atomically $ writeTVar tleft  $ setStatus left Kill
                 atomically $ writeTVar tright $ setStatus right Kill 
                 -- wait for children to terminate
                 atomically $ writeTVar ctxRef $ setStatus ctx Running 
                 repeat cnt
              else do
              ---------------
              -- more ticks
              ---------------
                 tickPSMVars ioRef ctx
                 atomically $ do -- could be separated, since sequentialised by program order; but keep bundled for didactic reasons
                    setPSMVarPot "PAUSE" ioRef tleft  upPred1
                    setPSMVarPot "PAUSE" ioRef tright upPred2
                 atomically $ writeTVar tleft  $ setStatus left  Start
                 atomically $ writeTVar tright $ setStatus right Start
                 atomically $ writeTVar ctxRef $ setStatus ctx Running 
                 repeat (cnt + 1)

        ----------------- DOWN PASS -----------
        Exiting -> do -- cannot be the root; we must be an inner node
        ---------------------------------------

          -- wait for ctx status to become Kill, set status to Running
          -- in addition and propagate Kill downwards
          atomically $ logStr ioRef "KILL\n"  
          start <- atomically $ waitForKill ctxRef -- $
          -- set status of children, unless the have terminated
          atomically $ trySetStatusKill tleft
          atomically $ trySetStatusKill tright
          atomically $ writeTVar ctxRef $ setStatus ctx $ Running
          repeat cnt

        -------------------------------------------------
        Kill -> do -- should not occur
        -------------------------------------------------
          error "syncTChildren: ThrdStatus Kill, should not occur"

        orelse -> repeat cnt
  
  repeat 0 
\end{code}

} 

\section{The PSM Language Combinators}
\label{appendix:partB}

The generic operators on \lhs{PSM} processes support the core functionality
for the programming of synchronous control flow. We can create local \lhs{PSMVar}s,
with \lhs{newPSMVar}, express instantaneous termination with \lhs{done}, \lhs{delay} or \lhs{kill} and
synchronise concurrent threads on clock ticks with \lhs{pause} and \lhs{wait}. The operators 
\lhs{ifte} and \lhs{switch} implement control flow branching.
Safe repeat loops can be constructed with \lhs{repUntil}. It permits the loop body to have instantaneous termination, i.e., we can generate data-dependent iterations within a single macro-step. 
The operator \lhs{var} provides policy-scheduled access to a shared \pvar and \lhs{val} access to a thread-local value. The operators \lhs{return} and \lhs{>>=} are the monad methods of \lhs{PSMVal} that embed Haskell conservatively inside \psm.  
The function \lhs{evalSTM} wraps arbitrary \stm transactions as \psm actions. 
Finally, there is fork-join 
parallel composition \lhs{(|||)} as well as sequential 
composition without value binding \lhs{(>>>)} and sequential composition with
value binding \lhs{(>>>=)}:

\begin{spec}
newPSMVar :: PSMVar a => PSMStatic a -> (PSMVar a -> PSM b) -> PSM b
done :: PSM ()
delay   :: Int -> PSM () 
pause   :: PSM () 
kill    :: PSM ()
wait    :: PSMVal a -> PSM a
ifte    :: PSMVal Bool -> PSM a -> PSM a -> PSM a 
switch :: PSMVal(Either a b) -> (PSMVal a -> PSM c) -> (PSMVal b -> PSM c) -> PSM c 
repUntil :: PSM Bool -> PSM ()
var     :: PSMType a => MethodId -> PSMVar a -> PSM a
val     :: PSMVal a -> PSM a
evalSTM :: PSM (STM a) -> PSM a
return  :: a -> PSMVal a 
(>>=)   :: PSMVal a -> (a -> PSMVal b) -> PSMVal b
(|||)   :: PSM () -> PSM () -> PSM () 
(>>>)   :: PSM a -> PSM b -> PSM b
(>>>=)  :: PSM a -> (PSMVal a -> PSM b) -> PSM b
\end{spec}

\noindent
We will discuss these operators one by one in the sequel.

\subsection{Generic Structure}

The applicative methods \lhs{pure} and \lhs{(<*>)} for the \psm context have been given in Sec.~\ref{sec:psm-language}.
The monad structure of \lhs{PSMVal} is seen in Sec.~\ref{sec:PSMVarVal}.
The evaluation of \stm transactions inside \psm is given in Sec.~\ref{sec:run-method}.

\subsection{Creating \lhs{PSMVar}s}

A process \lhs{newPSMVar stat (\\ v -> psm v)} constructs a fresh \lhs{PSMVar}
from the parameter \lhs{stat::PSMState a}, allocates it in memory, makes it available via reference \lhs{v::PSMVar a} and then evaluates \lhs{psm v}.  
The types \lhs{PSMState a} and \lhs{PSMVar a} must be inferable from the expressions \lhs{stat} and \lhs{psm}.

\lstset{numbers=left,numberstyle=\tiny}
\begin{spec}
newPSMVar :: PSMVar a => PSMStatic a -> (PSMVar a -> PSM b) -> PSM b
newPSMVar stat psm = PSM (pred, act) where
  PSM (pred, _) = psm Idle 
  act ioRef ctxRef cntPred = do
    vId <- atomically $ do -- $
      ctx <- readTVar ctxRef
      num  <- grabPSMVarInd ctx
      return $ PSMVarId num -- $
    -- assemble new instance of PSMVar in memory
    ctx <- atomically $ readTVar ctxRef -- $
    let TTreeNode (tId,_,pVarGlobMapRef,pVarLocMapRef) _ _ _ _ = ctx 
    -- get current global and local pvar predictions
    pVarGlobMap <- atomically $ readTVar pVarGlobMapRef
    pVarLocMap  <- atomically $ readTVar pVarLocMapRef
    tmust <- init stat
    tcan  <- atomically $ newTVar empty
    let newPSMVar = Active vId sId tmust tcan
    -- allocate pvar in memory and add reference into global and local pvar maps
    newPSMVarRef <- atomically $ newTVar (AnyPSMVar newPSMVar)
    atomically $ writeTVar pVarGlobMapRef (insert vId newPSMVarRef pVarGlobMap)
    atomically $ writeTVar pVarLocMapRef  (insert vId () pVarLocMap) -- $
    -- instantiate continuation process with new pvar
    let PSM (_, act2) = psm newPSMVar 
    -- run the action of continuation program
    res <- act2 ctxRef cntPred
    -- when finished terminate pVar and remove PSMVar in thread
    -- environment
    ctx <- atomically $ readTVar ctxRef -- $
    termPSMVars ctx
    atomically $ removePSMVar newPSMVar ctxRef -- $
    return res
\end{spec}
\lstset{numbers=none}

\begin{itemize}
  \item Since the \lhs{PSMVar} introduced by \lhs{newPSMVar stat (\\ v -> psm pv)} is local to the thread, it cannot be shared with the thread's concurrent environment. As a consequence, we do not need to publish any predictions about accesses to
  \lhs{v} made by the current thread. The prediction is obtained in line~3 from the prediction of the process \lhs{psm} with the \pvar set to \lhs{Idle}. This returns the potential accesses in \lhs{psm} to all \pvar{s} different from \lhs{v}. The following lines~4--32 define the action of process 
  \lhs{newPSMVar stat (\\ v -> psm v)} when it is evaluated in a tree context \lhs{ctxRef} and sequential continuation prediction \lhs{cntPred}.

  \item In lines~5--8 we pick a thread-locally fresh (unique) identifier \lhs{vID} for the new \lhs{PSMVar}. The \stm transaction \lhs{grabPSMVarInd} accesses the current \lhs{PSMVar} counter in the tree node \lhs{ctx} and increments it, atomically.
  
  \item In lines~10--19 we assemble the new \lhs{PSMVar} instance in memory. First, in lines~10--14, the current execution context \lhs{ctx} is accessed and the current global and local \lhs{PSMVar} references \lhs{pVarGlobMap} and \lhs{pVarLocMap} are extracted. Next, lines~15--19, we assemble the \lhs{newPSMVar} and allocate it in memory as \lhs{newPSMVarRef}.
  
  \item In lines~20--21 we add the new \lhs{PSMVar} identifier into the global and local \lhs{PSMVar} maps.

  \item Finally, in lines~23--32, the continuation process \lhs{psm} is instantiated with the new \lhs{newPSMVar} and its action executed in the current execution context \lhs{ctxRef}. When the thread returns, the \lhs{newPSMVar} is removed again from the thread environment in line~31. The function to do that is \lhs{removePSMVar}. 
  \begin{spec}
  removePSMVar :: PSMVar p => PSMVar p -> TVar ThrdEnv -> IO ()
  removePSMVar (Active vId _ state _) ctxRef = do 
    term state
    atomically $ do 
      ctx <- readTVar ctxRef
      let (TTreeNode (tId, _, pVarGlobMapRef, pVarLocMapRef) _ _ _ _) = ctx
      pVarGlobMap <- readTVar pVarGlobMapRef
      writeTVar pVarGlobMapRef (delete vId pVarGlobMap)
      pVarLocMap <- readTVar pVarLocMapRef
      writeTVar pVarLocMapRef (delete vId pVarLocMap)
  removePSMVar (Idle _) _ = error "removePSMVar: cannot remove Idle PSMVar"
  \end{spec}
  It first executes the generic \pvar method \lhs{term} (e.g., to release external resources, see Sec.~\ref{sec:pvar}) and then removes the \pvar from the global and local registers of the running \psm context.

\end{itemize}

An alternative would be to define the instantiation of a \lhs{PSMVar} as an operator \lhs{newPSMVar'} with the type 
\begin{spec}
  newPSMVar' :: PSMVar a => PSMState a -> PSM (PSMVar a)
\end{spec}
and then simply apply a parameterised process \lhs{psm::PSMVar a -> PSM b} by 
regular function application \lhs{proc (newPSMVar' stat)::PSM b}. This would make
\lhs{PSMVars} look more closely like \lhs{MVars} in the \lhs{IO} monad or \lhs{TVars} in the \lhs{STM} monad. 
However, \lhs{newPSMVar'}
would not allow us to hook up the clean-up function \lhs{removePSMVar} at the end of the life-time of the \lhs{PSMVar} when its scope is left. As a binding operator
\begin{spec}
  newPSMVar :: PSMVar a => PSMStatic a -> (PSMVar a -> PSM b) -> PSM b
\end{spec}
the \lhs{removePSMVar} clean-up action can be added, because the action context
\lhs{psm} hands back to the \lhs{newPSMVar} operator when it terminates.
This scopes \pvar{s} with strict lexical stack-like nesting, rather than closing them with the death of the thread. 

Note that we could define
\begin{spec}
  newPSMVar' :: PSMVar a => PSMStatic a -> PSM (PSMVar a)
  newPSMVar' stat = newPSMVar stat pure
\end{spec}
where \lhs{pure::PSMVar a -> PSM (PSMVar a)} is the applicative embedding. 
However, this is not creating a pinned \pvar for shared access in the \psm context. Instead the \pvar available with \lhs{newPSMVar' stat} is a pure (thread-local) \emph{value} that can be passed via \lhs{(>>>=)} as an input to an \psm action of type 
\begin{center}
   \lhs{psm::PSMVal(PSMVar a) -> PSM b}. 
\end{center} 
This does not give a \pvar reference for a \psm action which would have type \lhs{psm::PSMVar a -> PSM b}.

\ignore{

\begin{code}
newPSMVar :: PSMType p => String -> PSMStatic p -> (PSMVar p -> PSM b) -> PSM b
newPSMVar sId stat psmf = PSM (pred, trp, act) where
  PSM(pred, trp, _) = psmf (Idle sId) 
  act ioRef ctxRef cntPred = do
    vId <- atomically $ do -- $
      ctx <- readTVar ctxRef
      num  <- grabPSMVarInd ctx
      return $ PSMVarId num -- $
    -- assemble new instance of PSMVar in memory
    ctx <- atomically $ readTVar ctxRef -- $
    let TTreeNode (tId,_,numRef,_,pVarGlobMapRef,pVarLocMapRef) _ _ _ _ = ctx 
    -- get current global and local pvar predictions
    pVarGlobMap <- atomically $ readTVar pVarGlobMapRef
    pVarLocMap  <- atomically $ readTVar pVarLocMapRef
    tmust <- init stat -- this may have IO side effect (must be singleton)
    tcan  <- atomically $ newTVar empty
    let newPSMVar = Active vId sId tmust tcan
    -- allocate pvar in memory and 
    -- add reference into global and local pvar maps
    newPSMVarRef <- atomically $ newTVar (AnyPSMVar newPSMVar)
    atomically $ writeTVar pVarGlobMapRef (insert vId newPSMVarRef pVarGlobMap)
    atomically $ writeTVar pVarLocMapRef  (insert vId () pVarLocMap) -- $
    -- instantiate continuation process with new pvar
    let PSM (_, _, act2) = psmf newPSMVar 
    -- run the action of continuation program
    res <- act2 ioRef ctxRef cntPred
    -- when finished terminate pVar and remove PSMVar in thread
    -- environment; reading of ctxRef probably not needed
    ctx <- atomically $ readTVar ctxRef -- $
    removePSMVar newPSMVar ctxRef -- $
    return res
\end{code}

}

\ignore{

\begin{code}

removePSMVar :: PSMType p => PSMVar p -> TVar ThrdEnv -> IO ()
removePSMVar (Active vId _ state _) ctxRef = do 
  term state
  atomically $ do 
    ctx <- readTVar ctxRef
    let (TTreeNode (tId, _, _, _, pVarGlobMapRef, pVarLocMapRef) _ _ _ _) = ctx
    -- get current global and local pvar references
    pVarGlobMap <- readTVar pVarGlobMapRef
    writeTVar pVarGlobMapRef (delete vId pVarGlobMap)
    pVarLocMap <- readTVar pVarLocMapRef
    writeTVar pVarLocMapRef (delete vId pVarLocMap)
removePSMVar (Idle _) _ = error "removePSMVar: cannot remove Idle PSMVar"

\end{code}

}

\subsection{Prediction}

For the prediction we are interested only in the surface behaviour of a \psm program block.
This is the behaviour reachable via \emph{instantaneous} control-flow paths, i.e., paths that do not involve the execution of a \lhs{pause} (or \lhs{wait}) statement. 
The surface behaviour makes up the initial macro-step of the program block. 
Among all the method calls that are instantaneously reachable, called \emph{surface calls}, we distinguish the \emph{through} calls and the \emph{sink} calls. 
A \emph{through} call is instantaneously reachable and lies on a \emph{through path}, i.e., the termination point of the program block is reachable from the call via \emph{some} instantaneous path.   
A \emph{sink} call is instantaneously reachable but \emph{all} paths starting out from it must complete by pausing (if they complete at all). 
Since we use \emph{completion code} \lhs{CmplId 0} for termination and \lhs{CmplId 1} for pausing, the through calls are also called $0$-calls and the sink calls are referred to as $1$-calls. 
Note that every method call in the surface of a program block is either a $0$-call or a $1$-call. 

\smallskip 

The prediction of a program is a static over-approximation of the number of $0$, $1$ and $2$ calls, on every \pvar \lhs{vId::PSMVarId}, as follows.
Suppose \lhs{pred::ThrdPSMVarMap} is the prediction for a \psm expression \lhs{psm = PSM(pred,act)}.
Then \lhs{pred!CmplId 0!vId!mId == 5} means that there are at most $5$ distinct occurrences of a through call to the \pvar \lhs{vId::PSMVarId} with method identifier \lhs{mId::MethodId} in \lhs{psm}. 
Likewise, \lhs{pred!CmplId 1!vId!mId == 0} means that there are no sink calls of \pvar \lhs{vId} with method \lhs{mId} inside \lhs{psm}.    
When we compute \lhs{pred} then every occurrence of a surface call on \lhs{vId} in the program block \lhs{psm} is counted exactly once either with completion code \lhs{CmplId 0} or with code \lhs{CmplId 1}.

\smallskip

In order to permit instantaneous loops (\lhs{repUntil}) where method calls can be repeated for a statically unbounded (data-dependent) number of times, we introduce the \emph{recurrent} calls as a third category with completion code \lhs{CmplId 2}. 
A surface call is a \emph{recurrent} call, or a $2$-call, if it occurs in an instantaneous control-flow cycle, i.e., if it could potentially be repeated over and over in the intial macro-step. 
Obviously, the classification as a $2$-call expresses an extra property of a surface call. 
Each $0$-call or $1$-call in the surface may be a recurrent call or not. 
If a surface call does not lie in the scope of any loop, then it is counted as a \emph{transient} call. 
In the prediction \lhs{pred} of a program \lhs{psm} we actually count the number of \lhs{repUntil} scopes in which the method call occurs. 
So, for instance, \lhs{pred!CmplId 2!vId!mId == 2} means that there is either a single method call \lhs{mId} on \lhs{vId} that occurs inside two nested \lhs{repUntil} loops, or two occurrences of method calls \lhs{mId} that occur in the same or in different \lhs{repUntil} scopes. 
If \lhs{pred!CmplId 0!vId!mId == 1}, \lhs{pred!CmplId 2!vId!mId == 1} and \lhs{pred!CmplId 2!vId!mId == 1} then this tells us that we have two calls of \lhs{mId} on \lhs{vId} (one through and another sink) and one of them is transient and the other recurrent.

\smallskip

The predictions \lhs{map.(!vId) pred::PSMVarPred} are used to install and update the \emph{potentials} that the scheduling context maintains for each \pvar \lhs{v} to schedule the accesses to it.
For a given thread \lhs{tId::ThreadId} and potential \lhs{pot::PSMVarPotMap} the counters \lhs{potCnt::Prediction} and \lhs{recCnt::Prediction} in the pair
\begin{center} 
 \lhs{(potCnt, recCnt) = pot!tId} 
\end{center} 
bound the number of surface calls to \lhs{v} that are to be expected during the current macro-step.
The first, \lhs{potCnt}, is simply the sum of all possible \emph{remaining} through and sink calls. 
The second, \lhs{recCnt}, counts the number of outstanding loop scopes from which the potential for a method calls may be re-introduced. 
For instance, if \lhs{potCnt!mId == 0} and \lhs{recCnt!mId == 2} then this means that currently there are no more occurrences of \lhs{mId} on any direct forward control-flow path to completion. But we are inside two loop scopes with back-jumps that may potentially reintroduce a surface path to \lhs{mId}. 

\smallskip 

Note that at any moment, a method call is counted several times; Once for a forward surface path on which is its reachable (\lhs{potCnt!mId}) and again for each loop scope from which the method could be re-reachable (\lhs{recCnt!mId}). 
The potential counters \lhs{potCnt!mId} and \lhs{recCnt!mId} for a method \lhs{mId} is initiated from the prediction at the beginning of a macro-step. The former count \lhs{potCnt!mId} is decremented each time the method \lhs{mId} is executed.
The recurrent count \lhs{recCnt!mId}, however, is discharged only upon exit of a loop scope in which \lhs{mId} is instantaneously reachable. When the loop scope is repeated then the surface count of the loop body (i.e. all call instantaneously reachable) is again added to the surface potential \lhs{potCnt!mId}.

\smallskip 

To compute the initial potential of a process in a compositional way, we need to make predictions carry around the completion code as explained above. 
This completion status is propagated backwards in the collection of surface calls. Whenever these surface calls fall inside a loop body, then the through calls (code \lhs{CmplId 0}) among them are transformed into recurrent calls and we assign them the status \lhs{CmplId 2}.

\subsection{Skip and Thread Delay}

The statement \lhs{done::PSM()} 
\begin{spec}
done :: PSM ()
done = PSM (fromList [(CmplId 0, empty)], \ _ _ -> return ()) 
\end{spec}
terminates instantly. It is the same as \lhs{val}, i.e., 
\begin{spec}
done = val $ pure () 
\end{spec}
Note that the prediction publishes an empty map of method counters for completion code \lhs{CmplId 0}. The fact that 
\begin{spec} 
  keys $ fromList [(CmplId 0, empty)] == [CmplId 0]
\end{spec}
tells the context for prediction calculation that \lhs{done} is terminating instantaneously. A definition like 
\begin{center} 
  \lhs{done' = PSM (empty, \ _ _ -> return ())}
\end{center}
would be wrong, because then \lhs{keys.done' == []} would not carry any completion information.

\smallskip 

The operator \lhs{delay d::PSM()} is like \lhs{done} except it  
waits for a period of \lhs{d} milliseconds (GHC) before it terminates. It can be used to 
seed explicit delays into the control flow. This is useful for testing
the independence from the timing of individual 
micro-steps, or to enforce synchronisation of the clock tick with a real-time clock.
An extension of \psm could provide more sophisticated access to wall-clock time wrapping the \lhs{Data.UTC} module. 
\begin{spec}
delay :: Int -> PSM ()
delay d = PSM (fromList [(CmplId 0, empty)], \ _ _ -> threadDelay d) 
\end{spec}

\ignore{

\begin{code}

done :: PSM ()
done = PSM (pred, 0, \ _ _ _ -> return ()) where 
  pred = fromList [(CmplId 0, empty)]

delay :: Int -> PSM ()
delay d = PSM (pred, 0, \ _ _ _ -> threadDelay d) where
  pred = fromList [(CmplId 0, empty)]

\end{code}

}

\subsection{Pausing}

The \lhs{pause} operator 
\begin{spec}
pause :: PSM ()
pause = wait $ pure ()
\end{spec} 
delays a given \lhs{PSM} process by a single macro-step. It makes the \psm action wait for clock synchronisation. 
The more general construct \lhs{wait(e)::PSM a} completes the current macro-step (completion code \lhs{CmplId 1}) and returns a value $e::a$.  
Such a return value can be used to control the exit from a \lhs{repUntil} loop.

\lstset{numbers=left,numberstyle=\tiny}
\begin{spec}
wait :: PSMVal a -> PSM a
wait r = PSM (pred, act') where
  pred = fromList [(CmplId 1, empty)] -- we register a sink path with empty predictions
  act' ctxRef cntPred =
    case r of 
      Bot -> error "wait: -panic- trying to evaluate undefined value"
      PSMVal v -> do
        ctx <- atomically $ readTVar ctxRef -- $
        let TTreeNode (_, _, _, pVarLocMapRef) _ _ _ _ = ctx  
        -- get pvars local to this thread
        pVarLocMap <- atomically $ readTVar pVarLocMapRef -- $
        let -- obtain the prediction for the next tick
            -- remove local pvars from the downstream continuation 
          upPred = 
            Map.map (\p -> withoutKeys p (keysSet pVarLocMap)) cntPred
        atomically $ writeTVar ctxRef (setStatus ctx (Pausing upPred))
        -- depth behaviour (wait for start signal)
        start <- atomically $ waitForStartOrKill ctxRef
        if not start then do
          atomically $ writeTVar ctxRef (setStatus ctx Terminated)
          throwIO (PSMTrap "kill") -- kill continuation
        else do
          ctx <- atomically $ readTVar ctxRef -- $
          tickPSMVars ioRef ctx
          atomically $ writeTVar ctxRef $ setStatus ctx Running
          return v
\end{spec}
\lstset{numbers=none}

Since a \lhs{wait} completes instantly, its prediction \lhs{pred} for methods computed
in line~3 is empty at completion code \lhs{CmplId 0}.
The action \lhs{act'} of a \lhs{wait r} with \lhs{r::PMSVal a} is given in lines~4--26. If the value \lhs{r} is undefined we raise an error. This should never happen as long as integrity of the \psm library is preserved. If \lhs{r} contains a value \lhs{v} then this value is finally returned by the action in line~26. Until that happens, the action needs to complete the current macro-step and publish preditions for the next. 
It first extracts from the current execution context \lhs{ctx} the local object map \lhs{pVarLocMap} in lines~9--11. From this we obtain the prediction from the continuation process 
for the next macro-step. 
In general, the prediction contains method accesses both
for \lhs{PSMVars} introduced by the current thread (in which the \lhs{wait} is executed) as well as for objects introduced by the thread's ancestors. It is only for the latter that the \lhs{wait} thread can be in competition with its concurrent environment and hence needs to publish its predictions for the next tick. These predictions \lhs{upPred} are extracted from \lhs{pred} in lines~14--15. Then, in line~16
the \lhs{wait} thread sets its status to \lhs{Pausing} storing the next prediction
\lhs{upPred} in the thread's status. In line~18 the thread waits for the \lhs{Start} signal from the synchroniser. When it receives \lhs{Start}, it ticks the local
\lhs{PSMVars} (lines~23--24) and atomically sets the thread status to \lhs{Running}
(line~25), before returning in line (line~26).
If the signal arriving from the synchroniser is a \lhs{Kill} then the return value \lhs{start} is \lhs{False}.
In that case, our action set its own thread's status to \lhs{Terminated} and throws an \lhs{PSMTrap} exception to kill its continuation code (lines~20--21). This will immediately return to the \lhs{forkFinally} provided by the parent thread.

\subsection{Killing}

A thread can \emph{weakly} preempt the full \psm context at any moment by calling \lhs{kill}. This permits all other concurrent threads in the context to complete the current macro-step for one last time.
The thread executing \lhs{kill} completes with code \lhs{CmplId 1} and behaves like a \lhs{pause} for prediction, because it stops any possible instantaneous control-flow path. 
\lstset{numbers=left,numberstyle=\tiny}
\begin{spec}
kill :: PSM ()
kill = PSM (pred, act') where
  pred = fromList [(CmplId 1, empty)]
  -- end of affairs: we register a sink path with empty predictions
  act' ioRef ctxRef _ = do
    ctx <- atomically $ readTVar ctxRef -- $
    atomically $ writeTVar ctxRef (setStatus ctx Exiting)
    atomically $ waitForKill ctxRef 
    throwIO (PSMTrap "kill")
\end{spec}
\lstset{numbers=none}
In contrast to \lhs{pause} and \lhs{wait}, however, \lhs{kill} does not publish new predictions but instead sets its status to \lhs{Exiting} (line~7). 
This will make the synchroniser distribute the \lhs{Exiting} request across the thread tree and eventually pass down a \lhs{Kill} signal. 
When this arrives the \stm action in line~8 unblocks and the thread finishes itself by throwing \lhs{PSMTrap}.  

\ignore{ 

\begin{code}

-- pause with a return value
wait :: PSMVal a -> PSM a
wait r = PSM (pred, 0, act') where
  pred = fromList [(CmplId 1, empty)]
  -- we register a sink path with empty predictions
  act' ioRef ctxRef cntPred =
    case r of 
      Bot -> error "wait: -panic- trying to evaluate undefined value"
      PSMVal v -> do
        -- surface behaviour
        ctx <- atomically $ readTVar ctxRef -- $
        let TTreeNode (_, _, _, _, _, pVarLocMapRef) _ _ _ _ = ctx  
        -- get pvars local to this thread
        pVarLocMap <- atomically $ readTVar pVarLocMapRef -- $
        let -- obtain the prediction for the next tick
            -- remove local pvars from the downstream continuation 
          upPred = Map.map (\p -> withoutKeys p (keysSet pVarLocMap)) cntPred
        atomically $ writeTVar ctxRef (setStatus ctx (Pausing upPred))
        -- depth behaviour (wait for start signal)
        start <- atomically $ waitForStartOrKill ctxRef
        if not start then do
          atomically $ writeTVar ctxRef (setStatus ctx Terminated)
          throwIO (PSMTrap "kill") -- kill continuation
        else do
          ctx <- atomically $ readTVar ctxRef -- $
          tickPSMVars ioRef ctx
          atomically $ writeTVar ctxRef $ setStatus ctx Running
          return v

pause :: PSM ()
pause = wait $ pure ()

data Trap = PSMTrap String deriving Show
instance Exception Trap 

kill :: PSM ()
kill = PSM (pred, 0, act') where
  pred = fromList [(CmplId 1, empty)]
  -- end of affairs: we register a sink path with empty predictions
  act' ioRef ctxRef _ = do
        -- surface behaviour
        ctx <- atomically $ readTVar ctxRef -- $  
        -- set thread status
        atomically $ writeTVar ctxRef (setStatus ctx Exiting)
        atomically $ waitForKill ctxRef -- sets own status to Terminated
        throwIO (PSMTrap "kill") 
        -- kill any continuation of thread
        -- synchroniser is cleaning up
        

\end{code}

} 

\subsection{Forking}

Parallel composition \lhs{(|||)} is a binary infix operator that actually wraps the prefix function \lhs{fork}
\begin{code}
infixl 1 |||
p ||| q = fork p q
\end{code}
which itself uses the functions
\lhs{forkChild} and \lhs{syncTChildren} introduced above:
\lstset{numbers=left,numberstyle=\tiny}
\begin{spec}
fork :: PSM a -> PSM b -> IO ThreadId
fork psm1 psm2 = PSM (pred, act) where 
  ~(PSM(pred1, _)) = psm1
  ~(PSM(pred2, _)) = psm2
  pred = concPred pred1 pred2
  act ctxRef cntPredx = do 
    forkChild ctxRef psm1
    forkChild ctxRef psm2
    syncTChildren ctxRef cntPred
  myThreadId
\end{spec}
\lstset{numbers=none}
\noindent
The prediction \lhs{pred} of a parallel composition of \psm actions \lhs{psm1} and \lhs{psm2} is obtained as the concurrent composition of the predictions of the two sub-processes. This is computed in line~5 by the function \lhs{concPred}, see below. The action \lhs{act} of the parallel composition forks the actions of \lhs{psm1} and \lhs{psm2} as new child threads in lines~7--8 and then passes control to the synchroniser \lhs{syncTChildren} in line~9. 
The connection beteen the parent synchroniser and the children is maintained via the parent's thread node in the thread tree that is linked by \lhs{ctxRef}.


\begin{code}
concPred :: ThrdPSMVarMap -> ThrdPSMVarMap -> ThrdPSMVarMap 
concPred pred1 pred2 = unionsWith (unionWith maxPred) predMaps where 
  predMaps  = [predMap00, potCnt, predMap10, predMap11, predMap22]
  thr1Term  = member (CmplId 0) pred1 
  thr2Term  = member (CmplId 0) pred2 
  thr1Pause = member (CmplId 1) pred1 
  thr2Pause = member (CmplId 1) pred2
  predMap22 = unionWith (unionWith addPred) pred12 pred22 where 
    pred12 = restrictKeys pred1 $ Data.singleton (CmplId 2)
    pred22 = restrictKeys pred2 $ Data.singleton (CmplId 2)
  predMap00 = if thr1Term && thr2Term then -- if psm1 has a through path
       fromList [(CmplId 0, 
         unionWith addPred (pred1!(CmplId 0)) (pred2!(CmplId 0)))]
    else empty -- no through path
  predMap10 = 
    if thr1Pause && thr2Term then
       fromList [(CmplId 1, unionWith addPred (pred1!(CmplId 1)) (pred2!(CmplId 0)))]
    else empty
  potCnt = 
    if thr1Term && thr2Pause then
       fromList [(CmplId 1, unionWith addPred (pred1!(CmplId 0)) (pred2!(CmplId 1)))]
    else empty
  predMap11 = 
   if thr1Pause && thr2Pause then
      fromList [(CmplId 1, unionWith addPred (pred1!(CmplId 1)) (pred2!(CmplId 1)))]
   else empty
\end{code}

\ignore{

\begin{code}

forkPar :: TVar String -> String -> TVar ThrdEnv -> ThrdPSMVarMap -> PSM a -> PSM b -> IO ThreadId
forkPar = forkParN (\_ -> False)

forkParN :: (Int -> Bool) -> TVar String -> String -> TVar ThrdEnv -> ThrdPSMVarMap -> PSM a -> PSM b -> IO ThreadId
forkParN kill ioRef fName ctxRef cntPredx psm1 psm2  = do
  forkChild ioRef (fName ++ "-L") ctxRef psm1
  forkChild ioRef (fName ++ "-R") ctxRef psm2
  syncTChildren kill ioRef ctxRef cntPredx
  myThreadId

fork :: PSM a -> PSM b -> PSM ThreadId 
fork = forkN (\_ -> False) "root"

forkN :: (Int -> Bool) -> String -> PSM a -> PSM b -> PSM ThreadId 
forkN kill fName psm1 psm2 = PSM (predx, trp, act) where 
  ~(PSM(predx1, trp1, _)) = psm1
  ~(PSM(predx2, trp2, _)) = psm2
  predx = concPred predx1 predx2
  trp = max trp1 trp2
  act ioRef ctxRef cntPredx = do 
    forkChild ioRef (fName ++ "-L") ctxRef psm1
    forkChild ioRef (fName ++ "-R") ctxRef psm2
    syncTChildren kill ioRef ctxRef cntPredx
    myThreadId

xfork :: String -> PSM a -> PSM b -> PSM ThreadId 
xfork tName = forkN (\_ -> False) tName

\end{code}

}

\subsection{Sequencing}
\label{sec:sequencing}

Next we define sequencing with binding, which makes \lhs{PSM} behave almost
like a monad. A process \lhs{psm1 >>>= \\ x -> psm2 x} first
runs \lhs{psm1} until it terminates and then passes control to \lhs{psm2}.
The value \lhs{x} returned by \lhs{psm1} upon termination is passed as parameter to downstream \lhs{psm2}. 
The sequencing (if we monadic syntax was available for \psm) could viewed as a construct \lhs{let x <- psm1 in psm2 x} or even \lhs{do x <- run psm1; psm2 x}. 
Note that the parameter \lhs{x} to be bound must be a pure value,
i.e., have type \lhs{PSMVal a}. This prevents the process \lhs{psm2 x} from making its prediction depend on the run-time value \lhs{x}.

\lstset{numbers=left,numbersep=5pt}
\begin{spec}
infixl 1 >>>=

(>>>=) :: PSM a -> (PSMVal a -> PSM b) -> PSM b
(>>>=) psm1 fpsm2 = PSM (pred, act) where
  ~(PSM(pred1, act1)) = psm1
  ~(PSM(pred2, _))    = fpsm2 Bot
  pred  = addSeqPred pred1 pred2
  act ctxRef cntPred = do
    let cntPred1 = addSeqPred pred2 cntPred
    x <- act1 ioRef ctxRef cntPred1
    let PSM(_, act2) = fpsm2 (PSMVal x)
    act2 ioRef ctxRef cntPred
\end{spec}
\lstset{numbers=none}

\noindent
In lines~5--7 we compute the prediction of the composition \lhs{psm1 >>>=  fpsm2} from the prediction \lhs{pred1} of \lhs{psm1}and \lhs{pred2} of \lhs{fpsm2}. 
The undefined value \lhs{Bot} 
is used to extract the prediction from the program continuation \lhs{fpsm2}.
This value \lhs{Bot} is the only value we have available during the
prediction phase. The application \lhs{fpsm Bot} in line~6 
is guaranteed to return a
prediction \lhs{pred2}, while the action may not be defined and thus remains untouched by the wildcard pattern \lhs{_}.
In the action phase (lines~9--12), which executes the sequential composition, we obtain a
defined value \lhs{x} in line~10 which we feed forward as \lhs{PSMVal x} in line~11 to instantiate the downstream action \lhs{act2}. This action is finally executed in line~12.

\ignore{

\begin{code}
infixl 1 >>>=

(>>>=) :: PSM a -> (PSMVal a -> PSM b) -> PSM b
(>>>=) psm1 fpsm = PSM (pred, trp, act) where
  ~(PSM(pred1, trp1, act1)) = psm1
  ~(PSM(pred2, trp2, _)) = fpsm Bot -- lazy pattern to permit free guarded recursion
  thr' = member (CmplId 0) pred1
  pred2x = Map.foldr (unionWith addPred) empty $ delete (CmplId 2) pred2
  pred = addSeqPred pred1 pred2
  trp = if thr' then trp2 else trp1
  act ioRef ctxRef cntPred = do
    let cntPred1 = addSeqPred pred2 cntPred
    x <- act1 ioRef ctxRef cntPred1
    let PSM (_, _, act2) = fpsm (PSMVal x)
    act2 ioRef ctxRef cntPred
\end{code}

} 

\smallskip

The computation of the prediction is encapsulated in the function 
\lhs{addSeqPred} 
\begin{code}
addSeqPred :: ThrdPSMVarMap -> ThrdPSMVarMap -> ThrdPSMVarMap 
addSeqPred pred1 pred2 =
  let thr1Term = member (CmplId 0) pred1 in
    if thr1Term then -- upstream can terminate
      let thr2Term  = member (CmplId 0) pred2 -- downstream has a through path?
          predMap0 = 
            if thr2Term then 
              fromList [(CmplId 0, 
                unionWith addPred (pred1!(CmplId 0)) (pred2!(CmplId 0)))]
            else empty -- no through path
          predMap1 = 
            if not thr2Term then
              fromList [(CmplId 1, 
                  (unionWith addPred (pred1!(CmplId 0)) 
                    (pred2!(CmplId 1))))]
            else restrictKeys pred1 $ Data.singleton (CmplId 1)
          thr1Loop = member (CmplId 2) pred1
          thr2Loop = member (CmplId 2) pred2
          predMap2 = 
            if thr1Loop || thr2Loop then 
              fromList [(CmplId 2, 
                unionWith addPred (findDef pred1 (CmplId 2)) 
                  (findDef pred2 (CmplId 2)))]
            else empty
      in unions [predMap0, predMap1, predMap2]
    else pred1 -- upstream cannot terminate

findDef :: ThrdPSMVarMap -> CmplId -> Map PSMVarId Prediction
findDef predMap cmpl = findWithDefault empty cmpl predMap

\end{code}

The simple sequencing combinator \lhs{psm1 >>> psm2} without binding 
is a trivial case of \lhs{(>>>=)}:
\begin{spec}
infixl 1 >>>

(>>>) :: PSM a -> PSM b -> PSM b
psm1 >>> psm2 = psm1 >>>= \ x -> psm2
\end{spec}

\smallskip 

We note that the lazy pattern in code line~6 of \lhs{(>>>=)} is important to permit \emph{clock-guarded} \psm processes to be defined by simple recursion at Haskell level. For instance, the process
\begin{spec} 
halt = loop where 
  loop = pause >>> loop
\end{spec}
is a process that immediately pauses in every macro-step. 
The lazy pattern in line~6 makes the prediction calculation for \lhs{halt} stop and terminate at the \lhs{pause} construct. 
If we used an eager pattern, the evaluation for prediction would run round in an infinite loop. 
The use of recursion at Haskell level is only possible for \emph{guarded} \psm processes, i.e., where every recursion loop passes through a \lhs{pause} statement. For general, unguarded, \psm processes we provide the structured \lhs{repUntil} loop described below. 
The \lhs{halt} would then also be written 
\begin{spec}
halt = repUntil $ pause >>> val (pure False)
\end{spec}

\ignore{

\begin{code}

infixl 1 >>>

-- note the lazy pattern to permit pause-guarded tail recursions
-- without the need to use repUntil
(>>>) :: PSM a -> PSM b -> PSM b
(PSM psm1) >>> ~(PSM psm2) = PSM (predx, 0, act) where
 -- fix traps
  ~(predx1, _, act1) = psm1
  ~(predx2, _, act2) = psm2
  predx = addSeqPred predx1 predx2
  act ioRef ctxRef cntPredx = do
    let cntPredx1 = addSeqPred predx2 cntPredx
    act1 ioRef ctxRef cntPredx1
    act2 ioRef ctxRef cntPredx
  
\end{code}

} 

\subsection{Repeat Until Loop}

The repeat loop \lhs{repUntil psm} first runs \lhs{psm} for as many macro-steps as necessary until it terminates. It then repeats \lhs{psm} immediately if the return value is \lhs{False}. Otherwise, when \lhs{psm} terminates and returns \lhs{True}, \lhs{repUntil psm} terminates as well.
This implements arbitrary guarded or unguarded loops in \psm. 
\lstset{numbers=left,numbersep=5pt}
\begin{spec}
repUntil :: PSM Bool -> PSM ()
repUntil psm = PSM (pred, act) where
  ~(PSM(predBody, actBody)) = psm
  predRec  = Map.singleton (CmplId 2) $ findWithDefault empty (CmplId 0) predBody
  pred = unionsWith (unionWith addPred) [predBody, predRec]
  act ctxRef cntPred = do
    let cntPred1 = adjust (\_ -> empty) (CmplId 0) cntPred
    c <- actBody ctxRef cntPred1
    case c of 
      False -> do -- repeat
        atomically $ addPSMVarPot ctxRef $ 
          restrictKeys predBody $ Data.fromList [CmplId 0, CmplId 2]
        act ctxRef cntPred1
      True -> atomically $ exitPSMVarPot ctxRef predBody -- exit
\end{spec}
\lstset{numbers=none}
In the above code, the loop body has the prediction \lhs{predBody} and the action \lhs{actBody} (line~3). 
We book every method call predicted on a through path in the body (i.e., $0$-calls with \lhs{CmplId 0} in \lhs{predBody}) as a recurrent call ($2$-calls \lhs{CmplId 2}) for the loop. These $2$-calls are are computed as \lhs{predRec} in line~4 above. 
The prediction \lhs{pred} for the repeat loop is then obtained from \lhs{predBody} in line~5 by adding (for each completion code \lhs{CmplId} and \lhs{PSMVarId}) the recurrent count \lhs{predRec}. 
Note that any existing recurrent call in the body (calls in inner loops) keeps being accounted for as a $2$-call.

\smallskip 

The action \lhs{act} of \lhs{repUntil psm} is defined in lines~6--14. 
The action is called with a downstream prediction \lhs{cntPred} which contains the prediction for method calls to happen squentially \emph{after} the loop is exited. We do not pass any $0$-calls from this continuation of the loop to internal pauses and so compute an adjusted prediction \lhs{cntPred1} in which these are deleted. The body \lhs{actBody} is then started with \lhs{cntPred1} in line~8. 
The loop's body \lhs{act} will reduce the surface counts in the potential, i.e., consume all the counts for $0$, $1$ and $2$-calls that were added as part of \lhs{predBody}.   
In general, when a \psm action terminates, it will have removed all its original prediction from the potential. 
Depending on the return value \lhs{c} of the body's action (line~8) we now keep repeating or exit the loop. 
If we repeat (line~13), then we must recharge the surface calls of the body to the potential (line~11--12). 
If we exit, then we remove the body's through calls from the recurrent predictions of the potential in line~14 with the helper function \lhs{exitPSMVarPot}. These $0$-calls had been added at the start of the loop as $2$-calls. 

\ignore{

\begin{code}

repUntil :: PSM Bool -> PSM ()
repUntil psm = PSM (predLoop, trp1, act) where
  ~(PSM(predBody, trp, actBody)) = psm
  trp1 = trp + 1
  -- thr paths of body are transcribed to count as 2-calls of loop
  predRec = Map.singleton (CmplId 2) $ findWithDefault empty (CmplId 0) predBody
  predLoop = unionsWith (unionWith addPred) [predBody, predRec]
  act ioRef ctxRef cntPredx = do
    -- we do not pass any 0-calls from continuation of loop to internal pauses 
    let cntPredx1 = adjust (\_ -> empty) (CmplId 0) cntPredx
    c <- actBody ioRef ctxRef cntPredx1
    case c of 
      False -> do -- repeat
        -- recharge 0-calls and 2-calls of body for next iteration
        atomically $ addPSMVarPot "RECHARGE LOOP" ioRef ctxRef $ restrictKeys predBody $ Data.fromList [CmplId 0, CmplId 2]
        -- repeat loop
        act ioRef ctxRef cntPredx
      True -> do -- exit
        -- we discharge all 2-calls
        atomically $ exitPSMVarPot "EXIT LOOP" ioRef ctxRef predBody


-- for loop: entering into the loop
forLoop :: Int -> PSM a -> PSM a
forLoop cnt p = PSM (predLoop, trp, act) where
  (~(PSM(predBody, trp', _))) = p
  -- predBody is the prediction of the loop body for the current tick
  -- until completion, for a single unfolding of all inner loops
  -- predLoop is the prediction for the outer loop 
  predLoop = unionsWith (unionWith addPred) [predBody, pred2] where 
  -- predLoop = unionsWith (unionWith addPred) [pred01, pred2] where 
  -- thr/snk paths of body to be inherited as thr/snk paths of the outer loop
  -- thr paths of body are transcribed to become 2-calls of loop
  pred2 = Map.singleton (CmplId 2) $ findWithDefault empty (CmplId 0) predBody
  -- pred1 = Map.singleton (CmplId 2) $ findWithDefault empty (CmplId 1) predBody
  thrx = member (CmplId 0) predLoop
  trp = trp' + 1
  act ioRef ctxRef cntPredx = do
    -- get unique (thread-locally fresh) rId for new loop trap
    -- now start loop unfolding
    let PSM(_, _, act') = forLoopRun cnt p
    act' ioRef ctxRef cntPredx

-- unfolded version of the loop
forLoopRun :: Int -> PSM a -> PSM a
forLoopRun cnt p =
  if cnt <= 1 then PSM (predx1, trp1, actFinal) 
  else PSM (predx, trp1, actNext) where 
      ~(PSM(predx1, trp1, actBody)) = p
      ~(PSM(predx2, _, actRest)) = forLoopRun (cnt-1) p
      -- predictions are as for sequential composition
      predx = addSeqPred predx1 predx2
      pred2x = Map.foldr (unionWith addPred) empty $ delete (CmplId 2) predx2
      -- discharge at exit: transcribe 0-calls of loop body as 2-calls
      -- this is the loop exit; execute the loop body one final time 
      actFinal ioRef ctxRef cntPredx = do
        let cntPred = Map.foldr (unionWith addPred) empty $ delete (CmplId 2) cntPredx -- eventually remove
        -- execute the loop body
        x <- actBody ioRef ctxRef cntPredx
        -- now we discharge all 2-calls
        ctx <- atomically $ readTVar ctxRef
        atomically $ exitPSMVarPot "EXIT LOOP" ioRef ctxRef predx1 -- predL0
        -- return method result
        return x
      -- this is the continuation
      actNext ioRef ctxRef cntPredx = do
        let cntPredx1 = addSeqPred (adjust (\_ -> empty) (CmplId 0) predx1) cntPredx
        actBody ioRef ctxRef cntPredx1
        -- recharge 0-calls and 2-calls of body for next iteration
        atomically $ addPSMVarPot "RECHARGE LOOP" ioRef ctxRef $ restrictKeys predx1 $ Data.fromList [CmplId 0, CmplId 2]
        -- repeat loop
        actRest ioRef ctxRef cntPredx

\end{code}

} 

\subsection{Conditional Branching}

In a conditional branching the evaluation of a boolean pure value determines
which of two processes is to be executed. 
Note that the Boolean input \lhs{c} is of type \lhs{PSMVal Bool} to capture the fact that the prediction does not depend on the decision value. 
Since in the calculation of the prediction we do not know this value, we must over-approximate the \lhs{PSMVar} accesses of the two branches by taking the maximum of both.
The prediction of the conditional \lhs{ifte c psm1 psm2} is obtained as the maximum of the predcitions \lhs{pred1} and \lhs{pred2} of the two branches \lhs{psm1} and \lhs{psm2} in line~5.
\lstset{numbers=left}
\begin{spec}
ifte :: PSMVal Bool -> PSM a -> PSM a -> PSM a 
ifte c psm1 psm2 = PSM (pred, act) where
  ~(PSM(pred1, act1)) = psm1
  ~(PSM(pred2, act2)) = psm2 
  pred = unionWith (unionWith maxPred) pred1 pred2
  act ctxRef cntPred = do
    case c of 
      PSMVal True -> do
         let apred2 = unionWith union pred2 pred1
             delta1 = unionWith (unionWith (\x y -> diffBranchPred x y)) pred1 apred2 
         atomically $ subPSMVarPot ctxRef delta1
         act1 ioRef ctxRef cntPred
      PSMVal False -> do
        let apred1 = unionWith union pred1 pred2
            delta2 = unionWith (unionWith (\x y -> diffBranchPred x y)) pred2 apred1
        atomically $ subPSMVarPot ctxRef delta2
        act2 ioRef ctxRef cntPred
      Bot -> error "ifte: attempt to evaluate undefined PSMVal"  
\end{spec}
\lstset{numbers=none}
When the conditional is finally executed (action phase) we know the value and thus which branch is taken, either lines~8--12 or lines~13--17.
At this point, the predictions must be adjusted for each \lhs{PSMVar}. Specifically, they must be reduced if the branch that is taken has a smaller prediction than the branch not taken.
To this end, the conditional constructs two prediction deltas \lhs{delta1} (lines~9--10) and \lhs{delta2} (lines~14--15) from the predictions \lhs{pred1} and \lhs{pred2} of the two branches.
These are used to reduce the potential in lines~11 and~16 at the time the decision value is computed and the branch is decided.
The reduction is performed by the auxiliary function \lhs{subPSMVarPot}. It
subtracts, using \lhs{subPred}, a given prediction for a given \pvar in 
the current node and all ancestors of a given thread. 

\smallskip

More concretely, if \lhs{c} evaluates \lhs{True} the left branch \lhs{psm1} is taken and its action \lhs{act1} is executed in line~12. 
Before this happens, the potential is reduced in line~11 by the differential prediction 
\begin{spec}
delta1 = unionWith (unionWith (\x y -> diffBranchPred x y)) pred1 apred2 where 
  apred2 = unionWith union pred2 pred1
\end{spec}
which determines, for each completion code and \pvar, by what amount the method counts of the potentials must be reduced relative to the over-approximating maximum
\begin{spec}
pred = unionWith (unionWith maxPred) pred1 pred2
\end{spec}
if the first branch \lhs{psm1} is taken. 
We apply the difference function \lhs{diffBranchPred} between \lhs{pred1} and \lhs{pred2} to each such method count. 
Note that we need to extend \lhs{pred2} by the entry in \lhs{pred1} in case that \lhs{pred2} does not contain a completion code or a \pvar registered in \lhs{pred1}.
This adjustment is done by using  \lhs{unionWith union pred2 pred1} instead of \lhs{pred2} in the definition of \lhs{delta1}. 
A similar ajustment is made inside \lhs{diffBranchPred} for the \pvar indices.
 
\ignore{

\begin{code}

ifte :: PSMVal Bool -> PSM a -> PSM a -> PSM a 
ifte c psm1 psm2 = PSM (pred, trp, act) where
  ~(PSM(pred1, trp1, act1)) = psm1
  ~(PSM(pred2, trp2, act2)) = psm2 
  trp = max trp1 trp2
  pred = unionWith (unionWith maxPred) pred1 pred2
  act ioRef ctxRef cntPred = do
    case c of 
      PSMVal True -> do
         let apred2 = unionWith union pred2 pred1
             delta1 = unionWith (unionWith (\x y -> diffBranchPred x y)) pred1 apred2 
         atomically $ subPSMVarPot "LEFT BRANCH" ioRef ctxRef delta1
         act1 ioRef ctxRef cntPred
      PSMVal False -> do -- else do
        let apred1 = unionWith union pred1 pred2
            delta2 = unionWith (unionWith (\x y -> diffBranchPred x y)) pred2 apred1
        atomically $ subPSMVarPot "LEFT BRANCH" ioRef ctxRef delta2
        act2 ioRef ctxRef cntPred
      Bot -> error "ifte: attempt to evaluate undefined PSMVal"  


canDnPred :: TVar ThrdEnv -> Map PSMVarId Prediction -> Map PSMVarId Prediction -> PSMVarId -> IO()
canDnPred ctxRef deltaSurf deltaRec vId = do
  ctx <- atomically $ readTVar ctxRef -- $
  let (TTreeNode (_, _, _, _, pVarGlobMapRef, _) _ _ _ _) = ctx
  pVarGlobMap <- atomically $ readTVar pVarGlobMapRef 
  -- still to be checked: are indices defined in the map?
  pvar <- atomically $ readTVar (pVarGlobMap ! vId) -- $
  let AnyPSMVar(Active _ _ _ predMapRefx) = pvar
  (_, thrdPath) <- atomically $ 
    collectSpine (\ (x, _, _, _, _, _) -> x) ctxRef -- $
  let dSurf = findWithDefault empty vId deltaSurf
      dRec = findWithDefault empty vId deltaRec
      adjustFn (surf, rec) = 
        (subPred surf dSurf, subPred rec dRec) 
      adjustThrdPred tId = do
        potMapx <- atomically $ readTVar predMapRefx  -- $
        atomically $ writeTVar predMapRefx $ adjust adjustFn tId potMapx
  mapM_ adjustThrdPred thrdPath

\end{code}

} 

\smallskip 

The \lhs{switch} operator generalises \lhs{ifte} from Boolean conditions to choice types which also carry a value to be passed to the evaluated branch. 
\begin{spec}
switch :: PSMVal (Either a b) -> (PSMVal a -> PSM c) -> (PSMVal b -> PSM c) -> PSM c 
switch c psm1f psm2f = PSM (pred, act) where
  ~(PSM(pred1, _)) = psm1f Bot 
  ~(PSM(pred2, _)) = psm2f Bot
  pred = unionWith (unionWith maxPred) pred1 pred2
  act ctxRef cntPred = do
    case c of 
      PSMVal (Left a) -> do
         let PSM(_, act1) = psm1f (PSMVal a)
             apred2 = unionWith union pred2 pred1
             delta1 = unionWith (unionWith (\x y -> diffBranchPred x y)) pred1 apred2 
         atomically $ subPSMVarPot ctxRef delta1
         act1 ctxRef cntPred
      PSMVal (Right b) -> do
        let PSM(_, act2) = psm2f (PSMVal b)
            apred1 = unionWith union pred1 pred2
            delta2 = unionWith (unionWith (\x y -> diffBranchPred x y)) pred2 apred1
        atomically $ subPSMVarPot ctxRef delta2
        act2 ctxRef cntPred
      Bot -> error "switch: attempt to evaluate undefined PSMVal"  

\end{spec}

\ignore{

\begin{code}
switch :: PSMVal (Either a b) -> (PSMVal a -> PSM c) -> (PSMVal b -> PSM c) -> PSM c 
switch c psm1f psm2f = PSM (pred, trp, act) where
  ~(PSM(pred1, trp1, _)) = psm1f Bot 
  ~(PSM(pred2, trp2, _)) = psm2f Bot
  trp = max trp1 trp2
  pred = unionWith (unionWith maxPred) pred1 pred2
  act ioRef ctxRef cntPred = do
    case c of 
      PSMVal (Left a) -> do
         let PSM(_, _, act1) = psm1f (PSMVal a)
             apred2 = unionWith union pred2 pred1
             delta1 = unionWith (unionWith (\x y -> diffBranchPred x y)) pred1 apred2 
         -- adjust the predictions
         atomically $ subPSMVarPot "LEFT BRANCH" ioRef ctxRef delta1
         act1 ioRef ctxRef cntPred
      PSMVal (Right b) -> do -- else do
        let PSM(_, _, act2) = psm2f (PSMVal b)
            apred1 = unionWith union pred1 pred2
            delta2 = unionWith (unionWith (\x y -> diffBranchPred x y)) pred2 apred1
        atomically $ subPSMVarPot "LEFT BRANCH" ioRef ctxRef delta2
        act2 ioRef ctxRef cntPred
      Bot -> error "switch: attempt to evaluate undefined PSMVal"  

\end{code}

} 

\subsection{Closing the PSM Memory}

To execute a \psm process' action, our prototype provides the functions
\begin{spec}
runPSM    :: PSM a -> IO ThreadId
runStrIO  :: (PStrIO -> PSM a) -> IO ThreadId
runStrION :: Int -> (PStrIO -> PSM a) -> IO ThreadId
\end{spec}
They close a \lhs{PSM} process, i.e., remove it from the \psm environment,
and run it through an arbitrary number of macro-steps until it terminates. 
Specifically, \lhs{runStrIO (\\ io -> psm)} puts the process \lhs{psm} in parallel with a simple \lhs{IO} process called \lhs{ioLoop} that operates a string-based input-output object \lhs{io::PStrIO} as described in Sec.~\ref{sec:psmvars-assortment}:
\begin{spec}
ioLoop :: PStrIO -> PSM ()
ioLoop io = cycle where 
  cycle = 
    getStrIO io >>>
    rdStrIO io  >>>= \l ->
    ifte (do s <- l; return (s == ":q"))
      (kill)
      (putStrIO io >>> pause >>> cycle)

runStrIO :: (PStrIO -> PSM a) -> IO ThreadId
runStrIO fproc = 
  let psm = 
        newPSMVar "io" "0" $ \io ->
        -- test , second io object
        newPSMVar "ioX" "1" $ \ioX ->
        fork (ioLoop io) (fproc io) in do -- $
    rootId <- myThreadId
    pVarGlobMapRef <- atomically $ newTVar empty -- $
    pVarLocMapRef <- atomically $ newTVar empty -- $
    ctxRef <- atomically $ newRoot (rootId, Running, pVarGlobMapRef, pVarLocMapRef)
    let PSM (_, act) = psm
    act ctxRef (fromList [(CmplId 0, empty)])
\end{spec}
Inside the expression \lhs{psm} of \lhs{runStrIO (\\ io -> psm)} we may use the methods \lhs{wrStrIO} and \lhs{rdStrIO} to access \lhs{io} and connect to std-in and std-out. At every macro-step of \psm, the function \lhs{rdStrIO} makes available the contents of the command line that is read at the beginning of the step. The function \lhs{wrStrIO} appends a string to the outbuf which is printed to the command line at the end of the step. 

\smallskip 

The wrapper \lhs{runStrIO} adds the concurrent process \lhs{ioLoop} which uses an arbitrary \lhs{io::PStrIO}. 
This process drives the user interaction.
In each macro-step it first reads a string line from the keyboard with \lhs{getStrIO}. If the string is equal to \lhs{":q"} which abbreviates the command to quit execution, then the loop terminates.
Otherwise, the output buffer is written to the command line with \lhs{putStrIO} and we pause. 
Notice that since the \lhs{PStrIO} policy gives the writing action \lhs{wrStrIO} higher precedence over the printing \lhs{putStrIO}, it is guaranteed that the printing by \lhs{ioLoop} only happens when no thread in \lhs{psm} is trying any more to write to the command line.
Hence, the command line printout will contain all strings written by any 
thread during the current macro-step.
Similarly, the reading \lhs{rdStrIO} of the keyboard has lower priority than \lhs{getStrIO} so that no thread in \lhs{psm} will ever attempt to read the input buffer until it has been set by \lhs{ioLoop}.
Al threads connected to the same \lhs{PStrIO} object will see exactly the same sequence of command line strings. 

\smallskip

The function 
\begin{spec}
runPSM :: PSM a -> IO ThreadId
runPSM psm = runStrIO (\_ -> psm)
\end{spec}
is a wrapper for processes that do not use the standard \lhs{PStrIO} facility but handle input and output, if any, in other ways. 
Finally, the function \lhs{runStrION (\\ io -> psm)} behaves like \lhs{runStrIO (\\ io -> psm)} but iterates through a pre-specified number \lhs{n} of clock ticks:
\begin{spec}
runStrION :: Int -> (PStrIO -> PSM a) -> IO ThreadId
runStrION n fproc = 
  let prgm = 
        newPSMVar strIOInit $ \io ->
        fork (ioLoop io) (fproc io) in do -- $
    rootId <- myThreadId
    pVarGlobMapRef <- atomically $ newTVar empty -- $
    pVarLocMapRef <- atomically $ newTVar empty -- $
    ctxRef <- atomically $ newRoot (rootId, Running, pVarGlobMapRef, pVarLocMapRef)
    forkChild ctxRef done
    forkChild ctxRef prgm
    syncTChildren (\x -> x == n) ctxRef $ (fromList [(CmplId 0, empty)])
    myThreadId
\end{spec} 
Note the extra parameter \lhs{\\ x -> x == n} in the extended form of \lhs{synchTChildren} which defines a predicate on the macro-step index at which the \psm context is automatically and prematurely killed.  

\smallskip

\ignore{

\begin{code}

runPSM :: PSM a -> IO ThreadId
runPSM proc = runStrIO (\_ -> proc)

runStrIO :: (PStrIO -> PSM a) -> IO ThreadId
runStrIO fproc = 
  let prgm = 
        newStrIO "IO"  $ \io ->
        fork (ioLoop io) (fproc io) in do -- $
    rootId <- myThreadId
    pVarGlobMapRef <- atomically $ newTVar empty -- $
    pVarLocMapRef <- atomically $ newTVar empty -- $
    pVarCntRef <- atomically $ newTVar 0 -- $
    trapCntRef <- atomically $ newTVar 0 -- $
    ctxRef <- atomically $ -- $
      newRoot (ThrdId rootId "top", Running, pVarCntRef, trapCntRef, pVarGlobMapRef, pVarLocMapRef)
    ioRef <- atomically $ newTVar ""
    let PSM (_, _, act) = prgm
    act ioRef ctxRef $ fromList [(CmplId 0, empty)]

runStrION :: Int -> (PStrIO -> PSM a) -> IO ThreadId
runStrION max fproc = 
  let prgm = 
        newStrIO "IO" $ \io ->
        fork (ioLoop io) (fproc io) in do -- $
    rootId <- myThreadId
    pVarGlobMapRef <- atomically $ newTVar empty -- $
    pVarLocMapRef <- atomically $ newTVar empty -- $
    pVarCntRef <- atomically $ newTVar 0 -- $
    trapCntRef <- atomically $ newTVar 0 -- $
    ctxRef <- atomically $ -- $
      newRoot (ThrdId rootId "top", Running, pVarCntRef, trapCntRef, pVarGlobMapRef, pVarLocMapRef)
    ioRef <- atomically (newTVar "")
    -- create a dummy empty thread to install top-level synchroniser with kill
    forkChild ioRef "dummy" ctxRef done
    forkChild ioRef "top" ctxRef prgm
    syncTChildren (\x -> x == max) ioRef ctxRef $ (fromList [(CmplId 0, empty)])
    myThreadId

-- via kill and PSM-managed repeat ioLoopkr
runStrIOkr :: (PStrIO -> PSM a) -> IO ThreadId
runStrIOkr fproc = 
  let prgm = 
        newStrIO "IO" $ \io ->
        fork (ioLoopkr io) (fproc io) in do -- $
    rootId <- myThreadId
    pVarGlobMapRef <- atomically $ newTVar empty -- $
    pVarLocMapRef <- atomically $ newTVar empty -- $
    pVarCntRef <- atomically $ newTVar 0 -- $
    trapCntRef <- atomically $ newTVar 0 -- $
    ctxRef <- atomically $ -- $
      newRoot (ThrdId rootId "top", Running, pVarCntRef, trapCntRef, pVarGlobMapRef, pVarLocMapRef)
    ioRef <- atomically (newTVar "")
    let PSM (_, _, act) = prgm
    act ioRef ctxRef (fromList [(CmplId 0, empty)])

\end{code} 

} 

\ignore{

\begin{code}

ioLoop :: PStrIO -> PSM ()
ioLoop io = cycle where 
  cycle = 
    -- getStrIO io >>>
    rdStrIO io  >>>= \l ->
    ifte (do s <- l; return (s == ":q"))
      (kill)
      (-- putStrIO io >>> 
       pause >>> 
       cycle)

-- with kill and repeat
ioLoopkr :: PStrIO -> PSM ()
ioLoopkr io = cycle where 
  cycle = repUntil $
    -- getStrIO io  >>> 
    rdStrIO io >>>= \l ->
    -- putStrIO io >>> 
    printSLog >>> -- note that log is unsynchronised!
    wait (do s <- l; return (s == ":q"))

\end{code}
}

\smallskip

Observe that the execution is taking place inside the \lhs{IO} monad and
thus has access to the world environment. 
Due to the clock synchronisation and coherence of the \lhs{PSMVar} objects,
however, the external world (keyboard, mouse, etc) cannot influence
the evaluation of the \lhs{PSM} process other than through changes in the
memory at the clock tick. For instance, a \lhs{PStrIO} buffer \lhs{PSMVar} may 
read the external event queue at each clock tick through the \lhs{tick} method.
Since all threads run in lock-step with the clock tick, the evaluation
remains deterministic and reproducible from the sequence of memory states
at the clock ticks.

\section{Examples of PSMVars}
\label{appendix:partC}

\subsection{Input and Output}
\label{sec:IO}

Communication with the external world is not taking place in the \lhs{IO} monad, as for normal Haskell programs, but in the protected \lhs{PSM} context. In this context, all threads are synchronised in lock-step with each other via the clock tick method. The external world is connected through special \lhs{PSMVar} objects that prevent the volatile external environment from introducing data races.

\smallskip

In our experimental system we provide a race-free \lhs{PStrIO} object for string output and input, which has been described briefly in Sec.~\ref{sec:psmvars-assortment}. The objects are non-polymorphic, so their indexing type is a singleton, called \lhs{DynStrIO}, while \lhs{PStrIO} is the derived type of \psm scheduled objects, that wraps \lhs{DynStrIO} by the \lhs{PSMVar} type operator:
\begin{code}
data DynStrIO = DynStrIO
type PStrIO = PSMVar DynStrIO
\end{code}
The following declarations make up the implementation of the \lhs{PSMType} class methods for \lhs{DynStrIO}:

\lstset{numbers=left,numberstyle=\tiny}
\begin{spec}
instance PSMType DynStrIO where

  type PSMStatic DynStrIO = String
  data PSMState DynStrIO = DynStrIOState (TVar (Integer, String, String, String))

  data instance Method DynStrIO b c where 
    WrStrIO  :: Method DynStrIO String ()
    PutStrIO :: Method DynStrIO () ()
    GetStrIO :: Method DynStrIO () ()
    RdStrIO  :: Method DynStrIO () String

  methodId WrStrIO  = MethodId 1
  methodId PutStrIO = MethodId 2
  methodId GetStrIO = MethodId 3
  methodId RdStrIO  = MethodId 4
\end{spec}
\lstset{numbers=none}
Every \lhs{PStrIO} object is instantiated with a string to identify it as a specific IO-port, the associated type \lhs{PSMStatic DynStrIO} (Line~3) is \lhs{String}. The associated type \lhs{PSMState DynStrIO} (Line~4) stores in a \lhs{TVar} the current ordinal count\footnote{In the experimental system this is an \texttt{Integer}. In a serious application this should be replaced by \texttt{Int} to avoid memory overflow.} 
and three string buffers for the IO-port name and to connect with the GHC command line.
The four methods are specified Lines~6--15, with their argument and return types (Lines~6--10) and then their method indices (Lines~12--15). The precedence policy for the methods is given in Lines~17--20.  

\smallskip 

\noindent The policy for scheduling these method types, seen in Fig.~\ref{fig:ptsig-interface}, is declared as follows:
\begin{spec}
  prec _ (MethodId 1) = return $ Just [] -- WrStrIO,  writing to the outbuf
  prec _ (MethodId 2) = return $ Just [MethodId 1] -- PutStrIO, printing outbuf to std-out
  prec _ (MethodId 3) = return $ Just [MethodId 3] -- GetStrIO, acquiring a std-in line into the inbuf
  prec _ (MethodId 4) = return $ Just [MethodId 3] -- RdStrIO, reading the inbuf
\end{spec}
The global methods of the memory structure \lhs{DynStrIO} 
are the \lhs{init}, \lhs{tick} and \lhs{term} methods, as follows. 
There is no non-trivial information needed to create and instantiate a \lhs{DynStrIO}. The \lhs{init} method simply creates a fresh \lhs{TVar} with its port name, initial tick count $0$ and empty buffers:
\begin{spec}
  init name = do
    mustRef <- atomically $ newTVar (0, name, "", "")
    return $ DynStrIOState mustRef  -- $
\end{spec}
The \lhs{tick} method increments the counter for the macro-step and empties the outgoing buffer. Thus, the outbuf only stores the strings that are being sent to std-out during a given macro-step:
\begin{spec}
  tick (DynStrIOState mustRef) = do
    (cnt, name, inl, outl) <- atomically $ readTVar mustRef
    -- print buffer to stdout
    putStr $ "Tick " ++ show cnt ++ " @ Port " ++ name ++ ": " ++ outl
    -- read from stdin with prompt
    let lineBreak = "\n"
    putStr (lineBreak ++ "Port " ++ name ++ ">> ")
    line <- getLine
    -- refresh buffer
    atomically $ writeTVar mustRef (cnt+1, name, line, "")
\end{spec}
Finally, the \lhs{term} method is trivial. There is nothing to be done when the a \lhs{PStrIO} object reaches the end of its life time:
\begin{spec}
  term _ = return ()
\end{spec}

\smallskip 

\noindent Now we lift the methods as operations on \psm actions. First, to create a \lhs{PStrIO} the generic \lhs{newPSMVar} induces a function \lhs{newStrIO} as follows:
\begin{spec}
newStrIO :: (PStrIO -> PSM b) -> PSM b
newStrIO proc = newPSMVar proc
\end{spec}
Method \lhs{WrStrIO} (\lhs{MethodId 1}) appends a string to a std-in output buffer and \lhs{PutStrIO} (\lhs{MethodId 2}) prints the output buffer to std-out. A second string buffer in the \lhs{PStrIO} pipes to std-in. The method type \lhs{GetStrIO} (\lhs{MethodId 3}) overwrites the inbuf from the command line and \lhs{RdStrIO} (\lhs{MethodId 4}) extracts the inbuf as a \psm value.

\smallskip 

A method call \lhs{wrStrIO s} does not change the macro-step counter and the in-buffer, but appends the string $s$ to the out-buffer. It is implemented as follows:
\begin{spec}
  method WrStrIO (DynStrIOState mustRef) s = do -- $ 
      (cnt, name, inl, outl) <- readTVar mustRef
      writeTVar mustRef (cnt, name, inl, outl ++ s)

wrStrIO :: PStrIO -> PSMVal String -> PSM ()
wrStrIO = runSTMMtd WrStrIO $ method WrStrIO 
\end{spec}
The \lhs{method} \lhs{PutStrIO} prints the contents of the out-buffer to stdout preceeded by \lhs{"Tick <cnt>: "} where \lhs{<cnt>} is the current macro-step counter. Notice, for convenience, we declare the exported method \lhs{putStrIO::PStrIO -> PSM ()} without input argument. It is constant and fixed at \lhs{pure ()} which is the same as \lhs{PSMVal ()}:
\begin{spec}
  method PutStrIO (DynStrIOState mustRef) _ = do
      (cnt, inl, outl) <- atomically (readTVar mustRef)
      putStr $ "Tick " ++ show cnt ++ ": " ++ outl

putStrIO :: PStrIO -> PSM ()
putStrIO io = runIOMtd PutStrIO (method PutStrIO) io (pure ())
\end{spec}
The method \lhs{GetStrIO} (\lhs{MethodId 3}) reads the in-buffer and thus has return type \lhs{String}. 
The method prints a prompt \lhs{">> "} and copies the command line into the in-buffer: 
\begin{spec}
  method GetStrIO (DynStrIOState mustRef) _ = do
      (cnt, _, _) <- atomically (readTVar mustRef)
      let lineBreak = if cnt == 0 then "" else "\n" in do
        putStr (lineBreak ++ ">> ")
        line <- getLine
        atomically $ do
          (cnt, _, outl) <- readTVar mustRef
          writeTVar mustRef (cnt, line, outl)
        return line

getStrIO :: PStrIO -> PSM String
getStrIO io = runIOMtd PutStrIO (method PutStrIO) io (pure ())
\end{spec}
Finally, the method \lhs{RdStrIO} (\lhs{MethodId 4}) merely reads the current in-buffer, leaving the state of the \lhs{PStrIO} variable intact:
\begin{spec}
  method (DynStrIO mustRef) _ = do
      (_, inl, _) <- (readTVar mustRef)
      return inl

rdStrIO :: PStrIO -> PSM String
rdStrIO io = runSTMMtd RdStrIO (method RdStrIO) io (pure ())
\end{spec}

\ignore{

\begin{code}

instance PSMType DynStrIO where

  type PSMStatic DynStrIO = String
  data PSMState DynStrIO = 
    DynStrIOState (TVar (Integer, String, String, String))

  data instance Method DynStrIO b c where 
    WrStrIO  :: Method DynStrIO String ()
    RdStrIO  :: Method DynStrIO () String

  prec _ (MethodId 1 _) = return $ Just [] -- method 1 WrStrIO for writing to the out-buffer
  prec _ (MethodId 4 _) = return $ Just [] -- no need to wait

  init name = do
    mustRef <- atomically $ newTVar (0, name, "", "") -- $
    return $ DynStrIOState mustRef  -- $

  tick (DynStrIOState mustRef) = do
    (cnt, name, inl, outl) <- atomically $ readTVar mustRef
    -- print buffer to stdout
    putStr $ "Tick " ++ show cnt ++ " @ Port " ++ name ++ ": " ++ outl
    -- read from stdin with prompt
    let lineBreak = "\n"
    putStr (lineBreak ++ "Port " ++ name ++ ">> ")
    line <- getLine
    -- refresh buffer
    atomically $ writeTVar mustRef (cnt+1, name, line, "")

  term _ = return ()

  methodId WrStrIO  = MethodId 1 "wrt"
  methodId RdStrIO  = MethodId 4 "read"

  method WrStrIO (DynStrIOState mustRef) s = do -- $ 
      (cnt, name, inl, outl) <- readTVar mustRef
      writeTVar mustRef (cnt, name, inl, outl ++ s)

  method RdStrIO (DynStrIOState mustRef) _ = do
      (_, _, inl, _) <- (readTVar mustRef)
      return inl

newStrIO :: String -> (PStrIO -> PSM b) -> PSM b
newStrIO name proc = newPSMVar "io" name proc

wrStrIO :: PStrIO -> PSMVal String -> PSM ()
wrStrIO = runSTMMtd WrStrIO $ method WrStrIO 

rdStrIO :: PStrIO -> PSM String
rdStrIO io = runSTMMtd RdStrIO (method RdStrIO) io (pure ())

\end{code}

} 

The policy specification of the IO-object \lhs{PStrIO} of Fig.~6g exports concurrent independence between methods \lhs{wrtStrIO} and \lhs{rdStrIO}, i.e., $0 \Vdash \lhs{wrtStrIO}  \indep{} \lhs{rdStrIO}$. The obligation on the library author now is to verify that in any state of the encapsulated IO object any application of \lhs{wrtStrIO} commutes with any application of \lhs{rdStrIO}. I.e., if \lhs{io::PStrIO} is any memory state (abstracted as a single $0$ in Fig.~6g) and \lhs{str::PSMVal Str} any string then
  \lstinline[style=hs]{wrtStrIO io str >>> rdStrIO io >>>= $ \x -> ...}
which first writes \lhs{str} to \lhs{io} and then reads from \lhs{io}, is identical to 
  \lstinline[style=hs]{rdStrIO io >>>= $ \x -> wrtStrIO io str >>> ...}
which first reads value \lhs{x} from \lhs{io} and then writes \lhs{str}. The reason why this is true is that the \lhs{io} object maintains two independent buffers, one for incoming characters and one for outgoing characters. In other words, the memory regions on which the two methods \lhs{wrtStrIO} and \lhs{rdStrIO} operate are isolated.  
We envisage that Separation Logic and LiquidHaskell can be exploited for automated verification of such isolation properties. This will open a new playground for formal verification benchmarks by a novel class of problems: automatic verification of policy-coherence.

\subsection{Example: \lhs{TVars} as \lhs{PSMVars}}
\label{sec:mvars}

It is known how to encode the \lhs{TVar} reference types from \lhs{IO} in terms of \lhs{STM}.
Here we embed them into \psm and call them \lhs{PTVar}.
The raw memory structure behind \lhs{PTVar} is a \lhs{TVar} that holds a value of some type \lhs{a}:
\begin{spec}
data DynTVar a = DynTVar a
\end{spec}
\begin{spec}
instance PSMType (DynTVar a) where 

  type PSMStatic (DynTVar a) = a
  data PSMState (DynTVar a) = DynTVarState (TVar a)
\end{spec}
A \lhs{PTVar a} is initialised with an arbitrary value of type \lhs{a}, the \lhs{tick} and \lhs{term} methods are trivial. They have no effect. 
\begin{spec}
  init a = do
   mustRef <- atomically $ newTVar a
   return (DynTVar mustRef)

  tick _ = return ()
  term _ = return ()
\end{spec}
The methods are \lhs{readPTVar} with index \lhs{MethodId 1} and \lhs{writePTVar} with index \lhs{MethodId}:
\begin{spec}
  data instance Method (DynTVar a) b c where 
    ReadPTVar  :: Method (DynTVar a) () a
    WritePTVar :: Method (DynTVar a) a ()
\end{spec}
The policy makes \lhs{readPTVar} wait for \lhs{writePTVar} and the latter concurrently conflict with itself:
\begin{spec}
  prec _ (MethodId 1 _) = return $ Just [MethodId 2] -- readPTVar MethodId 1
  prec _ (MethodId 2 _) = return $ Just [MethodId 2] -- writePTVar MethodId 2

  methodId ReadPTVar  = MethodId 1
  methodId WritePTVar = MethodId 2
\end{spec}
What we call \lhs{PTVar} are the \psm-scheduled version of \lhs{DynTVar}, i.e.,
\begin{spec}
type PTVar a = PSMVar (DynTVar a)
\end{spec}
The methods \lhs{newPTVar}, \lhs{ReadPTVar} and  \lhs{WritePTVar} are specified as follows:
\begin{spec}
newPTVar :: a -> (PTVar a -> PSM b) -> PSM b  
newPTVar a = newPSMVar a 

  method ReadPTVar (DynTVarState tvarRef) _ = readTVar tvarRef
  method WritePTVar (DynTVarState tvarRef) = writeTVar tvarRef

readPTVar :: PTVar a -> PSM a
readPTVar tvar = runSTMMtd ReadPTVar (method ReadPTVar) tvar (PSMVal ())

writePTVar :: PTVar a -> PSMVal a -> PSM ()
writePTVar = runSTMMtd WritePTVar (method WritePTVar)
\end{spec}

\ignore{

\begin{code}
data DynTVar a = DynTVar a
 
instance PSMType (DynTVar a) where 

  type PSMStatic (DynTVar a) = a
  data PSMState (DynTVar a) = DynTVarState (TVar a)

  data instance Method (DynTVar a) b c where 
    ReadPTVar  :: Method (DynTVar a) () a
    WritePTVar :: Method (DynTVar a) a ()

  init a = do
   mustRef <- atomically $ newTVar a
   return (DynTVarState mustRef)

  tick _ = return ()
  term _ = return ()

  prec _ (MethodId 1 _) = return $ Just [MethodId 2 "wr"]
  prec _ (MethodId 2 _) = return $ Just [MethodId 2 "wr"]
    
  methodId ReadPTVar  = MethodId 1 "rd"
  methodId WritePTVar = MethodId 2 "wr"

  method ReadPTVar (DynTVarState tvarRef) _ = readTVar tvarRef
  method WritePTVar (DynTVarState tvarRef) v = writeTVar tvarRef v


type PTVar a = PSMVar (DynTVar a)

readPTVar :: PTVar a -> PSM a
readPTVar tvar = runSTMMtd ReadPTVar (method ReadPTVar) tvar (PSMVal ())

writePTVar :: PTVar a -> PSMVal a -> PSM ()
writePTVar = runSTMMtd WritePTVar (method WritePTVar)

newPTVar :: a -> (PTVar a -> PSM b) -> PSM b  
newPTVar a = newPSMVar "anonymous" a 

\end{code}
}

\subsection{Updateable \lhs{MVars} as \lhs{PSMVars}}
\label{sec:pmvars}

\begin{figure}[ht]
 \begin{subfigure}[l]{0.47\linewidth}
 \begin{center}
 \includegraphics[scale=0.5]{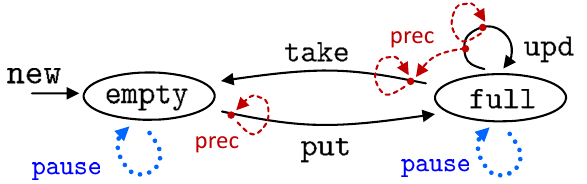}
 \end{center}
 \caption{Two-state \lhs{PMVar} policy.} 
  \label{fig:pmvar-policy-a}
 \end{subfigure}
\begin{subfigure}[l]{0.47\linewidth}
 \begin{center}
  \begin{spec}
   newPMVar  :: (PMVar a->PSM b)->PSM b
   updPMVar  :: PMVar a->PSMVal (a->a)->PSM ()
   putPMVar  :: PMVar a->PSMVal a->PSM ()
   takePMVar :: PMVar a->PSM a
  \end{spec}
 \end{center}
 \caption{\lhs{PMVar} method interface} 
  \label{fig:pmvar-methods-a}
 \end{subfigure}
\end{figure}

Although \lhs{MVars} implement synchronisation this is not enough for determinacy. 
We still need the single-reader and single-writer requirement.
{\lhs{MVars} need the single-reader and single-writer requirement for determinacy. The update method, if permitted, creates a race with reading the \lhs{MVar} which must be ordered by precedence.} 
Adding the \psm policy interface seen in Figs.~\ref{fig:pmvar-policy-a} and~\ref{fig:pmvar-methods-a} generates a race-free version which we call a \lhs{PMVar}.
A \lhs{PMVar} has two states, \lhs{empty} and \lhs{full}. A fresh \lhs{x::PMVar} is initially empty. \lvsv{Whenever it is empty it}{It} can be filled with a value \lhs{v::a} via method \lhs{putPMVar x v}. When it is full, the value can be update with \lhs{updPMVar}, or extracted by \lhs{takePMVar x} which empties the variable again. Like for regular Haskell \lhs{MVars}, the admissibility of methods in the policy make sure that reading will block until a value is available and writing will block until \lhs{x} is empty. 
Obviously, a race occurs when two concurrent threads aim to put a value into an empty cell or 
both want to read from a full cell.  
\lvsv{To eliminate the races we can give \lhs{putPMVar} and \lhs{takePMVar} precedence over themselves.}{To eliminate write-write and read-read races we give all methods \lhs{putPMVar}, \lhs{takePMVar} and \lhs{updPMVar} precedence over themselves.} 
This 
protects the \lhs{PMVars} from write-write and read-read races.
The precedences have the effect that concurrent accesses such as \lhs{putPMVar x v1 ||| putPMVar x v2} and \lhs{takePMVar x ||| takePMVar x} will block each other from creating non-deterministic behaviour. 
In this fashion we replace non-determinism by blocking. 
Since this is robustly observable it may be a better choice when predictability is paramount.
When integrity is more important than availability, the forced deadlock is preferable than unpredictable behaviours. Deadlock analysis is a standard problem for which analysis tools are available, and a routine part of static program analysis in synchronous programs. On the other hand the standard way to detect non-determinism is by trial and error.
The race between \lhs{updPMVar} and \lhs{takePMVar} is resolved by giving the former precedence over the latter.   
Hence in a parallel \lhs{takePMVar x ||| updPMVar x f}, the update will be executed first. 
Both these features of \lhs{PMVar}s properly extend the expressiveness over regular \lhs{MVar}s in \stm.  
\onlylv{
As an aside, note that Haskell's \lhs{IVars}, which can only be written once, are obtained if we make the \lhs{take} transition in Fig.~\ref{fig:mvar-policy} point back to state \lhs{full} and remove the ``prec'' self-loop on it.
}

Let us see how the policy-wrapped version \lhs{PMVar} of updateable \lhs{MVars}
is obtained in \psm.
Recall the method interface from Fig.~\ref{fig:pmvar-methods-a} and the policy from Fig.~\ref{fig:pmvar-policy-a}.
First, we define the core type \lhs{PMVar_ a} for our memory objects that we want to become an instance of the \lhs{PSMVar} class:
\lstset{numbers=left,numbersep=5pt}
\begin{code} 
data PMVar_ a = PMVar_ a 

instance PSMType (PMVar_ a) where 

  type PSMStatic (PMVar_ a) = ()
  data PSMState (PMVar_ a) = PMVarState (TVar (Maybe a)) 

  init _ = fmap PMVarState $ atomically $ newTVar Nothing
  tick _ = return ()
  term _ = return ()

  data instance Method (PMVar_ a) b c where 
    TakePMVar :: Method (PMVar_ a) () a
    PutPMVar  :: Method (PMVar_ a) a ()
    UpdPMVar  :: Method (PMVar_ a) (a -> a) ()

  methodId TakePMVar = MethodId 1 "take"
  methodId PutPMVar  = MethodId 2 "put"
  methodId UpdPMVar  = MethodId 3 "upd"

  prec (PMVarState mustRef) (MethodId 1 _) = do    -- takePMVar = MethodId 1
    must <- readTVar mustRef
    case must of 
      Just _  -> return (Just [MethodId 1 "take", MethodId 3 "upd"])
      Nothing -> return Nothing

  prec (PMVarState mustRef) (MethodId 2 _) = do    -- putPMVar MethodId 2
    must <- readTVar mustRef
    case must of 
      Just _  -> return Nothing
      Nothing -> return (Just [MethodId 2 "put"]) 

  prec (PMVarState mustRef) (MethodId 3 _) = do    -- updPMVar = MethodId 3
    must <- readTVar mustRef
    case must of 
      Just _  -> return (Just [MethodId 3 "upd"])
      Nothing -> return Nothing 

  method TakePMVar (PMVarState rVarRef) _ = do
      rvar <- readTVar rVarRef
      case rvar of 
        Nothing -> error "panic: taking empty PMVar"
        Just v -> do 
         writeTVar rVarRef Nothing
         return v

  method PutPMVar (PMVarState rVarRef) v = do
      rvar <- readTVar rVarRef
      case rvar of 
        Nothing -> writeTVar rVarRef (Just v)
        Just v  -> error "panic: putting into full PMVar"

  method UpdPMVar (PMVarState rVarRef) f = do
      rvar <- readTVar rVarRef
      case rvar of 
        Nothing -> error "panic: updating empty PMVar" 
        Just v  -> writeTVar rVarRef (Just $ f v)
\end{code}
The contents of a \lhs{PMVar} are stored in a \lhs{TVar} which may be \lhs{Nothing} when the \lhs{PMVar} is empty or \lhs{Just v} if it is full holding value \lhs{v::a} (Line 6). The global methods \mi{init}, \mi{tick} and \mi{term} are trivial (Lines~8--10). 
\lhs{PMVars} are initialised as empty, so the static type \lhs{PSMStatic (PMVar_ a)} from which \lhs{newPMVar} will instantiate is the trivial type \lhs{()} (line~5) and the \lhs{init} function creates an empty cell (line~8). The \lhs{tick} and \lhs{term} actions have no effect on a \lhs{PMVar}. 

\smallskip 

The methods are \lhs{takePMVar} indexed by \lhs{MethodId 1}, \lhs{putPMVar} by \lhs{MethodId 2} and \lhs{updPMVar} by \lhs{MethodId 3} (Lines~17--19) with input-output types as specified in Lines~13--15.

\smallskip 

Next we define the policy for our methods, in Lines~21--37. 
Notice that the precedence relation \lhs{prec} introduces a precedence loop each for \lhs{MethodId 1} (line~24) and \lhs{MethodId 3} (line~36) when the \lhs{TMVar_} is full and for \lhs{MethodId 2} (line~31) when it is empty. In state \lhs{full} also \lhs{MethodId 3} takes precedence over \lhs{MethodId 1} (line~24). In the other cases, the methods are not admissible.

\medskip 

The last step is to install the methods.  
This works analogously for both methods, so we only describe the definitions for \lhs{takePMVar}.
The generic function \lhs{method} that implements the  \lhs{TakePMVar} action is seen in Lines~40--46.
It reads the contents of the memory which is called \lhs{rVarRef} in line~41 and returns the content if it is non-empty. The case that it is empty throws a (data corruption) error. This should never happen, because the action \lhs{takePMVar} is not admissible in this case. 
Finally, we use the generic \lhs{method} function from the \lhs{PSMVarMtd} class to export the methods under \psm control:
\lstset{numbers=none}
\begin{code}
type PMVar a = PSMVar (PMVar_ a)

takePMVar :: PMVar a -> PSM a
takePMVar rvar = runSTMMtd TakePMVar (method TakePMVar) rvar (PSMVal ())

putPMVar :: PMVar a -> PSMVal a -> PSM ()
putPMVar rvar v = runSTMMtd PutPMVar (method PutPMVar) rvar v

updPMVar :: PMVar a -> PSMVal (a -> a) -> PSM ()
updPMVar rvar f = runSTMMtd UpdPMVar (method UpdPMVar) rvar f

newPMVar :: String -> (PMVar a -> PSM b) -> PSM b  
newPMVar name = newPSMVar name () 
\end{code}

Let us look at the coherence requirement: What is the library designer supposed to verify? 
The scheduling precedences \lhs{prec} between the methods declared with the \lhs{PSMType} instance declaration above, establishes the scheduling policy. In particular, it decrees that \lhs{takePMVar} takes precedence over itself when the variable is full, i.e., the state \lhs{must} of the PSMVar is of the form \hlst{Just _}. In the notation of Sec.~\ref{sec:psm-scheduling} we have \hlst{Just _} $\Vdash$ \lhs{takePMVar} $\prec$ \lhs{takePMVar}.
This has the effect that the PSM scheduler will refuse to execute \lhs{takePMVar} concurrently with itself.  
This is necessary since \lhs{takePMVar} 
does not commute with itself: if we execute \lhs{takePMVar} on state \hlst{Just _} it will take the value and empty the \lhs{PMVar}. Any other concurrent thread also executing \lhs{takePMVar} would now be blocked. If the programmer had left out the self-precedence in the policy, writing 
\begin{center} 
  \hlst{Just _ -> return (Just [MethodId 1 "take", MethodId 3 "upd"])}
\end{center}
in line~24 of the instance declaration for \hlst{PSMType (PMVar_ a)}, then the method \lhs{takePMVar} would be concurrently independent from itself, i.e., \hlst{Just _} $\Vdash$ \lhs{takePMVar} $\indep{}$ \lhs{takePMVar}.
In that case, the requirement of policy coherence would oblige the programmer to verify that any \emph{two} applications of \lhs{takePMVar} commute; Specifically, that after the first application of \lhs{takePMVar} another application remains admissible. The attempt to verify that \lhs{method TakePMVar} (as implemented in lines~40--46 is isolated from itself would not be possible in this case. In synchronous programming the compiler checks this statically and will refuse to compile.

\subsection{Example: Esterel Temporary Signals as PSMVars}
\label{sec:tmp-sig}

Another important type of \lhs{PSMVar}s
are Esterel-style temporary signals, as described in Sec.~\ref{sec:psmvars-assortment}.
Each signal has a static and a dynamic part. The static information
consists of a default value for initialisation and a
commutative and associative combination function. The dynamic
part is a \lhs{TVar} memorising a value and a presence status:
\begin{code}
data StatSig a = StatSig a (a -> a -> a)
\end{code}
A special instance of a signal could use $0$ for initial value and \lhs{+} as the combination functions. 
\begin{code}
zpSigInit :: StatSig Int
zpSigInit = StatSig 0 (+)
\end{code}
The signals are called \emph{temporary} because their value computed in
a macro-step is not preserved for the next macro-step. Instead, the default value is used.
This models the full valued temporary signals of Esterel V7 
which are useful to model hardware in which the values are
stored in wires, and hence do not need to be memorised across macro-steps. The type
\begin{code}
newtype TSig a  = TSig a
\end{code}
is the raw data type for signals encapsulating only the 
memory structure needed to manipulated signals (the \emph{must} information).
For determinate scheduling we must add prediction (the \emph{can} 
information).
The signal variables then have the following type:   
\begin{code}
type PTSig a = PSMVar (TSig a)
\end{code}

In order for \lhs{TSig a} to be useable for predictive scheduling we must make
it an instance of the \lhs{PSMType} class.
For this we identify the methods to be put under policy control.
The policy distinguishes four method indices,
called \lhs{MethodId 1} (\lhs{DInit}), \lhs{MethodId 2} (\lhs{DEmit}), \lhs{MethodId 3} (\lhs{DPres}) and \lhs{MethodId 4} (\lhs{DRead}).
These methods are locking each other in a way specific to the signal type, implementing the policy in Fig.~\ref{fig:ptsig-methods}.

\begin{spec}

instance PSMType (TSig a) where

  type PSMStatic (TSig a) = StatSig a
  data PSMState (TSig a)  = TSigState a (a -> a -> a) (TVar (a, Bool))

  init (StatSig v0 cmb) = do
    mustRef <- atomically $ newTVar (v0, False) -- $
    return $ TSigState v0 cmb mustRef  -- $

  tick (TSigState def _ mustRef) =
    atomically $ writeTVar mustRef (def, False) -- $

  term _ = return ()

  data instance Method (TSig a) b c where 
    DInit :: Method (TSig a) a ()
    DEmit :: Method (TSig a) a ()
    DPres :: Method (TSig a) () Bool
    DRead :: Method (TSig a) () a

  methodId DInit = MethodId 1
  methodId DEmit = MethodId 2
  methodId DPres = MethodId 3
  methodId DRead = MethodId 4

  prec _ (MethodId 1 _) = return $ Just [MethodId 1]
  prec _ (MethodId 2 _) = return $ Just [MethodId 1]
  prec (TSigState _ _ mustRef) (MethodId 3 _) = do
    (_, status) <- readTVar mustRef
    return $ if status then Just [MethodId 1] 
                       else Just [MethodId 1, MethodId 2]
  prec _ (MethodId 4 _) = return $ Just [MethodId 1, MethodId 2]


\end{spec}

\ignore{
\begin{code}

instance PSMType (TSig a) where

  type PSMStatic (TSig a) = StatSig a
  data PSMState (TSig a)  = TSigState a (a -> a -> a) (TVar (a, Bool))

  data instance Method (TSig a) b c where 
    DInit :: Method (TSig a) a ()
    DEmit :: Method (TSig a) a ()
    DPres :: Method (TSig a) () Bool
    DRead :: Method (TSig a) () a

  init (StatSig v0 cmb) = do
    mustRef <- atomically $ newTVar (v0, False) -- $
    return $ TSigState v0 cmb mustRef  -- $

  tick (TSigState def _ mustRef) =
    atomically $ writeTVar mustRef (def, False) -- $

  term _ = return ()

  prec _ (MethodId 1 _) = return $ Just [MethodId 1 "init"]
  prec _ (MethodId 2 _) = return $ Just [MethodId 1 "init"]
  prec (TSigState _ _ mustRef) (MethodId 3 _) = do
    (_, status) <- readTVar mustRef
    return $ if status then Just [MethodId 1 "init"] 
                       else Just [MethodId 1 "init", MethodId 2 "emit"]
  prec _ (MethodId 4 _) = return $ Just [MethodId 1 "init", MethodId 2 "emit"]

  methodId DInit = MethodId 1 "init"
  methodId DEmit = MethodId 2 "emit"
  methodId DPres = MethodId 3 "pres"
  methodId DRead = MethodId 4 "read"

  method DInit (TSigState _ _ tmust) v = 
      writeTVar tmust (v, False)

  method DEmit (TSigState _ cmb tmust) v = do
      (val, _) <- readTVar tmust
      writeTVar tmust (cmb val v, True)

  method DPres (TSigState _ cmb mustRef) v = do
      (_, stat) <- readTVar mustRef
      return stat

  method DRead (TSigState _ _ tmust) _ = do
      (val, _) <- readTVar tmust
      return val

\end{code}
} 

The four method indices \lhs{MethodId 1}, \lhs{MethodId 2}, \lhs{MethodId 3} and
\lhs{MethodId 4} are associated with four methods. The implementation of these methods are presented
in the sequel. 

\paragraph{Signal Initialisation.}
\label{sec:sig-init}

The first method is \lhs{DInit} with index \lhs{MethodId 1}. It initialises
the signal in each tick. This is useful if the static default
value is to be overridden. 
\begin{spec}
method DInit (TSigState _ _ tmust) v = 
      writeTVar tmust (v, False)
\end{spec}
The \lhs{Init} method is exported for the
\lhs{PSM} context with the generic \lhs{runSTMMtd} wrapper:
\begin{code}
initPTSig :: PTSig a -> PSMVal a -> PSM ()
initPTSig = runSTMMtd DInit $ method DInit 
\end{code}

\paragraph{Signal Emission.}
\label{sec:sig-emit}

The signal emission type has the method index \lhs{MethodId 2} and is called \lhs{DEmit}. When this method is called a value is emitted on the signal and its status is set present.
\begin{spec}
  method DEmit (TSigState _ cmb tmust) v = do
      (val, _) <- readTVar tmust
      writeTVar tmust (cmb val v, True)
\end{spec}

\begin{code}
emitPTSig :: PTSig a -> PSMVal a -> PSM ()
emitPTSig = runSTMMtd DEmit $ method DEmit 
\end{code}

\paragraph{Present Test.}
\label{sec:sig-pres}

The method with index \lhs{MethodId 3} is the present test, called \lhs{DPres}. It returns the boolean status of the signal. This method must wait for all concurrent initialisations or emissions.
\begin{spec}
  method DPres (TSigState _ cmb mustRef) v = do
      (_, stat) <- readTVar mustRef
      return stat
\end{spec}

\begin{code}
presPTSig :: PTSig a -> PSM Bool
presPTSig sig = runSTMMtd DPres (method DPres) sig (PSMVal())
\end{code}

\paragraph{Signal Reading.}
\label{sec:sig-read}

The value of a signal is obtained by the method \lhs{DRead} with index \lhs{MethodId 4}. Like the present test \lhs{DPres} is must wait for all concurrent \lhs{Dinit} and \lhs{Demit}.

\begin{spec}
  method DRead (TSigState _ _ tmust) _ = do
      (val, _) <- readTVar tmust
      return val
\end{spec}

\begin{code}
readPTSig :: PTSig a -> PSM a
readPTSig sig = runSTMMtd DRead (method DRead) sig (PSMVal ())
\end{code}

\paragraph{Example.}
\label{sec:esterel-example}

\begin{figure}[ht]
  \begin{center}
    \includegraphics[scale=0.7]{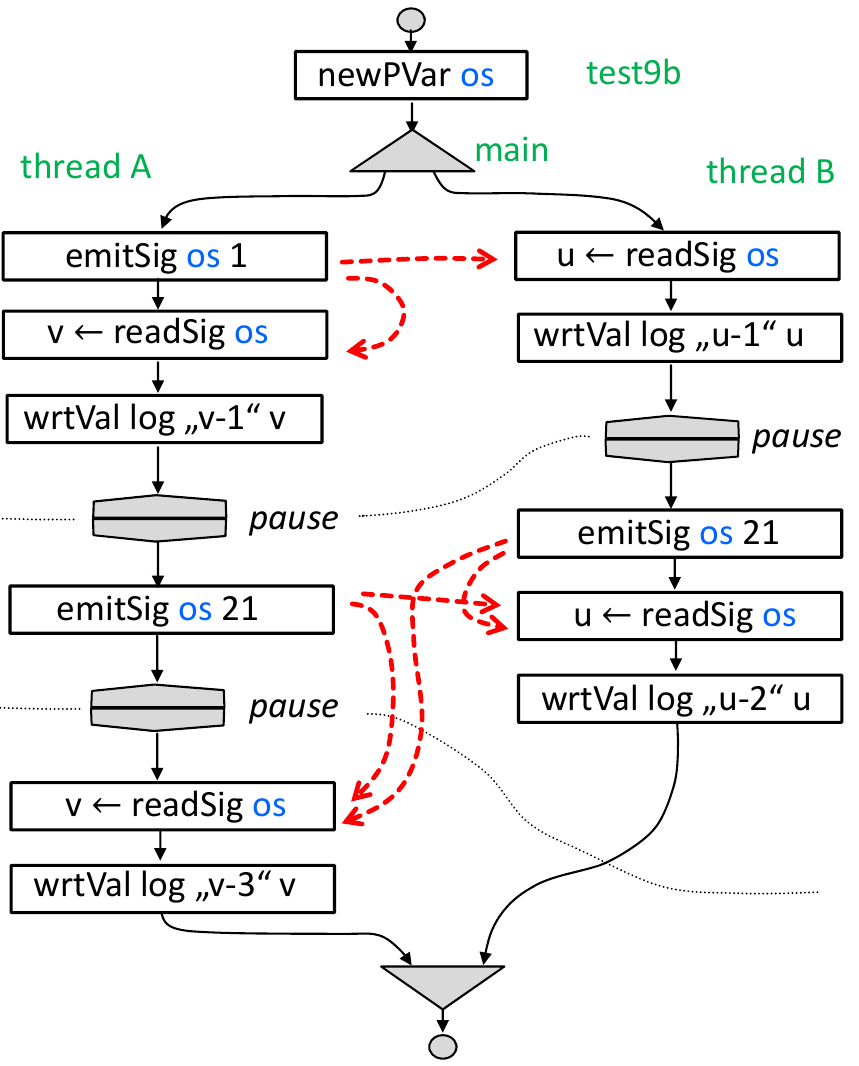}
  \end{center} 
\caption{Example \psm program using Esterel temporary signals.}
\label{fig:esterel-example}
\end{figure}

Consider example program illustrated in Fig.~\ref{fig:esterel-example}. 
It is a parallel composition of two worker threads \lhs{A} and \lhs{B} that communicate with each other via a shared (temporary) Esterel signal \lhs{os::PTSig Int} carrying integer values. 
The \psm code is as follows:
\lstset{numbers=left,numbersep=5pt}
\begin{code}
test9b = runStrIO $ \log ->
 newPSMVar "os" zpSigInit $ \ os -> 
  xfork "main"
    ( -- Thread A
      emitPTSig os (pure 1) >>>
      readPTSig os >>>= \ v -> 
      wrtVal log "v-1" v >>>
      pause >>>
      emitPTSig os (pure 21) >>>
      pause >>>
      readPTSig os >>>= \ v -> 
      wrtVal log "v-3" v
    )
    ( -- Thread B
      readPTSig os >>>= \ u ->
      wrtVal log "u-1" u >>>
      pause >>>
      delay 1000000 >>>
      emitPTSig os (pure 21) >>>
      readPTSig os >>>= \ u ->
      wrtVal log "u-2" u
    ) 
\end{code}
\lstset{numbers=none}

The thread \lhs{A} lives through $3$ and thread \lhs{B} lives for $2$ macro-cycles which are separated by the \lhs{pause} statements in the control flow, in lines~8 and~10 for \lhs{A} and line~17 for \lhs{B}. These clock barriers are illustrated by a grey boxes in the control-flow graph. 

In the first cycle, \lhs{A} emits value $1$ on signal \lhs{os}, then reads the value of \lhs{os} into the value variable $v$ and prints it to the IO interface object \lhs{log::PStrIO} (see Sec.~\ref{sec:IO}) that is introduced by the \lhs{runStrIO} wrapper function. The output logging is encapsulated in the auxiliary function 
\begin{code}
-- logging a value
wrtVal :: Show a => PStrIO -> String -> a -> PSM ()
wrtVal log s n = 
  let line = s ++ " = " ++ show n ++ " " in 
    wrStrIO log (pure line)
\end{code}
which also prints a message string. 
After the pause, in the second macro step, the left thread \lhs{A} only emits value $21$ to the signal \lhs{os} and pauses. In the third macro step, thread \lhs{A} only reads the value of \lhs{os} and terminates.
The right thread \lhs{B} is active for only two cycles: First, it reads \lhs{os} and logs its values and in the second macro step it emits value $21$ and also reads and logs.  

As can be noted, in this scenario we have a dynamically varying number of readers and writers on the shared signal \lhs{os}. In the first tick we have one writer \lhs{A} while both threads read; In the second tick we have two writers and only one reader \lhs{B}; in the third tick there is no no writer and one reader \lhs{A}. The data flow of read and write accesses are visualised by thick dashed arrows in Fig.~\ref{fig:esterel-example}.

Now if we look at the output generated by this program (under the control of \lhs{runStrIO}, we see the following output:
\begin{spec} 
  *PSM> test9b   
  Tick 0 @ Port IO: v-1 = 1 u-1 = 1 
  Port IO>>
  Tick 1 @ Port IO: u-2 = 42 
  Port IO>>
  Tick 2 @ Port IO: v-3 = 0 
  Port IO>>
  Tick 3 @ Port IO: 
  Port IO>> :q
  ThreadId 71
\end{spec} 
As can be noted, in the first \lhs{Tick} both threads receive the same value $1$ which is the value emitted by \lhs{A}. In the second \lhs{Tick} thread \lhs{B} reads value $42$ which is the combined value of $21$ emitted by \lhs{A} and a value of $21$ emitted by \lhs{B}. Recall that the signal \lhs{os} is created with the initialisation constant \lhs{zpSigInit} which defines addition as the  combination function and $0$ as the default value. This default value is then what is read by thread \lhs{A} in the third \lhs{Tick} where there is no signal emission on  \lhs{os}.      

Notice that we have added a timing constant of \lhs{1000000} to delay the emission of $2$ by \lhs{B} in the second tick. This verifies that that the reading of the signal by \lhs{A} is indeed waiting for \emph{all} signal emissions to have taken place before it gets access to the value.  

\ignore{

\begin{code}
test9c = runStrIO $ \log ->
 newPSMVar "os" zpSigInit $ \os -> 
  xfork "main"
    (
      pause >>>
      -- delay 200000 >>>
      readPTSig os >>>= \v -> 
      wrtVal log "v-2" v
    )
    ( 
      pause >>>
      delay 100000 >>>
      emitPTSig os (pure 21)
    ) 

{-
*PSM13> test9c
>> 
Tick 0: 
>>      
Tick 1: v-2 = 21 
>>
Tick 2: 
>>      
Tick 3: 
>>      
Tick 4: 
>> :q
-}

\end{code}

}

\ignore{%

\section{Ancillary Code}
\label{appendix:partD}

\begin{code}

-- for debugging only: clear all predictions to unblock
unsafeClearPred :: PSMType a => PSMVar a -> PSM()
unsafeClearPred (Idle sId) = PSM (predx, 0, act) where
  predx = fromList [(CmplId 0, empty)]
  act _ _ _ = error $ "unsafeClearPred: cannot execute on idle pvar" ++ sId

unsafeClearPred (Active vId _ _ predMapRef) = 
  PSM (predx, 0, act) where
    predx = fromList [(CmplId 0, empty)]
    act ioRef ctxRef cntPredx = do
          ctx <- atomically $ readTVar ctxRef
          let (TTreeNode (tId, _, _, _, pVarGlobMapRef, pVarLocMapRef) _ _ _ _) = ctx
          pVarGlobMap <- atomically $ readTVar pVarGlobMapRef
          pVarLocMap  <- atomically $ readTVar pVarLocMapRef
          (_, tIds) <- atomically $ 
            collectSpine (\ (tId, _, _, _, _, _) -> tId) ctxRef
          let adjustPSMVarPred vId tId = do
                AnyPSMVar (Active _ _ _ predMapRefx) <- atomically $ readTVar (pVarGlobMap!vId) 
                predMapx <- atomically $ readTVar predMapRefx
                let clear = Map.map (\x -> 0)
                    decPredx thrd (pMap, pMapx) = 
                       if thrd == tId then (clear pMap, clear pMapx) 
                       else (pMap, pMapx) 
                atomically $ writeTVar predMapRefx $
                 Map.mapWithKey decPredx predMapx
          mapM_ (adjustPSMVarPred vId) tIds

\end{code}

\ignore{

\begin{spec}
bindVarOrig :: PSMType a => MethodId -> PSM(a -> b) -> PSMVar a -> PSM b
bindVarOrig mId (PSM(predx, trp, fact)) (Active vId _ pvar predMapRefx) =
  PSM (predx', trp, act') where
    pred = Map.foldr (unionWith addPred) empty $ delete (CmplId 2) predx
    predx' = empty
    act' ioRef ctxRef cntPredx = do
      -- check enabledness (thrdPath must not change)
      thrdPath <- atomically (waitForEnabled mId vId pvar predMapRefx ctxRef)
      -- execute function call
      -- WARNING: this is only getting the function, not executing it.
      f <- fact ioRef ctxRef cntPredx
      -- adjust prediction       
      -- WARNING: should be after the function call to pvar!
      let !x = f pvar in do
        atomically $ do -- $ 
          predMapx <- readTVar predMapRefx
          let canDnx (pMap, pMapx) = Just $ 
                (adjust dec mId pMap, pMapx)
              -- if we are inside a loop nest do not decrement
              -- the decrement is done by the loop exit
              dec = \x -> x - 1
              -- nothing for the moment on extended predictions FIX 
              -- decPredx _ = id
          writeTVar predMapRefx $ -- $
            Prelude.foldr (update canDnx) predMapx thrdPath
        -- return method result
        return x -- $ f pvar
bindVarOrig _ _ (Idle sId) = PSM (empty, 0, act) where
    act _ _ _ = error $ "bindVar: cannot execute Idle argument" ++ sId

\end{spec}

\begin{code}

bindVarXX :: PSMType a => Method a b c -> PSM((PSMState a) -> b) -> PSMVar a -> PSM b
bindVarXX mtd f =
  bindVar mtd (callVal f)

bindVarY :: PSMType a => Method a b c -> 
   PSM (PSMState a -> b -> IO d) -> PSMVar a -> PSMVal b -> PSM d

bindVarY mtd (PSM(predx, trp, fact)) 
      (Active vId _ pvar predMapRefx) (PSMVal v) =
  PSM (predx', trp, act') where
    pred' = empty
    predx' = empty
    act' ioRef ctxRef cntPredx = do
      f <- fact ioRef ctxRef cntPredx
      -- check enabledness
      thrdPath <- atomically 
        (waitForEnabled mtd vId pvar predMapRefx ctxRef) 
      -- execute function call
      x <- f pvar v
      -- adjust prediction       
      atomically $ do -- $ 
        predMapx <- readTVar predMapRefx
        let canDnx (pMap, pMapx) = Just $ 
                (adjust dec (methodId mtd) pMap, pMapx) -- (adjust dec mId pMap, pMapx)
              -- if we are inside a loop nest do not decrement
              -- the decrement is done by the loop exit
            dec x = x - 1
        writeTVar predMapRefx $ -- $
          Prelude.foldr (update canDnx) predMapx thrdPath
        -- return method result
        return x

bindVarY _ _ (Idle sId) _ = PSM (predx, 0, act) where
  predx = fromList [(CmplId 0, empty)]
  act _ _ _ = error $ "bindVar: cannot execute Idle argument" ++ sId -- $

\end{code}

} 

\begin{code}

type Prediction = Map MethodId Int
type PSMVarPred = Map CmplId Prediction

showPotential :: Prediction -> String
showPotential pVarPred = 
  snd (Map.foldlWithKey assemble (False, "[") pVarPred)  ++ "]" where
    assemble (comma, s) c p = (comma || True, 
      s ++ (if comma then "," else "") ++ show c ++ "/" ++ show p) 

show2Potential :: (Prediction, Prediction) -> String 
show2Potential (potCnt, predMap2) = 
  showPotential potCnt ++ "surf, " ++ showPotential predMap2 ++ "rec"

showPSMVarPred :: PSMVarPred -> String 
showPSMVarPred predMap = 
  snd (Map.foldlWithKey assemble (False, "[") predMap) ++ "]" where 
    assemble (comma, s) c p = (comma || True, 
     s ++ (if comma then ", " else "") ++ showPotential p ++ show c)

\end{code}

\begin{code}

printPred s ctxRef pred = do
  TTreeNode (tId,_,_,_,_,_) _ _ _ _ <- atomically $ readTVar ctxRef 
  print $ show tId ++ " prediction " ++ s ++ ": " ++ show pred

-- log all current predictions registered in pvars
prtAllPSMVarPred :: TVar String -> String -> TVar ThrdEnv -> IO ()
prtAllPSMVarPred ioRef s tvt = do 
  vt <- atomically $ readTVar tvt
  let TTreeNode (tId, _, _, _,pVarGlobMapRef, pVarLocMapRef) _ _ _ _ = vt
  pVarGlobMap <- atomically $ readTVar pVarGlobMapRef
  let pVars = Map.keys pVarGlobMap
      prtPred vId = do 
        AnyPSMVar (Active _ name _ predMapRef) <- atomically $ readTVar (pVarGlobMap ! vId)
        predMap <- atomically $ readTVar predMapRef
        io <- atomically $ readTVar ioRef      
        atomically $ writeTVar ioRef $
          io ++ "\n" ++ s ++ " Predictions for " 
          -- ++ name ++ " = " 
          ++ show vId ++ " = " ++ show predMap 
  mapM_ prtPred pVars 


-- dump string into log
logStr :: TVar String -> String -> STM ()
logStr ioRef s = do 
  io <- readTVar ioRef      
  writeTVar ioRef $ io ++ "\n" ++ s


-- log predictions of local PSMVars only, atomically
logLocPSMVarPreds :: String -> TVar String -> TVar ThrdEnv -> STM()
logLocPSMVarPreds s ioRef ctxRef = do
  ctx <- readTVar ctxRef
  let TTreeNode (tId, _, _, _, pVarGlobMapRef, pVarLocMapRef) _ _ _ _ = ctx
  pVarLocMap <- readTVar pVarLocMapRef
  pVarGlobMap <- readTVar pVarGlobMapRef
  io <- readTVar ioRef
  writeTVar ioRef $ io ++ "\n" ++ 
    "Local PSMVars @" ++ s ++ " by " ++ show tId ++ ": "
  let message = "logLocPSMVarPreds: local PSMVar not in global pVarGlobMap" 
      logPSMVarId vId = do
        pvar <- readTVar (findWithDefault (error message) vId pVarGlobMap)
        let AnyPSMVar(Active _ name _ predMapRefx) = pvar 
        io <- readTVar ioRef 
        predMapx <- readTVar predMapRefx              
        logStr ioRef $ "Predictions for " ++ show vId ++ " = \n   " ++ showThrPredMapxx predMapx
  mapM_ logPSMVarId (Map.keys pVarLocMap)


-- log predictions of global PSMVars, atomically
logGlobPSMVarPreds :: String -> TVar String -> TVar ThrdEnv -> STM ()
logGlobPSMVarPreds s ioRef ctxRef = do
  ctx <- readTVar ctxRef
  let TTreeNode (tId, _, _, _, pVarGlobMapRef, pVarLocMapRef) _ _ _ _ = ctx
  -- pVarLocMap <- readTVar pVarLocMapRef
  pVarGlobMap <- readTVar pVarGlobMapRef
  io <- readTVar ioRef
  writeTVar ioRef $ io ++ "\n" ++ 
    "Global PSMVars @" ++ s ++ " by " ++ show tId ++ ": "
  let message = "logGlobPSMVarPreds: PSMVar not in pVarGlobMap" 
      logPSMVarId vId = do
        pvar <- readTVar (findWithDefault (error message) vId pVarGlobMap)
        let AnyPSMVar(Active _ name _ predMapRefx) = pvar 
        io <- readTVar ioRef 
        predMapx <- readTVar predMapRefx              
        logStr ioRef $ "Predictions for " ++ show vId ++ " = \n   " ++ showThrPredMapxx predMapx
  mapM_ logPSMVarId (Map.keys pVarGlobMap)

\end{code}

\begin{code}

printLog :: String -> String -> Int -> TVar String -> IO () 
printLog header trail cnt ioRef = do
  io <- atomically $ readTVar ioRef
  putStr $ header ++ show cnt ++ ": " ++ io ++ trail
  atomically $ writeTVar ioRef "" -- $

printSLog ::  PSM ()
printSLog = PSM (predx, 0, act) where 
  predx = fromList [(CmplId 0, empty)]
  act logRef _ _ = do
    ts <- atomically $ readTVar logRef
    putStrLn $ "\n LOG: " ++ ts
    atomically $ writeTVar logRef "" -- $

addSLog ::  String -> TVar String -> IO ()
addSLog line ioRef = do
  ts <- atomically $ readTVar ioRef
  atomically $ writeTVar ioRef (ts ++ line)

\end{code}

\begin{code}

-- prints prediction for debugging
xxpause :: String -> PSM () 
xxpause s = PSM (predx, 0, act) where
  predx = fromList [(CmplId 1, empty)]
  -- we register sink path with empty prediction
  act ioRef ctxRef cntPredx = do 
    -- surface behaviour
    ctx <- atomically $ readTVar ctxRef -- $
    case ctx of 
      TTreeNode (_, _, _, _, _, pVarLocMapRef) _ _ _ _ -> do
        pVarLocMap <- atomically $ readTVar pVarLocMapRef -- $
        let -- obtain the prediction for the next tick 
          upPredx = Map.map (\p -> withoutKeys p (keysSet pVarLocMap)) predx in do
          putStrLn $ s ++ " => " ++ show upPredx
          -- store the next prediction in thread status
          atomically $ writeTVar ctxRef (setStatus ctx (Pausing upPredx))
    -- depth behaviour (wait for start signal)
    start <- atomically $ waitForStartOrKill ctxRef -- $
    if not start then return ()
    else do
      ctx <- atomically $ readTVar ctxRef -- $
      tickPSMVars ioRef ctx
      atomically $ writeTVar ctxRef $ setStatus ctx Running
      return ()

\end{code}

} 

\end{document}

\typeout{get arXiv to do 4 passes: Label(s) may have changed. Rerun}



